\documentclass[twocolumn,useAMS,usenatbib]{mn2e}
\usepackage{graphicx,natbib}
\usepackage{url}
\usepackage{color}
\usepackage{rotating}
\newcommand{\aap}{{Astron.~Astrophys.}}
\newcommand{\apjl}{{Astrophys.~J.~Lett.}}

\newcommand{\apjs}{{Astrophys.~J.~Supp.}}

\newcommand{\zphot}{$z_{\rm phot}$}
\newcommand{\zspec}{$z_{\rm spec}$}
\newcommand{\photoz}{photo-$z$}
\newcommand{\Photoz}{Photo-$z$}

\title{Estimating the Redshift Distribution of Photometric Galaxy Samples II.  
Applications and Tests of a New Method}

\author[Cunha et al.]
{Carlos E. Cunha$^{1,2,3}$\thanks{{\tt cunha@uchicago.edu}},
Marcos Lima$^{2,4,5}$,
Hiroaki Oyaizu$^{1,2}$,
Joshua Frieman$^{1,2,6}$,
\newauthor
Huan Lin$^{6}$
\\
${}^{1}$Department of Astronomy and Astrophysics, University of Chicago, Chicago, IL 60637 \\
${}^{2}$Kavli Institute for Cosmological Physics, University of Chicago, Chicago, IL 60637 \\
${}^{3}$Department of Physics, University of Michigan, 450 Church St., Ann Arbor, MI 48109 \\
${}^{4}$Department of Physics, University of Chicago, Chicago, IL 60637 \\
${}^{5}$Department of Physics and Astronomy, University of Pennsylvania, Philadelphia, PA 19104 \\
${}^{6}$Center for Particle Astrophysics, Fermi National Accelerator Laboratory, Batavia, IL 60510 \\
}

\date{\today}

\begin{document}         
\maketitle

\begin{abstract}

In \cite{lim08} we presented a new method for estimating the redshift
distribution, $N(z)$, of a photometric galaxy sample, using photometric 
observables and weighted sampling from 
a spectroscopic subsample of the data. In this paper, 
we extend this method and explore various applications of it, using 
both simulations of and real data from the SDSS. In addition to 
estimating the redshift distribution for an entire sample, 
the weighting method 
enables accurate estimates of the redshift probability distribution, $p(z)$, for 
{\it each} galaxy in a photometric sample. Use of $p(z)$ in cosmological 
analyses can substantially reduce biases associated with traditional 
photometric redshifts, in which a single redshift estimate is associated 
with each galaxy. The weighting procedure also naturally indicates 
which galaxies in the photometric sample are expected to have accurate 
redshift estimates, namely those that lie in regions of photometric-observable 
space that are well sampled by the spectroscopic subsample. In addition 
to providing a method that has some advantages over standard {\photoz} estimates, 
the weights method can also be used in conjunction {\it with} {\photoz} estimates, e.g., 
by providing improved estimation of $N(z)$ 
via deconvolution of $N(z_{\rm phot})$ and improved 
estimates of {\photoz} scatter and bias.
We present a publicly available p(z) catalog for $\sim 78$ million SDSS DR7
galaxies.
\end{abstract}

\vspace{-0.3in}

\section{Introduction}\label{sec:int}

Optical and near-infrared wide-area surveys planned for the next 
decade will increase the size of photometric galaxy samples by an order 
of magnitude, delivering measurements of billions of galaxies.
Much of the utility of these samples for astronomical and cosmological studies will rest on 
knowledge of the redshift distributions of the galaxies they contain. 
For example, surveys aimed at probing dark energy via clusters, 
weak lensing, and baryon acoustic oscillations (BAO) will rely on the 
ability to coarsely bin galaxies by redshift, enabling approximate 
distance-redshift measurements as well as study of the growth of 
density perturbations. The power of these surveys to constrain 
cosmological parameters will be limited in part by the accuracy 
with which the galaxy redshift distributions can be determined
\citep{hut04,hut06,zha06a,zha06b,ma06,lim07}.

Photometric redshifts---approximate estimates of  
galaxy redshifts based on their broad-band 
photometric observables, e.g., magnitudes or colors---offer one 
set of techniques for approaching this problem. 
However, {\photoz} estimators are typically biased to some degree, 
and they can suffer from catastrophic failures in certain regimes.
These problems motivate the development of potentially more robust methods.

In \cite{lim08} we presented a new, empirical technique aimed not 
at estimating individual galaxy redshifts but instead at estimating 
the  redshift distribution, $N(z)$, for an entire photometric galaxy sample or 
suitably selected subsample. The method is 
based upon matching 
the distributions of photometric observables (e.g., magnitudes, colors, etc) 
of a spectroscopic subsample to those of the photometric sample.
The method assigns weights to galaxies in the spectroscopic 
subsample (hereafter denoted the training set, in analogy with 
machine-learning methods of {\photoz} estimation), such 
that the weighted distributions of observables for these galaxies 
match those of the photometric sample. The weight for 
each training-set galaxy is computed by comparing the local ``density''  
of training-set galaxies
in the multi-dimensional space of photometric 
observables to the density of the 
photometric sample in the same region. 
We estimate the densities using a nearest-neighbor approach that 
ensures that the density estimates are both local and stable in sparsely 
occupied regions of the space. 
The use of the nearest neighbors ensures optimal binning of the data, which 
minimizes the requisite size of the spectroscopic subsample. 
After the training-set galaxy weights are derived, we sum them 
in redshift bins to estimate the redshift distribution for the photometric sample. 

As \cite{lim08} show, this weighting method provides a precise and nearly unbiased 
estimate of the underlying redshift distribution for a photometric sample 
without recourse to {\photoz} estimates for individual galaxies. 
Moreover, the spectroscopic training set does {\it not} have to be 
representative of the photometric sample, 
in its distributions of magnitudes, colors, or redshift, 
for the method to work. (By contrast, the performance of training-set-based {\photoz} estimators 
generally degrades as the training set becomes less representative of the photometric 
sample.) 
The only requirement is that the spectroscopic training set {\it covers}, even 
sparsely, the range of photometric observables spanned by the photometric 
sample. 
The weighting method can be applied to different combinations of photometric 
observables that correlate with redshift---here, 
we confine our analysis to magnitudes and colors. 

In this paper we present additional applications of the weighting method, test 
its performance on simulated data sets, and show results of those applications 
using data from the SDSS. The applications of the weighting method naturally 
fall into two categories, those that enhance {\photoz} estimators and those that 
(potentially) replace {\photoz} estimation. In the first category, 
we show that the weighting method can be used to improve estimates of the 
scatter and bias of training-set-based {\photoz} estimates
as functions of (true) spectroscopic redshift. Knowledge of 
such errors are very important, since uncertainties in  
{\photoz} bias and scatter are nuisance parameters that 
significantly degrade the power of cosmological probes \citep[eg.][]{hut04,ma06,lim07}.
We also show that the weights can be used to obtain improved estimates of the error 
distribution of the {\photoz}'s, $P(z_{\rm phot}|z_{\rm spec})$, and thereby
improve the deconvolution procedure used to infer the underlying redshift 
distribution, $N(z)$, from the distribution of {\photoz}'s  
\citep{pad05}.

In the second category of applications, we consider the weighting technique on its own, 
independently of `traditional' {\photoz} estimates.  
The accuracy of the weighting method in directly reconstructing $N(z)$ is affected by photometric errors  
and by sparse or incomplete coverage by the training set of the space of photometric observables 
spanned by the photometric data. 
We develop and test a bootstrap technique to estimate random errors in the weighted $N(z)$ estimate  
and present a technique for detecting systematic errors in it as well. 
We also discuss the effects of training-set non-representativeness on the $N(z)$ estimate.
Perhaps most importantly, we show that the weighting procedure can be used to estimate not only 
the redshift distribution for the (entire) photometric sample, $N(z)$, but also a 
redshift probability distribution, $p(z)$, for each galaxy in the photometric sample. Such 
a distribution contains much more information than a discrete {\photoz} estimate, {\zphot}.  
Use of $p(z)$ instead of {\zphot} in cosmological analyses 
can potentially greatly 
reduce the biases arising from {\photoz}'s.

The paper is organized as follows.
In \S \ref{sec:wei} we review and extend the weighting 
method for estimating the redshift distribution and the redshift 
probability distribution, focusing in particular on sources and 
estimates of errors in the method. In \S \ref{sec:cat} we describe 
the actual and simulated SDSS galaxy catalogs that we use to test the 
weighting method and its alternatives. We demonstrate how the weighting method 
improves upon photometric-redshift estimates in the mock catalog in 
\S \ref{sec:app1}, and we demonstrate its effectiveness in estimating 
$N(z)$, in comparison with \photoz-based methods, in \S \ref{sec:app2}. 
We apply the new methods to the real SDSS DR6 in \S \ref{sec:res.real}.
We present our conclusions in \S \ref{sec:conclusion} and include 
some technical details of the analysis in the Appendices.

\section{The Weighting Method} \label{sec:wei}

In this section, we briefly review and extend the weighting method 
introduced in 
\cite{lim08}. We define the weight, $w$, of a 
galaxy in the spectroscopic training set 
as the normalized ratio of the density of galaxies in the photometric sample to 
the density of 
training-set galaxies around the given galaxy.
These densities are calculated in a 
local neighborhood in the space of photometric observables, e.g., 
multi-band magnitudes.
More formally, given a training-set galaxy, we define its weight by

\begin{equation}
w\equiv\frac{1}{N_{\rm P,tot}}\frac{\rho_{\rm P}}{\rho_{\rm T}} ~,\label{eqn:wei}
\end{equation}
\noindent where $N_{\rm P,tot}$ is the total number of galaxies 
in the photometric sample,  
and $\rho_{\rm P}$ and $\rho_{\rm T}$ are the local number
densities in the space of observables for the photometric and training sets,
\begin{equation}
\rho_{\rm P,T}\equiv\frac{N_{\rm P,T}}{V_{\rm P,T}}~, \label{eqn:den_m_gen_def}
\end{equation} 
\noindent where $N_{\rm P(T)}$ is the number of photometric (training) set 
galaxies within volume $V_{\rm P(T)}$.

We adopt a nearest-neighbor approach to estimating the density of 
galaxies in magnitude space, because it enables control of 
statistical errors (shot noise) 
while also ensuring adequate ``locality'' of
the volume in magnitude space. We define 
the distance $d_{\alpha\beta}$ in magnitude space 
between the $\alpha^{\rm th}$ and $\beta^{\rm th}$ galaxies in a 
(photometric or spectroscopic) sample using a Euclidean metric,

\begin{eqnarray}
(d_{\alpha\beta})^2 \equiv ({\bf m}_\alpha-{\bf m}_\beta)^2
                       =\sum_{a=1}^{N_m} (m_{\beta}^{a}-m_{\alpha}^{a})^2 \, ,
\label{eqn:dist_gen_def}
\end{eqnarray}

\noindent where $N_m$ denotes the number of magnitudes (i.e., different passbands) 
measured for each galaxy. 
We use this distance to find the set of 
{\it nearest neighbors} to the $\alpha^{\rm th}$ object, i.e., 
the set of galaxies with the smallest $d_{\alpha\beta}$. 
For a fixed number of nearest neighbors
$N_{\rm nei}$, if we order the neighbors by their 
distance from the $\alpha^{\rm th}$ galaxy, then we can define the hypervolume 
in terms of the distance from galaxy $\alpha$ to the $N_{\rm nei}^{\rm th}$ nearest  
neighbor, indexed by $\gamma$, i.e., $V_m = (d_{\alpha\gamma})^{N_m}$. 

Estimating the local density in the spectroscopic training set 
using a fixed value for $N({\bf m}_\alpha)_{\rm T}=N_{\rm nei}$ 
ensures that the density estimate is positive-definite and that the resulting  
weight is well defined. To estimate the corresponding density in the 
photometric sample, we simply count the number of galaxies in the 
photometric sample, $N({\bf m}_\alpha)_{\rm P}$, that occupy the {\it same} 
hypervolume $V_m$ around the point ${\bf m}_\alpha$. 
Since the densities are estimated in the spectroscopic and photometric sets 
using the same hypervolume, the ratio of the densities in Eqn. \ref{eqn:wei} is  
simply the ratio of the corresponding numbers of objects within the volume,  
and the weight for the $\alpha^{\rm th}$ training-set galaxy is 
therefore given by

\begin{eqnarray}
w_{\alpha} =  \frac{1}{N_{\rm P,tot}} 
              \frac{ N({\bf m}_\alpha)_{\rm P} }{ N({\bf m}_\alpha)_{\rm T} } \,.
\label{eqn:wei_def_num} 
\end{eqnarray}
\noindent $N_{\rm nei}$ can be chosen to balance locality, which favors 
small $V_m$, against statistical errors, which favor large $N_{\rm nei}$. 

\subsection{Weights and the Redshift Distribution $N(z)_{\rm P}$}

As shown in \cite{lim08}, by construction the {\it weighted} spectroscopic training set has 
essentially identical distributions of multi-band magnitudes and 
colors as the photometric sample 
from which it is drawn, even though the spectroscopic set is in general not 
representative of the photometric sample. The weighting procedure 
in effect corrects for that non-representativeness, provided the training set 
adequately spans the range of the photometric-observable space covered by the 
photometric sample. Since the weighted training set has identical distributions 
of photometric observables as the photometric sample, it is reasonable to assume that 
the former also provides an accurate 
estimate of the binned redshift distribution of the photometric 
sample, 

\begin{equation}  
N(z)_{\rm wei} \equiv \hat{N}(z_1<z<z_2)_{\rm P} = \sum_{\beta=1}^{N_{\rm T,tot}} w_\beta 
N(z_1<z_\beta<z_2)_{\rm T},
\label{eqn:Nzest}
\end{equation}
\noindent where the weighted sum is over all galaxies
in the training set. \cite{lim08} show that this indeed provides 
a nearly unbiased estimate of the redshift distribution of the 
photometric sample, $N(z)_{\rm P}$, under suitable conditions. 
Examples of this application will be discussed in \S \ref{sec:appdist}.

\subsection{Weights and the Redshift Probability Distribution $p(z)$}
\label{sec:pz}

Although knowledge of the redshift distribution for a photometric sample, $N(z)_{\rm P}$, 
is sufficient for many applications, there are of course instances in which one would like 
redshift information about individual galaxies in the sample. As noted 
in the Introduction, {\photoz} estimators provide one approach to this problem. 
However, {\photoz} estimates are limited by the fundamental assumption that there 
is a functional relationship between the photometric observables and redshift. 
In fact, galaxies occupying a small cell in the space of photometric observables 
will have a range of redshifts. One can therefore associate that cell with 
a redshift probability distribution function (PDF), $p(z|{\rm observables})$. 
The shape of the PDF 
is determined by the choice of observables, the size of the cell, the 
photometric errors, and the range of spectral energy distributions of the galaxies.
If the PDF is narrowly peaked, 
{\photoz} estimates can be both precise (small scatter) and accurate (small 
bias). However, if the distribution is broad, skewed, or multiply peaked, then {\photoz} 
estimates will suffer large scatter, bias, and potentially catastrophic failures.
The ubiquitous positive bias of {\photoz} estimates for low-redshift galaxies 
and negative bias for high-redshift galaxies are consequences of this 
fundamental assumption. Low- and high-redshift objects can in some cases occupy 
the same cell of magnitude space, but {\photoz} estimators will assign them 
all essentially the same redshift.

To overcome these problems and avoid the biases 
intrinsic to {\photoz} estimates, 
it is preferable to use the full redshift PDF for the  
galaxies in a small cell in the space of photometric observables, 
$p(z) \equiv p(z|{\rm observables})$. This PDF encodes all the information 
available about the redshift of an individual galaxy in a photometric sample. 
One can choose to extract a single redshift estimate from the PDF, e.g., 
its mean, median, or mode, but often that is not necessary in applications. 

The weighting method described above can be straightforwardly applied  
to estimate $p(z)$ using a spectroscopic training set. The estimator 
$\hat{p}(z)$ for a galaxy in the photometric sample is given by the weighted 
redshift distribution of its $N_{\rm nei}$ nearest neighbors in the training set, 
using the metric of Eqn. \ref{eqn:dist_gen_def},

\begin{equation}
\hat{p}(z) = \sum_{\beta=1}^{N_{\rm nei}} w_\beta \delta(z-z_\beta)~,
\label{eqn:pzest}
\end{equation}
\noindent where, as before, $N_{\rm nei}$ can be determined from simulations 
by minimizing 
the sum of the shot-noise and ``non-locality'' errors. In practice, 
we estimate $p(z)$ in redshift bins. 
This estimate for 
$p(z)$ was used in a study of galaxy-galaxy lensing by \cite{man07} and 
was shown to yield significantly smaller lensing calibration bias than 
use of {\photoz} estimates.

We can also construct a new estimator for $N(z)_{\rm P}$ by summing the 
$\hat{p}(z)$ distributions for all galaxies in the photometric sample,

\begin{equation}
\hat{N}(z)_{\rm P} = \sum_{i=1}^{N_{\rm P,tot}}\hat{p}_i(z)~.
\label{eqn:Nhat2}
\end{equation}
\noindent This estimator is similar but not identical to that of Eqn. \ref{eqn:Nzest}. 
We will see in \S \ref{sec:weires} that these two are comparable in recovering the true 
redshift distribution of a photometric sample.

\subsection{Sources of Errors in the Weighting Method}\label{sec:wei.err}

The errors arising in the weights method can be considered the errors in
estimating $p(z|{\rm observables})$ for a galaxy in the photometric sample 
from the information in the training set. 
Any differential selection effect between the spectroscopic 
and photometric samples will lead to errors in $\hat{p}(z)$. 
There are several kinds of selection effects: 
(1) statistical effects, (2) large-scale structure (LSS), 
(3) spectroscopic failures in the training set, 
(4) survey selection in the photometric 
observables, (5) survey selection in non-photometric observables, 
and (6) non-locality of the weights. 

Statistical errors arise because the training set is just a subsample of the 
photometric survey and is subject to statistical fluctuations. 
These fluctuations can be significant in regions of magnitude space 
where the training set is very sparse. In such regions, the shot-noise 
errors in $\hat{p}(z)$ will either be large or else the nearest-neighbor 
volume must be made large,
leading to increased non-locality (see below).
Statistical errors can be estimated by bootstrap resampling the training and 
photometric sets. 
If the magnitude errors are well known, one can further Monte Carlo resample 
the magnitudes.
We present results of bootstrap error estimation in \S \ref{subsubsec:errest}.

Errors due to LSS can be significant if certain regions of the 
space of photometric observables are only represented in the training set 
by a spectroscopic survey that covers a small solid angle, in which one or a few large 
structures dominate. In this case, $p(z|{\rm observables})$ for the 
training set will comprise one or a few redshift spikes rather than a 
smooth distribution. 
If these effects occur in regions of magnitude space where the true redshift 
PDF is broad or multiply peaked, they can potentially cause systematic 
errors in the estimates of $p(z)$ or $N(z)$ for the photometric sample. 
The resulting errors may be large if the linear size of the training-set 
volume is not large compared to the galaxy clustering correlation length. The 
errors from LSS can in principle be estimated by constructing 
mock training-set volumes using N-body simulations of structure formation.  

Spectroscopic failures, i.e., targeted objects in the training set for which redshifts 
could not be obtained, can also lead to systematic errors in $\hat{p}(z)$ 
if the failures happen systematically, for instance, if 
they occur preferentially for a particular galaxy spectral type and 
if that type has a different redshift PDF from other 
galaxy types in the same region of magnitude space. 
Since such spectroscopic failures will tend to 
occur in specific and identifiable regions of magnitude space, however, one can at 
minimum excise or down-weight those regions in estimating 
quantities for the photometric sample (see \S \ref{sec:select}), at the 
cost of incompleteness. 

The severity of these systematic errors is regulated by the 
width of the redshift PDF. In the limit of a large number of 
photometric observables with very small measurement errors and 
a large spectroscopic training set, 
the redshift PDF in a small cell in magnitude space approaches 
a delta function. In this regime, the effects of LSS and of spectroscopic failures would 
be simply to increase the statistical errors in certain regions of 
observable space, an effect accounted for in the bootstrap error 
estimate. As one moves away from this ideal limit, the systematic 
errors grow, in the sense that one can no longer reliably 
estimate $p(z|{\rm observables})$ for a galaxy in the photometric 
sample from its training-set neighbors. That 
effect is not captured by the bootstrap and must be 
estimated by other means, e.g., using simulations.
The mock SDSS DR6 catalog we have constructed for this 
paper (see \S \ref{sec:mock}) does not simulate LSS or spectroscopic 
failures; we plan to study such effects in the future. Some of the 
surveys that comprise the training set for the real DR6 data  
are individually affected by LSS effects. 
Having a combination of them helps to alleviate the problem, though
more testing is required to quantify the possible systematics. 

LSS and spectroscopic failures lead to unavoidable differences in 
the selection functions for the photometric and spectroscopic 
samples. In addition, there are differential selection effects 
that are built in by those designing the spectroscopic 
survey. For example, one typically makes magnitude and color 
cuts in selecting spectroscopic targets from a photometric sample. 
In this case, where the selection is made explicitly in the 
photometric observables, there will be regions of observable 
space where the weights cannot be used to reliably estimate redshift distributions.
Again, such regions are known from the target selection cuts and 
can be safely excised from the photometric sample (see \S  \ref{sec:select}). 
If, on the other hand, there 
are differences in spectroscopic and photometric selection based 
on non-photometric observables, then systematic errors in $\hat{p}(z)$ can occur.

A variant of this problem arises when the training set is selected using  
photometric observables that are different from the ones measured in 
the photometric sample. For example, for the SDSS DR6 photometric catalog, 
the spectroscopic target selection for the DEEP2 sample in the training set 
used a different magnitude system (coming from different photometric 
samples) from the SDSS.
Similarly, the selection of the 2SLAQ spectroscopic 
catalog made use of  photometric observables that were not used 
in the photo-z estimation.
Whether such cuts will cause systematic errors depends on how well the 
selection in those systems can be approximated using the SDSS $ugriz$ filters.

Finally, the non-locality of the weights solution is a 
source of systematic error.
Here, non-locality refers to the fact that, in the nearest-neighbor 
approach, we are 
using information from a finite volume to estimate the density at a point in 
observable space, and the density varies over the space. This procedure 
corresponds to applying a smoothing kernel to the density field. Non-locality 
becomes a problem if the volume occupied by the neighbors (or the scale 
of the smoothing kernel) becomes comparable to or larger than the scale 
over which the density changes appreciably. 
In this limit, the shape of the volume used to select the nearest neighbors 
may be important.
Non-locality errors are reduced by choosing a smaller neighbor volume 
for the density estimate, but at the cost of increasing the shot-noise errors. 
Ultimately, the combined errors can be reduced by increasing the density of the training 
set in a particular region of observable space, i.e., by measuring more 
spectra.

\subsection{Selecting the ``Recoverable'' Part of a Photometric Sample}
\label{sec:select}

One of the necessary conditions for the weights procedure to work
is that the spectroscopic training set covers the same region of 
photometric observables as the photometric sample. That is, 
the weights can only recover the redshift PDF of a galaxy in the photometric 
sample if it lies in the region of intersection
of the redshift-observables hypersurfaces of the training and photometric sets.
Defining this region of intersection is not always trivial, especially 
given the high number of dimensions that may be involved. To do so, 
we count how many times a galaxy in the photometric sample 
is used in the weights calculation for all members of 
the training set. By definition, photometric 
galaxies that are never counted in the weights procedure are 
not in the region of intersection, hence the redshift
distribution of those galaxies will not be accurately recovered by the weighting  
procedure. We make use of this criterion below.
If one does not require the photometric sample to be complete, 
one can choose to excise such galaxies from consideration. 
Using several real and mock catalogs, we have found empirically 
that using $\sim 5$ nearest neighbors in the weights calculation is optimal 
for determining the intersection region for the mock catalog. 

As examples, consider the mock and real SDSS DR6 catalogs 
of \S \ref{sec:cat}. 
From Figs. \ref{dist.rmag}b, \ref{dist.col}b, and \ref{dist.dndz}b, one might expect 
that the combined 
training set covers the same region of observables as the photometric sample.
However, using the definition of the previous paragraph, 
more than  $\sim 43\%$ of the 
mock photometric-sample galaxies are not used in the weights 
calculation, i.e., 
they are not well represented in the training set. 
Fortunately, in the real SDSS DR6 catalog, by the same criterion 
we find that $\sim 98\%$ of photometric-sample  
galaxies with $r<22$ are well represented in the training set.
It is important to apply such a recoverability test whenever a training-set 
method is used. 

\section{Catalogs} \label{sec:cat}

To test the performance of the weighting method and compare it 
with standard {\photoz} estimates, we employ two kinds of catalogs. The first 
is drawn from the SDSS DR6 \citep{dr6} photometric sample and various 
spectroscopic subsamples of it and allows us to display results of the weighting 
method on real data. The second is a mock catalog 
constructed to have properties similar to the SDSS DR6 photometric and spectroscopic 
samples. The goal of the mock catalog is not to precisely reproduce all features 
of the SDSS catalog but to have a sample with realistic spectroscopic 
and photometric features and for which 
we have ground truth (i.e., redshifts and galaxy types) for all galaxies. 
In this section, we describe the relevant features of the 
real and mock catalogs, relegating the details to Appendix A.

\subsection{SDSS DR6 Data}
\label{sec:real}

The SDSS DR6 photometric and spectroscopic data samples are 
drawn from those used by \cite{oya08a}  to produce a neural 
network {\photoz} catalog.  

\subsubsection{Photometric Sample}\label{sec:real.phot}

We use a random $1\%$ subset of the galaxies in the SDSS DR6 Photoz2 catalog described 
in \cite{oya08a} as our photometric sample. This subset contains approximately 
769,582 galaxies with $r<22$. 
The catalog is approximately flux-limited at this magnitude limit.
For details 
of the parent sample, see Appendix A and \cite{oya08a}.
The $r$ magnitude, $g-r$, and $r-i$ color distributions 
are shown in the bottom right panel of Fig.~\ref{dist.rmag}a and the 
bottom panels of Fig.~\ref{dist.col}a.

\subsubsection{Spectroscopic Training Set}\label{sec:real.spec}

The spectroscopic training sample we use for SDSS DR6 is drawn 
from a number of spectroscopic galaxy catalogs that overlap 
with SDSS DR6 imaging. 
We impose a magnitude limit of $r<23$ on the spectroscopic 
samples as well as
additional cuts based on the quality of the spectroscopic
redshifts reported by the different surveys (see Appendix A).
The SDSS spectroscopic sample 
provides $531,594$ redshifts, principally from the MAIN and 
Luminous Red Galaxy (LRG) samples. The remaining redshifts are:
$20,381$ from the Canadian Network for Observational Cosmology (CNOC) 
Field Galaxy Survey \cite[CNOC2;][]{yee00},
$1,531$ from the Canada-France Redshift Survey \cite[CFRS;][]{lil95},  
$11,040$ from the Deep Extragalactic Evolutionary Probe \cite[DEEP;][]{deep2}
and  DEEP2 \citep{wei05}, 
$654$ from the Team Keck Redshift Survey \cite[TKRS;][]{wir04}, and 
$52,762$ LRGs from the 2dF-SDSS LRG and QSO Survey 
\cite[2SLAQ;][]{can06}.

\begin{figure}
  \begin{minipage}[t]{85mm}
    \begin{center}
      \resizebox{85mm}{!}{\includegraphics[angle=0]{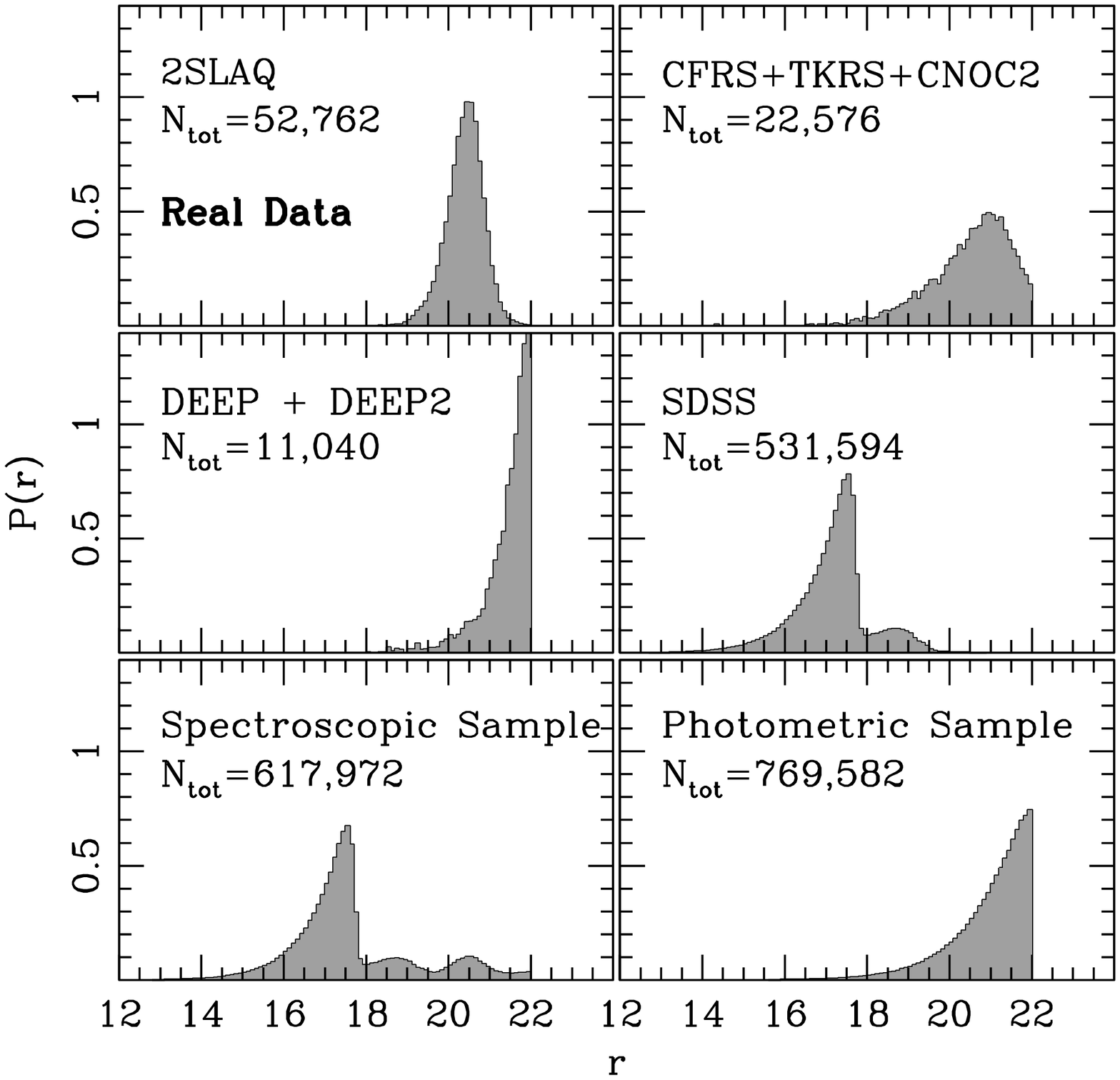}}
    \end{center}
  \end{minipage}
  \begin{minipage}[t]{85mm}
    \begin{center}
      \resizebox{85mm}{!}{\includegraphics[angle=0]{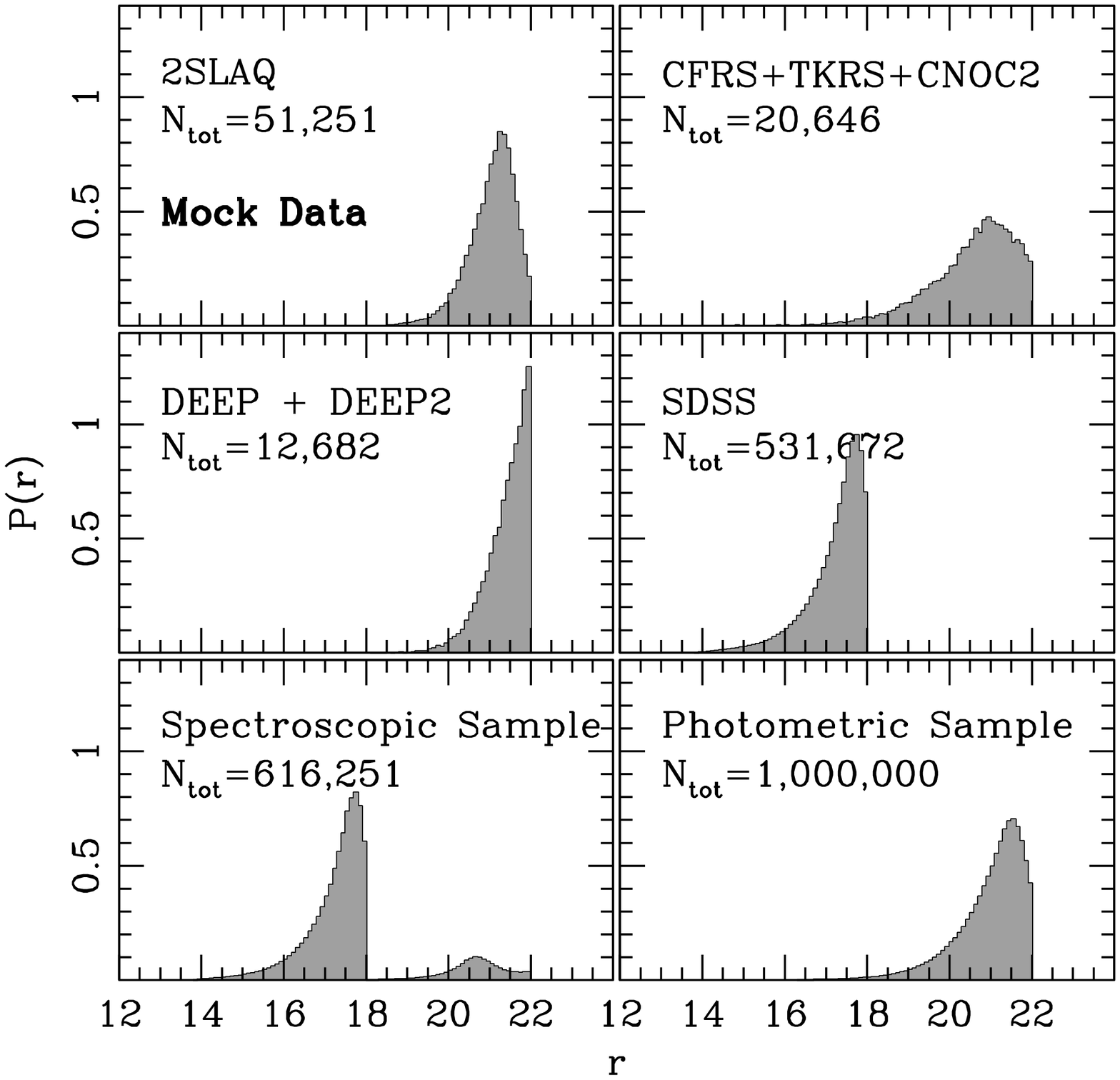}}
    \end{center}
  \end{minipage}
  \caption{Normalized $r$ magnitude distributions for the catalogs comprising 
the real SDSS DR6 ({\it top: figure a}) and the mock SDSS DR6 ({\it bottom: figure b}) catalogs. 
In each figure, the top four panels indicate the distributions 
for the different spectroscopic subsamples (see text), bottom left panels indicate flux 
distributions for the combined spectroscopic samples, and bottom right panels 
distributions for the photometric samples. In each panel, $N_{tot}$ denotes the 
total number of galaxy measurements used in each sample.
}\label{dist.rmag}
\end{figure}

\begin{figure}
  \begin{minipage}[t]{85mm}
    \begin{center}
      \resizebox{85mm}{!}{\includegraphics[angle=0]{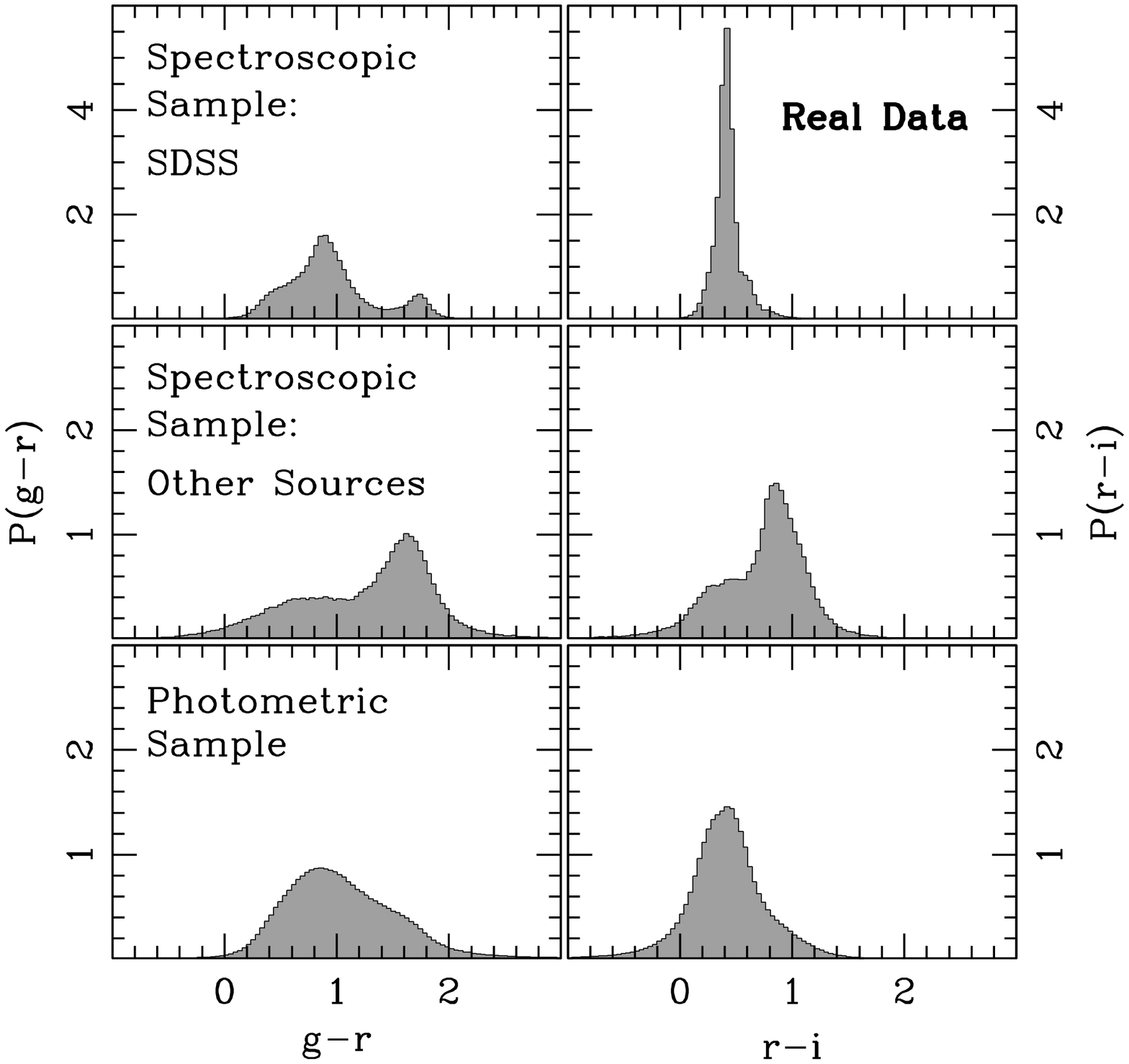}}
    \end{center}
  \end{minipage}
  \begin{minipage}[t]{85mm}
    \begin{center}
      \resizebox{85mm}{!}{\includegraphics[angle=0]{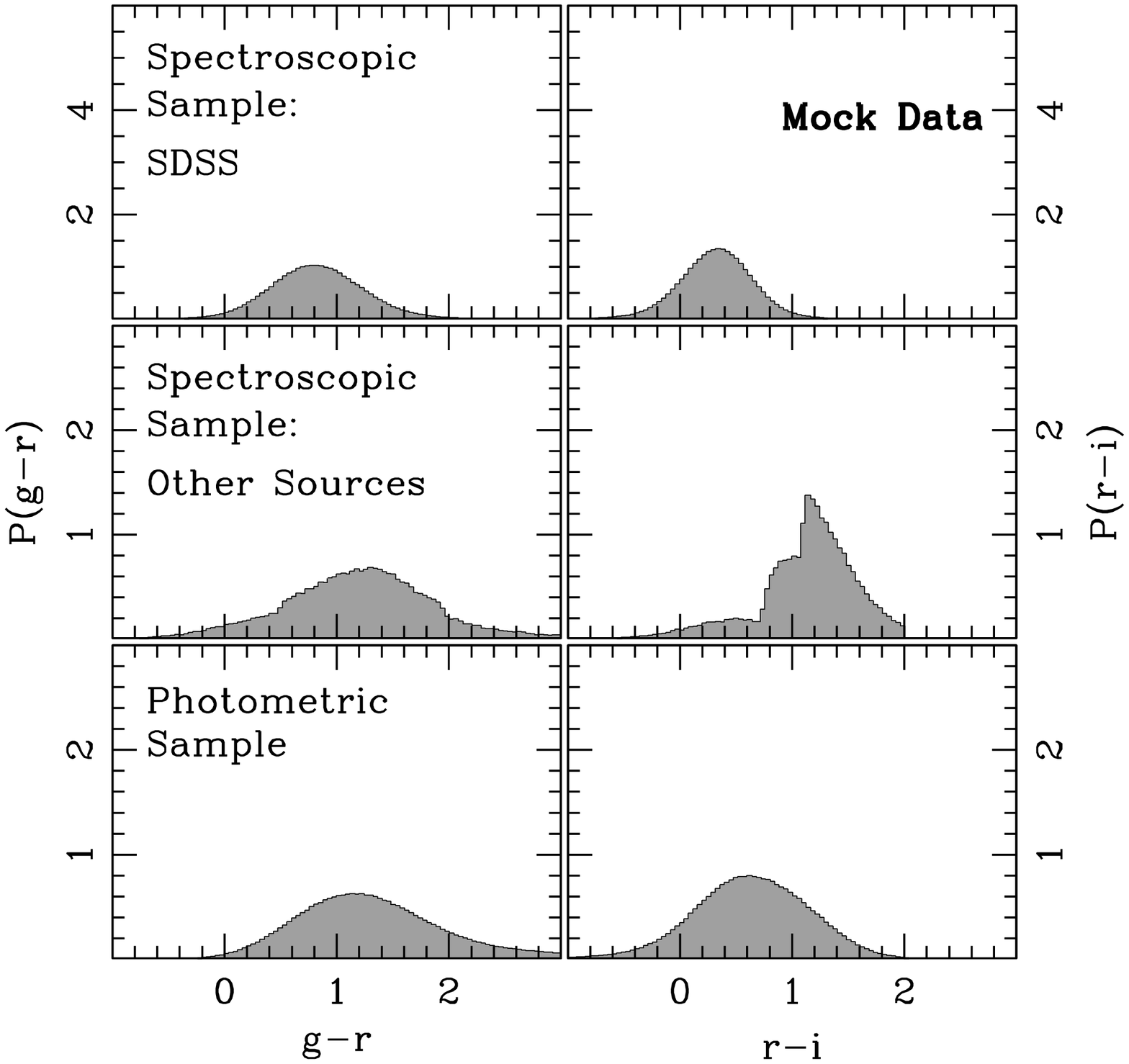}}
    \end{center}
  \end{minipage}
  \caption{
Distributions of $g-r$ and $r-i$ colors for the catalogs comprising spectroscopic 
training and photometric sets for the real SDSS DR6 ({\it top: figure a}) and the mock 
SDSS DR6 ({\it bottom: figure b}). Top rows give distributions for the SDSS 
spectroscopic sample, middle rows the distributions for the other  
spectroscopic samples, bottom rows the distributions for the photometric samples.
The real and mock SDSS spectroscopic color distributions differ  
primarily because the latter does not include LRGs.
}\label{dist.col}
\end{figure}

\begin{figure}
  \begin{minipage}[t]{85mm}
    \begin{center}
      \resizebox{85mm}{!}{\includegraphics[angle=0]{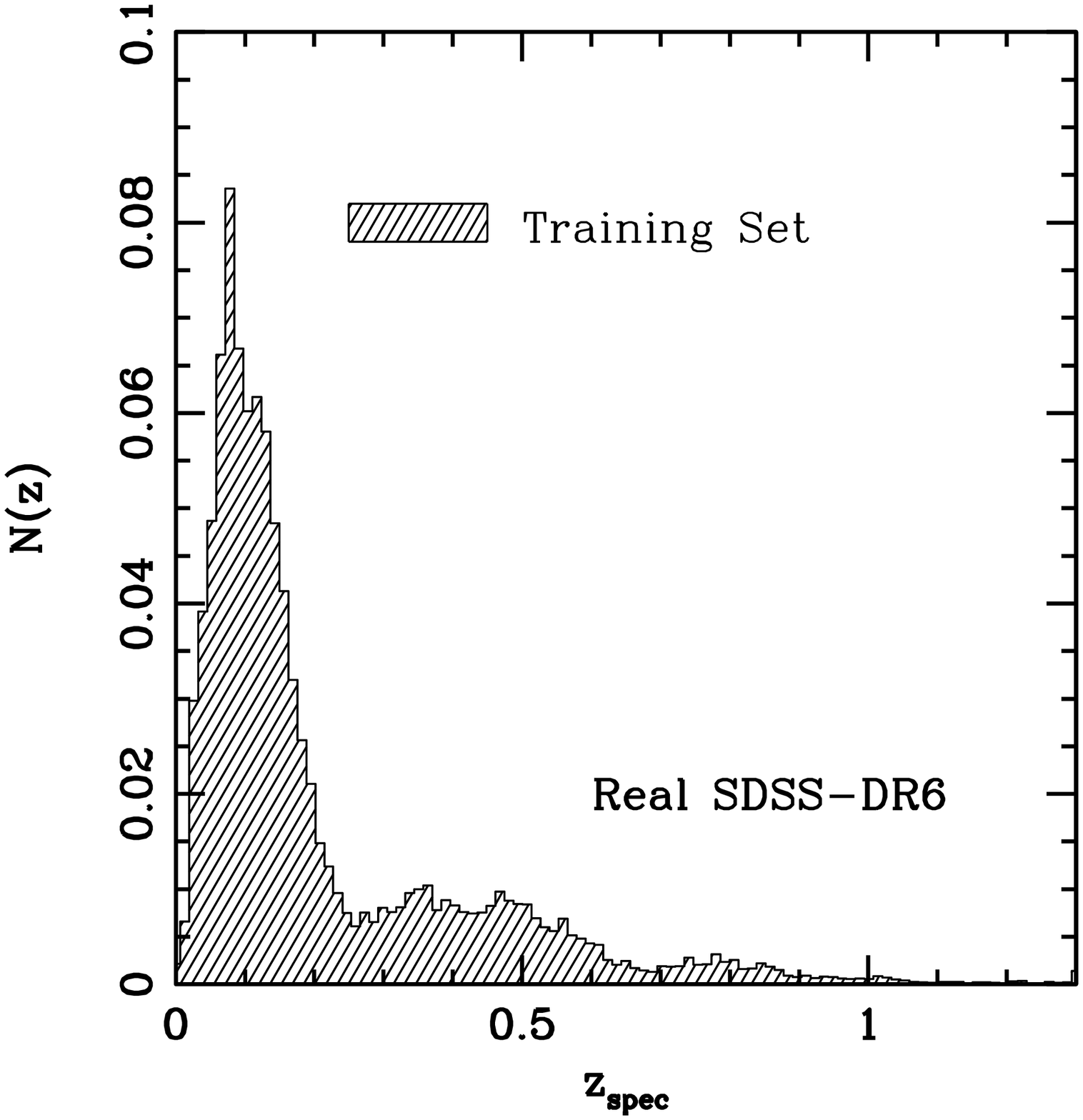}}
    \end{center}
  \end{minipage}
  \begin{minipage}[t]{85mm}
    \begin{center}
      \resizebox{85mm}{!}{\includegraphics[angle=0]{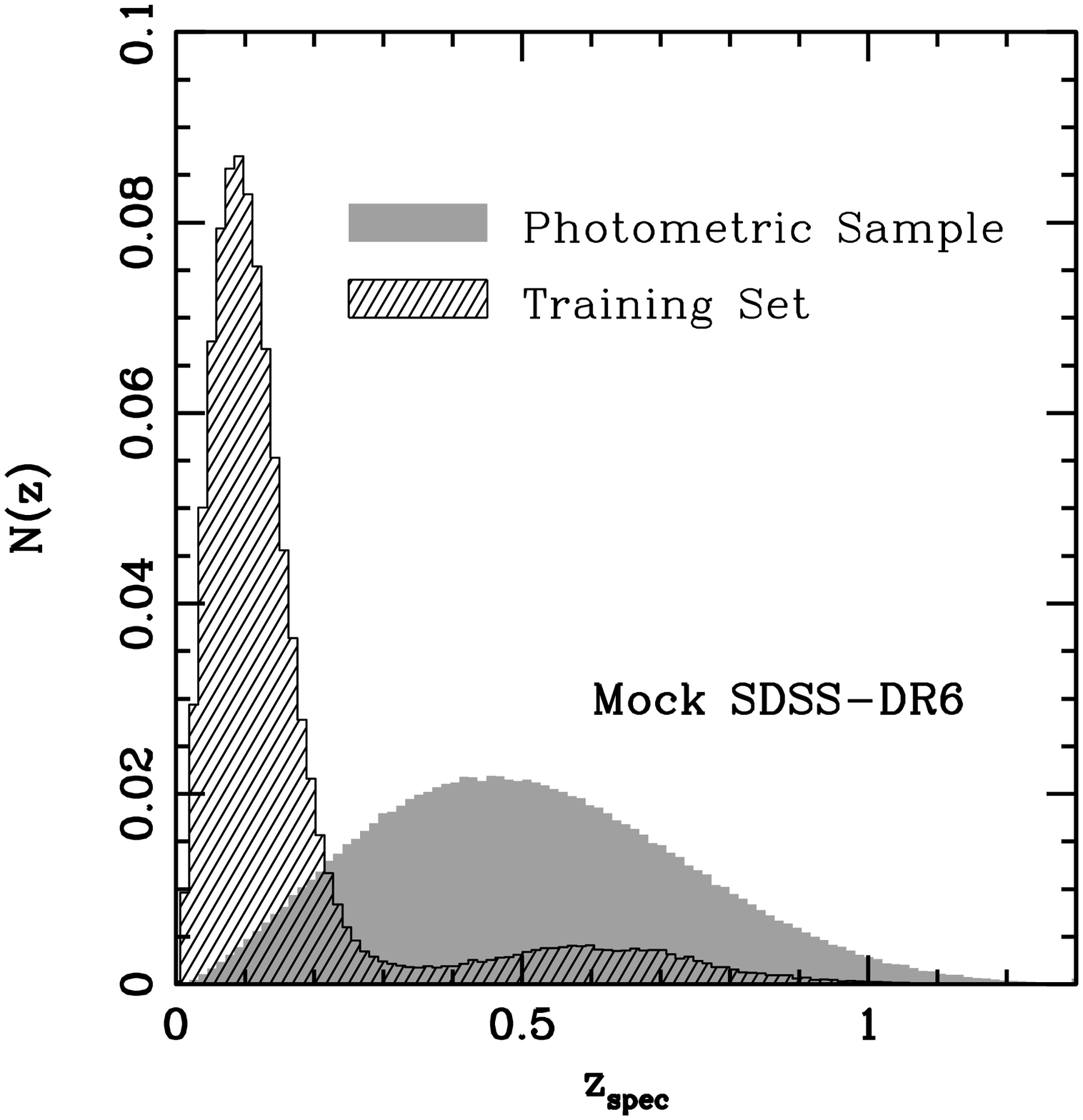}}
    \end{center}
  \end{minipage}
  \caption{({\it Top, figure a}) Spectroscopic redshift distribution for the combined SDSS DR6 
spectroscopic training set.
({\it Bottom, figure b}) Spectroscopic redshift distributions for the mock SDSS DR6 training and photometric sets.
}\label{dist.dndz}
\end{figure}

The $r$-magnitude and color ($g-r$ and $r-i$) 
distributions for the spectroscopic samples are shown 
in Figures \ref{dist.rmag}a and \ref{dist.col}a.  
Although the magnitude and color distributions of 
the combined spectroscopic sample are not 
identical to those of the photometric sample, the 
spectroscopic sample does span the  
ranges of apparent magnitude and colors of the photometric sample.
Fig. \ref{dist.dndz}a gives the spectroscopic redshift distribution 
for the combined spectroscopic sample.

\subsection{SDSS DR6: Mock Catalog}
\label{sec:mock}

Using spectral template libraries and observational data on the redshift-dependent 
luminosity functions of galaxies of different types, 
we have constructed mock photometric and spectroscopic samples that 
reproduce the main features of the real SDSS DR6 samples. We describe 
these briefly below. 

\subsubsection{Mock Photometric Sample}
\label{subsubsec:mockphot}

The simulated SDSS catalog contains $10^7$ galaxies 
with redshift $z < 2.0$ and magnitude $14 < r < 22$.
We use the {\tt lf\_mock\_schechter} code from the kcorrect 
package \citep{bla03c} to generate redshift, type, and $i$-magnitude relations.
The inputs to the code are the redshift range, Schechter 
luminosity function parameters, and the ranges 
of absolute and apparent $r$-magnitudes. 
The code outputs a list of redshifts and apparent r-magnitudes. 
We set the range of absolute $i$-band magnitudes to $(-24,-14)$. Using 
data from the VVDS survey, \cite{zuc06} estimated galaxy luminosity 
functions and Schechter-function fits thereto for different galaxy 
types in redshift bins of size $\Delta z=0.2$ from $z_{min}=0.2$ to $z_{max}=1.5$. 
We fit simple polynomial functions to the Schechter parameters of 
\cite{zuc06} to derive a continuous relationship between the Schechter parameters
$M^*, \alpha, \phi^*$, redshift $z$, and galaxy type $T$, using 
the centroid of each redshift bin for the fit.
To regularize the fits, we visually extrapolated the results of \cite{zuc06} to 
the $z=(0, 0.2)$ bin and, where needed (for certain galaxy types), 
for the $(1.2, 1.5)$ bin. The detailed fits 
are given in Appendix B.

Galaxy colors are generated using the four Coleman, Wu, \& Weedman 
spectral templates \citep{col80}---E, 
Sbc, Scd, Im---extended to UV and NIR wavelengths using synthetic templates 
from \citet{bru93}. 
These templates are mapped to galaxy SED type $T$ (used by \citet{zuc06}) as 
$($E, Sbc, Scd, Im$)$ $\rightarrow T=(1,2,3,4)$. 
To improve the sampling and coverage of color space, 
we have created additional 
templates by interpolating between adjacent templates.
The redshift, $r$-magnitude, and type relations are first
generated without photometric errors; 
errors are then added to produce observed magnitudes.
Magnitude errors are modeled as sky-background dominated errors 
approximated as Gaussians that are uncorrelated between the different 
SDSS filters.

The resulting magnitude and color distributions for 
the mock photometric sample 
are shown in the lower right panel of Fig.~\ref{dist.rmag}b and the 
bottom panels of Fig.~\ref{dist.col}b. The redshift distribution for the 
sample is shown as the dark grey region in Fig. \ref{dist.dndz}b. 
The $r$-magnitude distribution of the mock photometric sample 
peaks at slightly brighter magnitude 
than for the actual DR6 photometric sample, and the $r-i$ distribution 
is slightly less peaked than that of the real data, but overall the 
real and mock distributions are quite similar in their photometric 
properties. As noted above, the goal of the mocks is {\it not} to 
exactly reproduce the real data distributions. 

\subsubsection{Mock Spectroscopic Training Set}\label{sec.mock.phot}

\begin{table*}
\caption{Mock Spectroscopic Training Set Properties: Number of galaxies and photometric selection cuts applied.}
\begin{center}
\leavevmode
\begin{tabular}{ l c c l l l }\hline \hline
\multicolumn{1}{c}{Catalog} & \multicolumn{1}{c}{Unique objects} & \multicolumn{1}{c}{All Objects} & \multicolumn{1}{c}{Selection Cuts} \\
\hline
mockSDSS &531,672 &531,672 &$r\leq 18.0$ \\
mockDEEP+DEEP2 &2,419&31,716&$g-r < 2.35(r-i)-0.45$,\\
               &&&$g-r<1.95$, $1.1<r-i<2$, \\
               &&&$r<22$\\
\hline
mockTKRS+CFRS+CNOC2 &1,827&23,681& $u<23$, $g<23$, \\
                    &&&$r<22$, $i<22$ \\
\hline
mock2SLAQ &11,082&51,251& $((r-i)-(g-r))/8 \geq 0.55$, \\
          &&& $0.7(g-r)+1.2(r-i-0.18) \geq 1.6$, \\
          &&& $17.5 \leq i \leq 19.8$, \\
          &&& $0.5 < g-r <3$, $r-i<2$ \\
\hline \hline
\end{tabular}
\end{center}
\label{tbl:mockselec}
\end{table*}

We construct the mock spectroscopic training
set by piecing together a variety of different catalogs with 
different selection
functions, each meant to qualitatively represent one of the spectroscopic 
training samples 
described above in \S \ref{sec:real.spec}.
We obtain each component catalog 
of the training set by generating an independent realization 
of the mock photometric sample and applying the selection cuts of the 
spectroscopic catalog to the realization. The selection cuts we use 
for each component spectroscopic catalog are given in 
Table~\ref{tbl:mockselec}. 
As discussed in Appendix \ref{sec:real.train}, many of 
the real training set galaxies are located in the southern celestial 
stripe, which was imaged repeatedly by the SDSS. In the real 
training set, multiple photometric measurements of the same galaxy were 
treated as independent. We have simulated this effect in the mock 
training set by 
regenerating the magnitudes of each galaxy in the mock training sets as needed. 
The number of unique mock galaxies and total number of galaxies 
(counting all realizations of the same galaxy as different objects) are shown 
in the second and third columns of Table~\ref{tbl:mockselec}.
For comparison, we have also generated spectroscopic catalogs with the same 
total number of objects but using only unique objects. 
We found no discernible differences in the resulting {\photoz}'s or weights.

The $r$-magnitude, color ($g-r$ and $r-i$), and spectroscopic redshift 
distributions of the spectroscopic samples for the mock SDSS DR6 data 
are shown in Figs. \ref{dist.rmag}, \ref{dist.col}, and \ref{dist.dndz}.
As is evident from comparison of the {\it a} and {\it b} 
components of Figs. \ref{dist.rmag} and \ref{dist.col}, 
there are some noteworthy differences between the selection cuts used for 
the mock training set 
and the actual target selection cuts applied in constructing the 
spectroscopic surveys described in 
\S \ref{sec:real.spec} and Appendix \ref{sec:real.train}. 
For example, for the SDSS spectroscopic catalog, the mock sample 
is flux limited at $r=18$, while the actual spectroscopic 
catalog comprises the MAIN sample, with a flux limit of $r=17.7$, and 
the LRG sample, with red colors and a flux distribution that peaks around $r\approx 19$. 
For the other spectroscopic surveys, the actual photometric selection cuts were typically 
made in non-SDSS passbands, while our mock data and selection cuts 
were generated using the SDSS $ugriz$ bands. Therefore, the 
mock photometric cuts do not exactly match the actual cuts used. 
As a result of this mismatch, e.g., 
the peak of the $r$-magnitude distribution of the mock 2SLAQ 
sample is about one magnitude fainter than the corresponding peak 
in the real data, as shown in the upper left panels of Figs. \ref{dist.rmag}a and b.

\section{Applications of the Weighting Method I: Improving Photometric Redshift 
Measures}

\label{sec:app1}

With the mock and real galaxy catalogs in hand, we can now test the performance 
of the weighting method in different applications. In this section, we describe 
the utility of the weighting method in improving the performance of traditional 
{\photoz} estimates. In the next section, we use the weighting method to 
directly estimate $N(z)$ and compare the results with \photoz-based estimates.


\subsection{Estimating Photo-$z$ Bias and Scatter}\label{sec:biasig.mock}

We have applied an Artificial Neural Network (ANN) {\photoz} estimator, described in 
Appendix \ref{app:neu} and in more detail in \citet{oya08a}, 
to the SDSS DR6 mock catalog of \S \ref{sec:mock}. 
Despite the fancy name, an ANN is simply a function 
which relates redshifts to photometric observables. 
The training set is used to determine the best-fit value 
for the free parameters of the ANN.
The best-fit parameters are found by minimizing the overall scatter 
(see definition below) of the photo-z's determined
for the training set galaxies.
The ANN configurations are not unique in the sense that different sets of 
parameters can result in the same overall scatter. 
The best-fit parameters found after minimizing the scatter depend on where in 
parameter space the optimization run begins.
Hereafter we refer to an ANN function using a given set of best-fit 
parameters as a neural network solution.
The network is trained on the 
mock spectroscopic training set described in \S \ref{sec.mock.phot} and 
used to estimate redshifts for the mock 
photometric sample of \S \ref{subsubsec:mockphot}. We have also trained 
and applied the network using the real DR6 data described in \S \ref{sec:real}.

\begin{figure*}
  \begin{minipage}[t]{43mm}
    \begin{center}
      \resizebox{43mm}{!}{\includegraphics[angle=0]{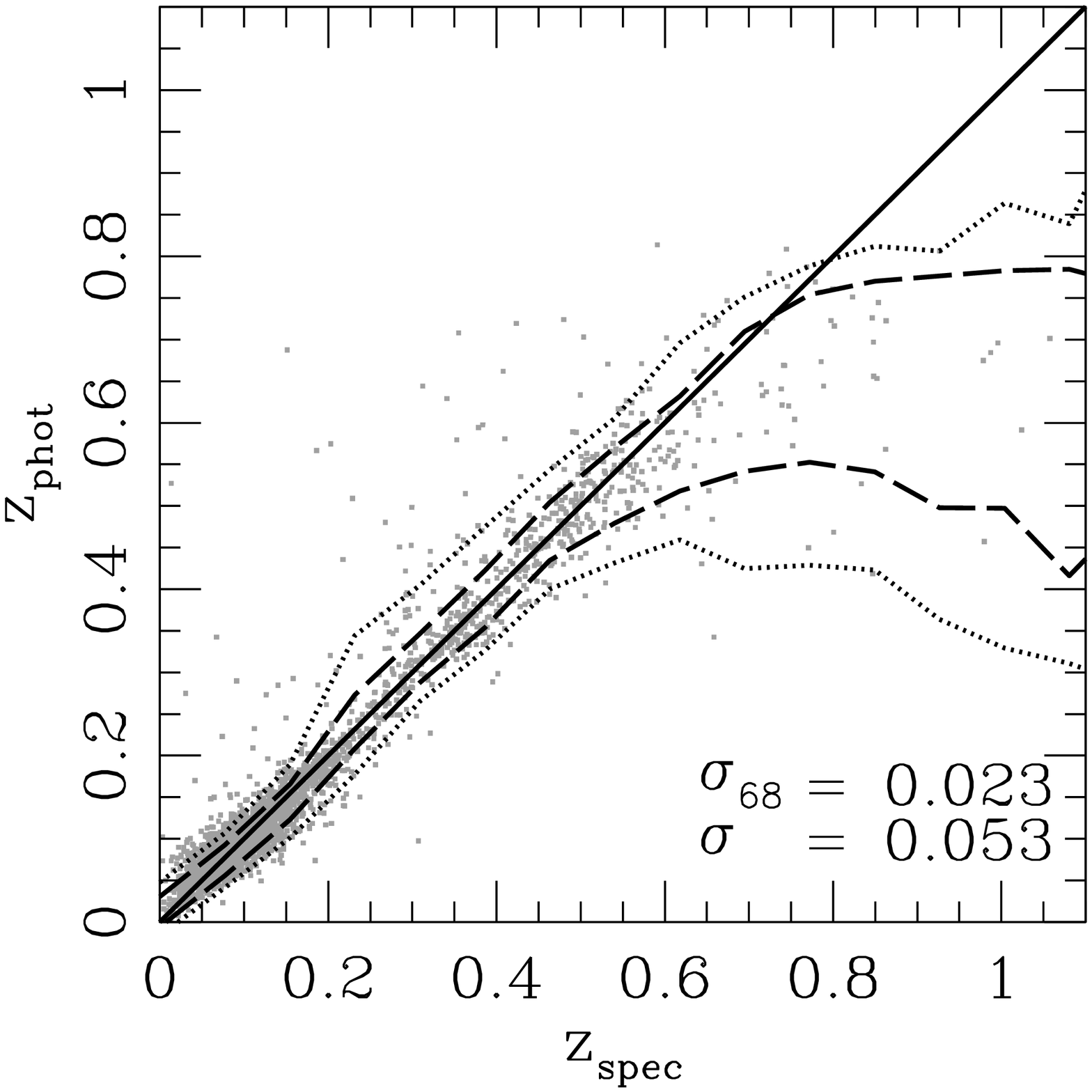}}
    \end{center}
  \end{minipage}
  \begin{minipage}[t]{43mm}
    \begin{center}
      \resizebox{43mm}{!}{\includegraphics[angle=0]{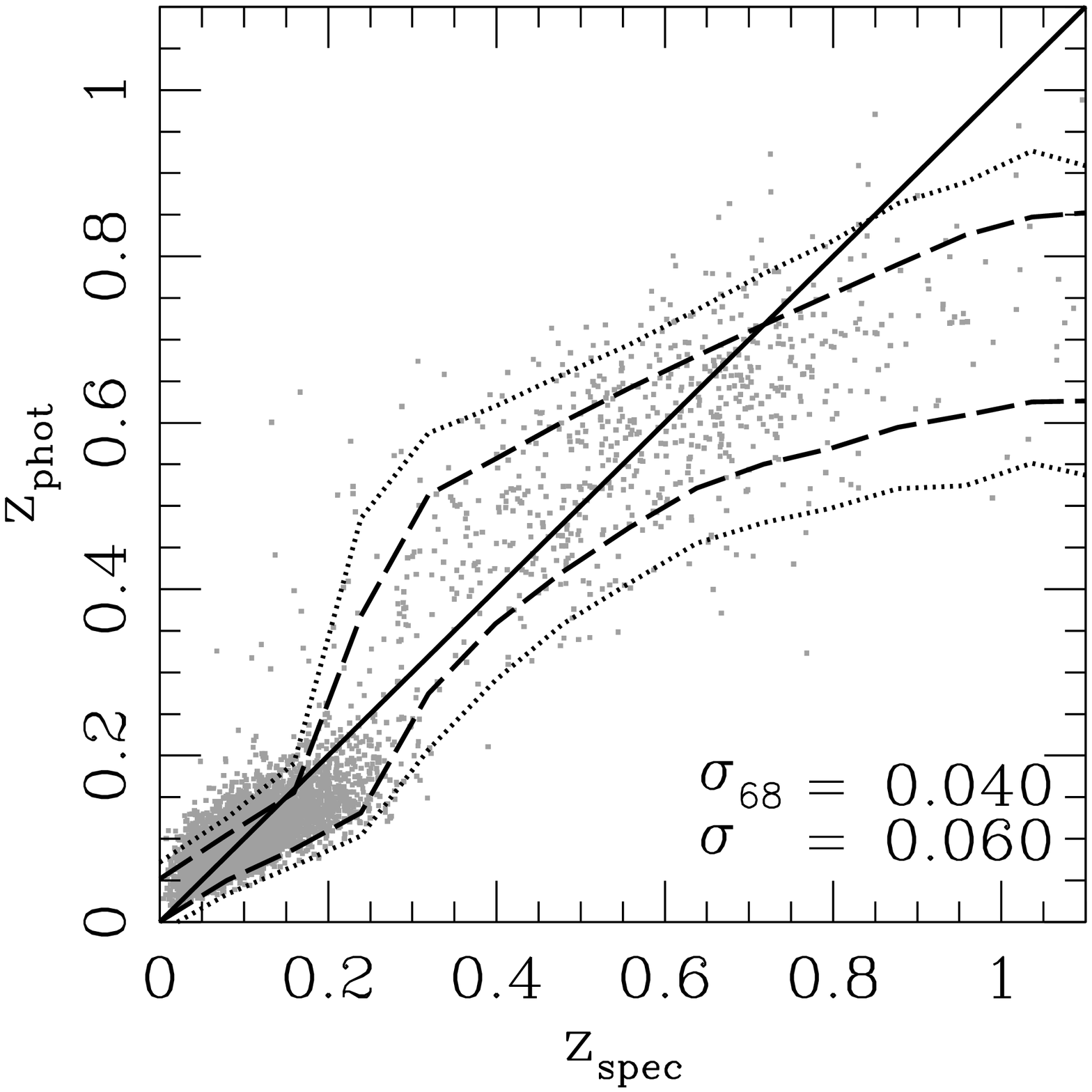}}
    \end{center}
  \end{minipage}
  \begin{minipage}[t]{43mm}
    \begin{center}
      \resizebox{43mm}{!}{\includegraphics[angle=0]{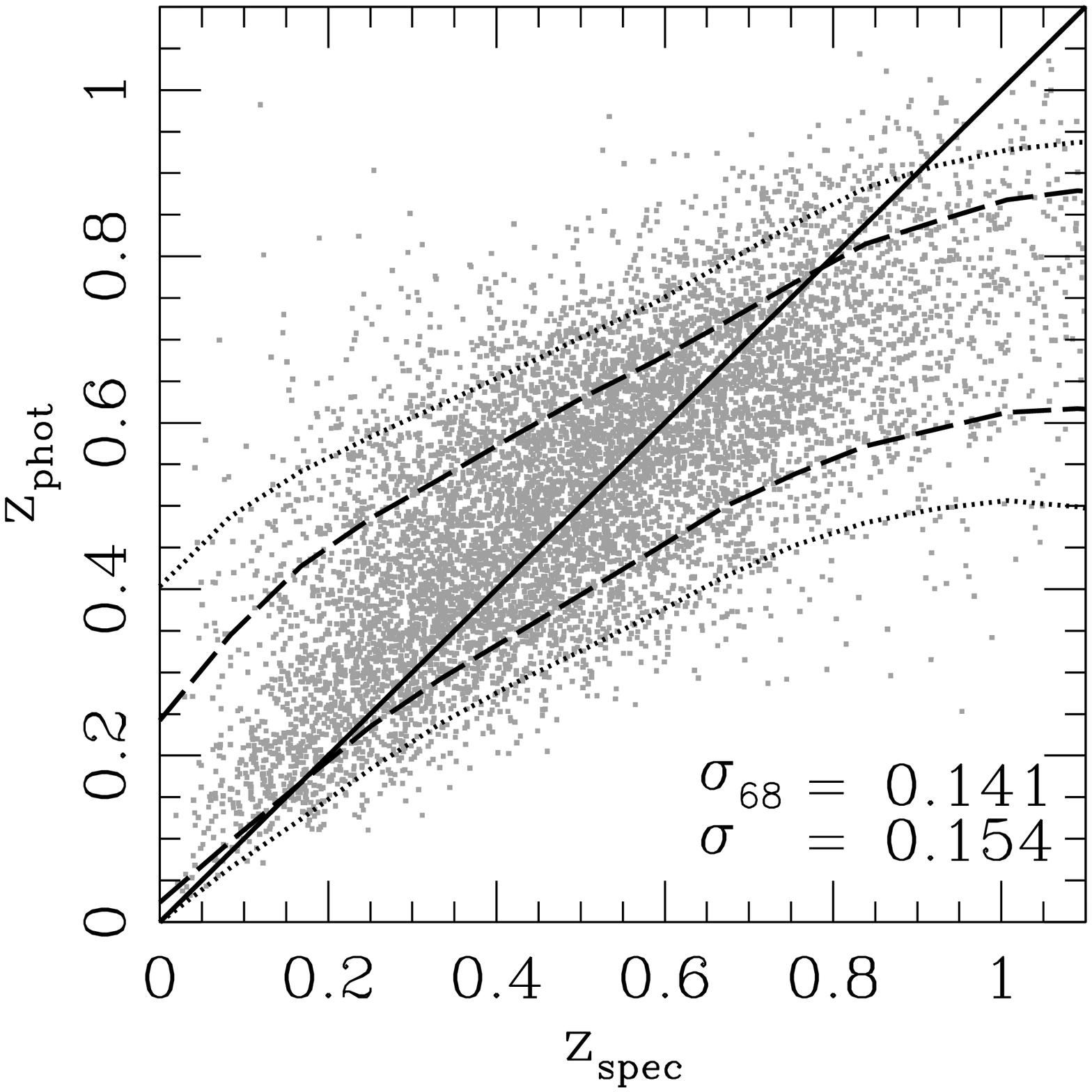}}
    \end{center}
  \end{minipage}
  \begin{minipage}[t]{43mm}
    \begin{center}
      \resizebox{43mm}{!}{\includegraphics[angle=0]{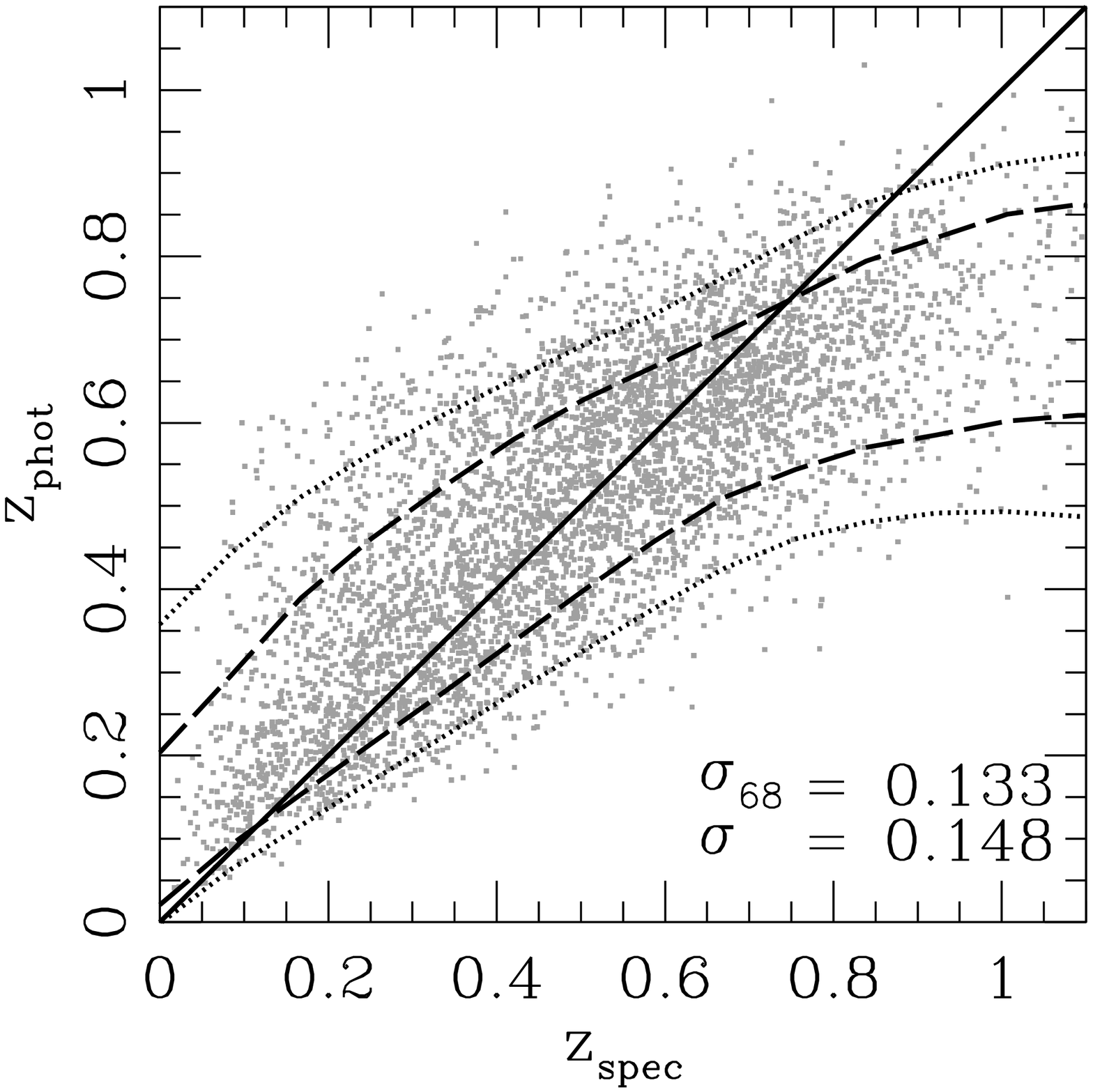}}
    \end{center}
  \end{minipage}
  \caption{$z_{\rm phot}$ vs.\ $z_{\rm spec}$ for ({\it from left to right}): (a) the 
real SDSS DR6 training set, (b) the mock SDSS training set, 
(c) the full mock photometric set, and (d) the recoverable mock photometric set, i.e.
the part of the mock photometric set that is well represented in the training set.
The dashed and dotted curves enclose  68\% and 95\% of the points in each \zspec~ bin. 
In the lower right of each panel, $\sigma$ is the {\it rms} photo-$z$ scatter averaged 
over all objects in the catalog, and $\sigma_{68}$ is the range containing 68\% of the  
objects in the distribution of $z_{\rm phot}-z_{\rm spec}$.
}\label{fig.zpzs}
\end{figure*}

The results of the ANN {\photoz} estimator are displayed in 
Fig. \ref{fig.zpzs}, which shows the inferred redshift {\zphot} vs. 
true redshift {\zspec}. Panel (b) shows the results for the mock spectroscopic 
training set, while panel (c) shows the results for the mock photometric 
sample. For comparison, panel (a) shows results for the real SDSS DR6 training 
set data. As was seen in Fig. \ref{dist.dndz}b, the redshift 
distribution of the mock photometric sample is considerably deeper than 
that of the mock training set. Not surprisingly, the {\photoz} errors 
as a function of redshift for the 
mock photometric sample are somewhat larger than one would estimate based 
on the training set (compare the 68 and 95\% contours in panels (b) and (c)). 
This is a problem since, for real (as opposed to mock) galaxy catalogs,  
one does not have the information necessary to make panel (c), i.e., one can only  
estimate {\photoz} performance using the training set. Since the training set is, 
as in this mock example,  
usually not representative of the photometric sample, 
the statistics of {\photoz} quality for the training set are not 
accurate indicators of {\photoz} quality for the photometric sample.

\begin{figure*}
  \begin{minipage}[t]{85mm}
    \begin{center}
      \resizebox{85mm}{!}{\includegraphics[angle=0]{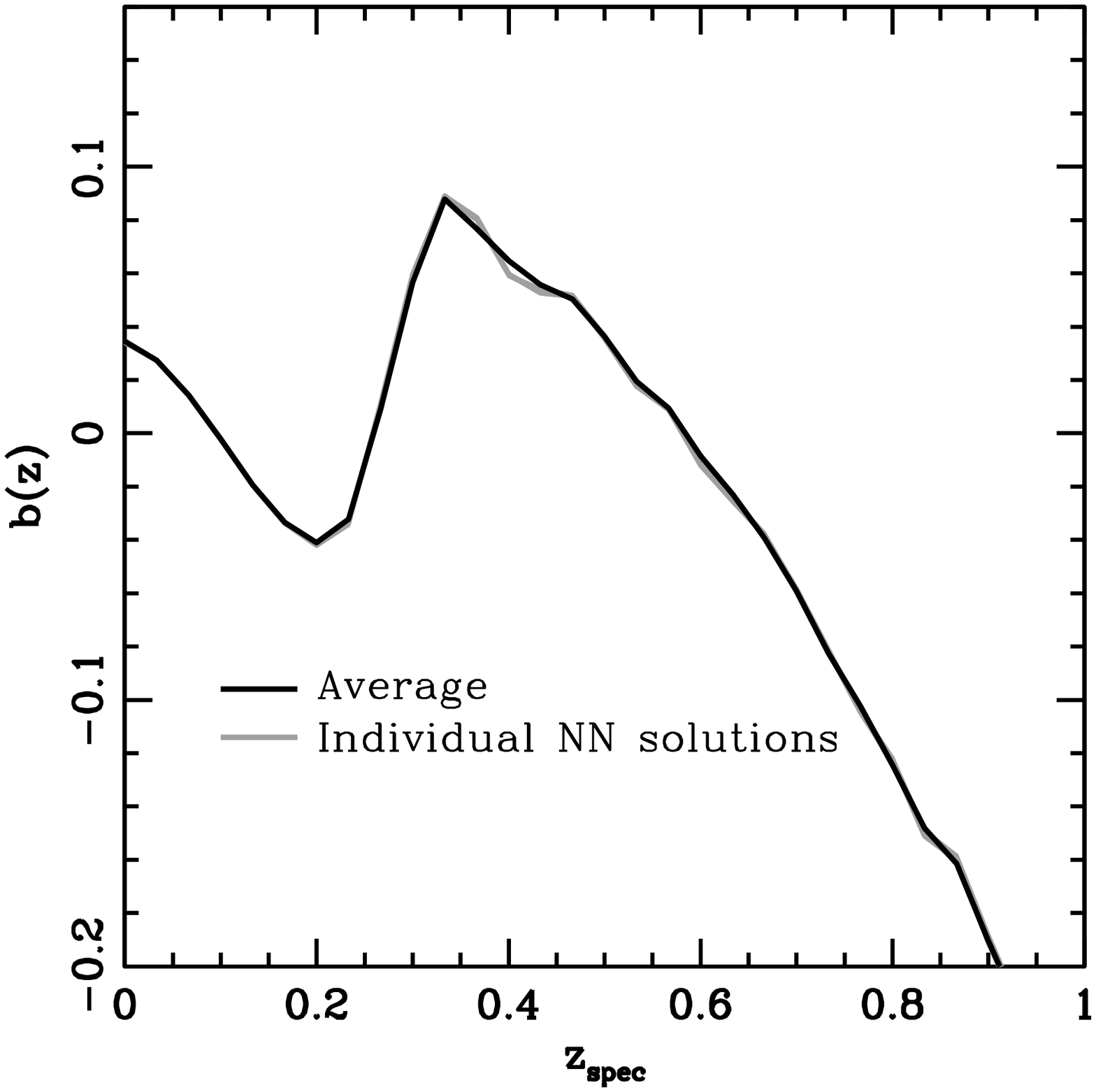}}
    \end{center}
  \end{minipage}
  \begin{minipage}[t]{85mm}
    \begin{center}
      \resizebox{85mm}{!}{\includegraphics[angle=0]{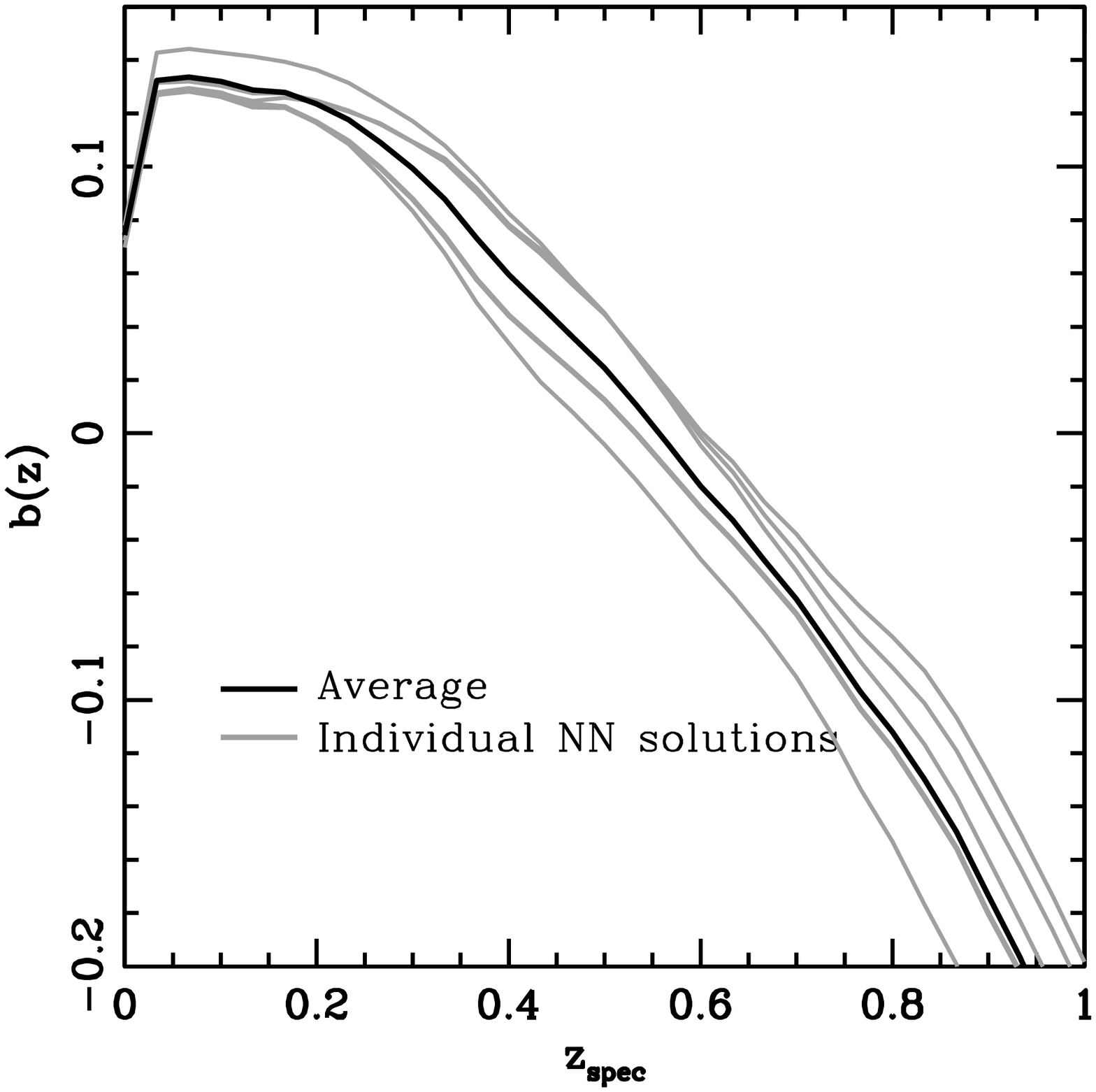}}
    \end{center}
  \end{minipage}
  \begin{minipage}[t]{85mm}
    \begin{center}
      \resizebox{85mm}{!}{\includegraphics[angle=0]{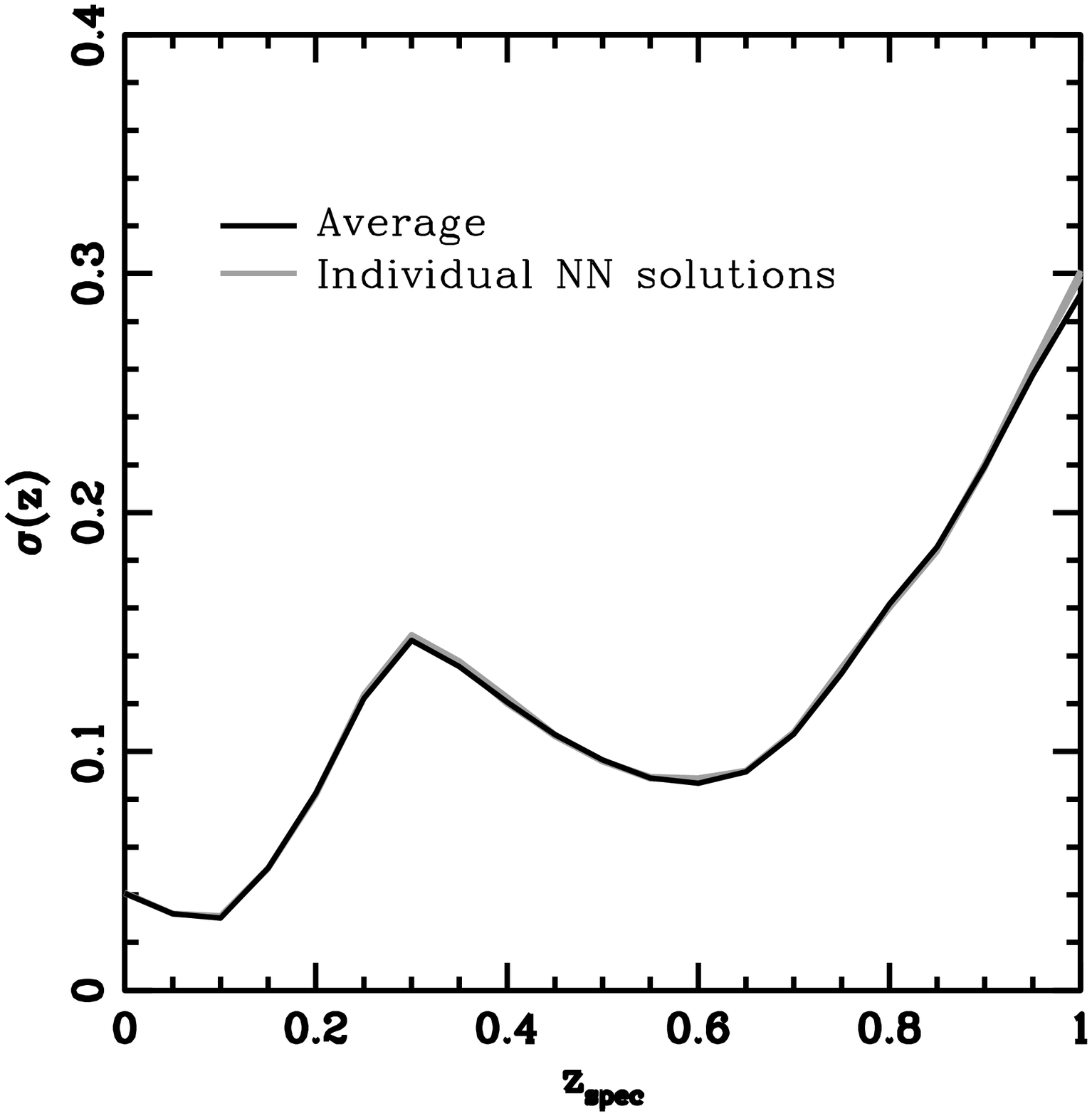}}
    \end{center}
  \end{minipage}
  \begin{minipage}[t]{85mm}
    \begin{center}
      \resizebox{85mm}{!}{\includegraphics[angle=0]{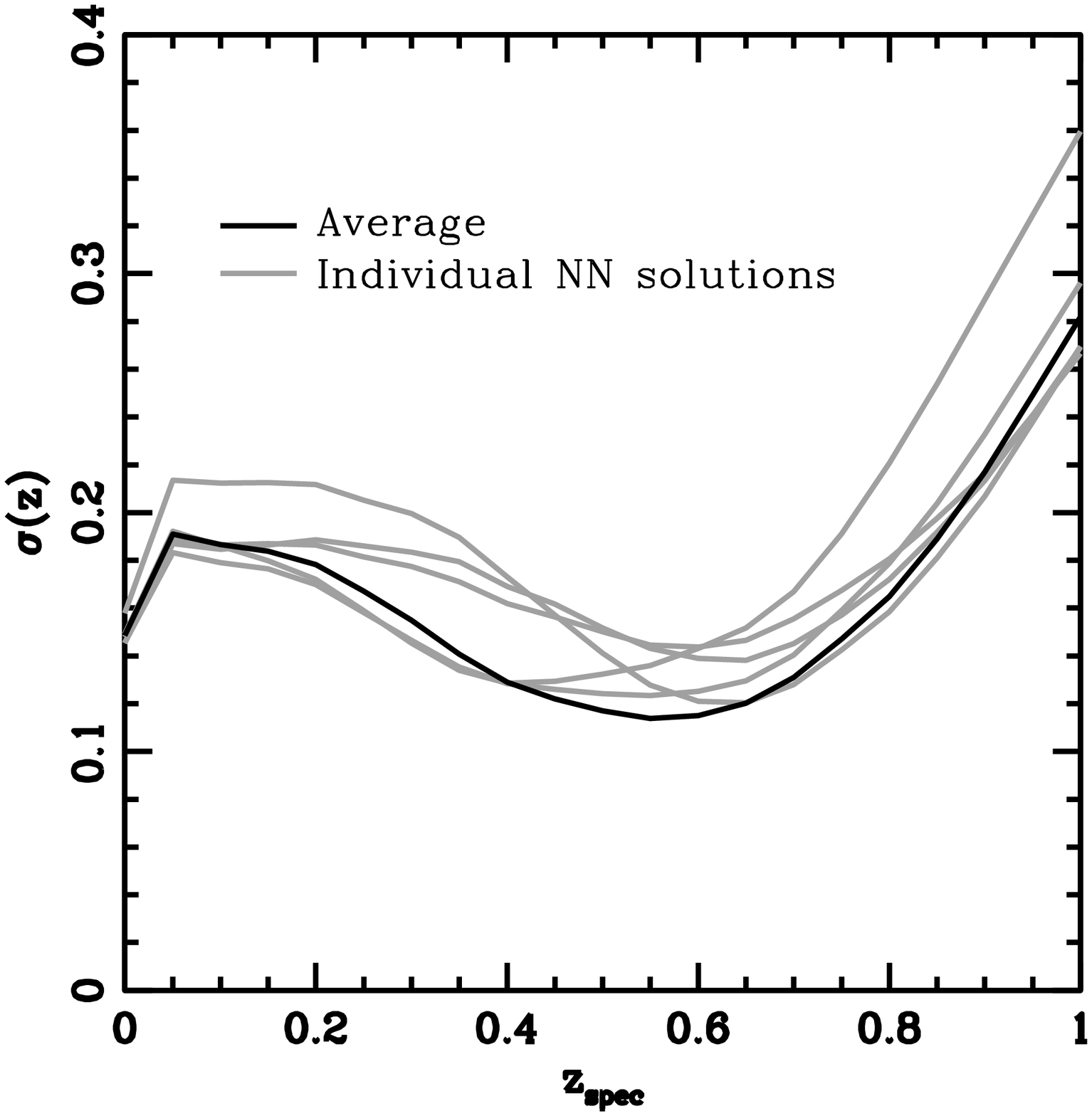}}
    \end{center}
  \end{minipage}
  \caption{{\it Upper panels}: {\Photoz} bias $b$ vs. $z_{spec}$ for the 5 neural network  
{\photoz} solutions for the mock SDSS sample: ({\it top left}) 
training set (unweighted) and ({\it top right}) photometric set. {\it Lower panels}: 
{\Photoz} scatter $\sigma$ vs. $z_{spec}$ for the 5 NN {\photoz} solutions of the ({\it bottom left})
training set and ({\it bottom right}) photometric set. 
}
\label{fig.bias.sig.variability}
\end{figure*}

To make this point more quantitative, we consider two standard statistical 
measures of {\photoz} quality, the scatter and bias 
as functions of spectroscopic redshift, 
\begin{equation}
\sigma^2(z_j)\equiv(1/N_j)\Sigma_{i=1}^{N_j}|z_{{\rm phot},i} - z_{{\rm spec},i}|^2,
\end{equation}
\begin{equation}
b(z_j)\equiv(1/N_j)\Sigma_{i=1}^{N_j}(z_{{\rm phot},i} - z_{{\rm spec},i}),
\end{equation}
where $N_j$ is the number of objects in the $j$th \zspec \ bin, i.e., with true 
redshifts in the interval $z_j \pm \Delta z$.
Fig. \ref{fig.bias.sig.variability} shows these measures for the mock training sample 
(left panels) and photometric sample (right panels) for five 
different neural network solutions. These five solutions come from networks with 
the same structure (same number of layers and nodes per layer, see Appendix \ref{app:neu}) but 
with different initial values for the network weights $w_{i \alpha \beta}$. The 
left panels of Fig. \ref{fig.bias.sig.variability} 
show that the different solutions yield essentially identical results 
for the scatter and bias for the training set, but the right panels show a dispersion of 
quality measures for the photometric sample.  
We can address this issue by working with the average of the five {\photoz} solutions 
for each galaxy. 
The solid (black) curve in the top right panel 
of Fig. \ref{fig.bias.sig.variability} shows that the 
average {\photoz} solution results in a $b(z)$ that is the average of the biases of the individual 
neural net solutions, as may be expected. 
The bottom right panel of Fig. \ref{fig.bias.sig.variability} shows a more interesting 
result, that the scatter of the average {\photoz} solution is considerably
smaller than the average scatter of the individual neural net solutions.

\begin{figure*}
  \begin{minipage}[t]{85mm}
    \begin{center}
      \resizebox{85mm}{!}{\includegraphics[angle=0]{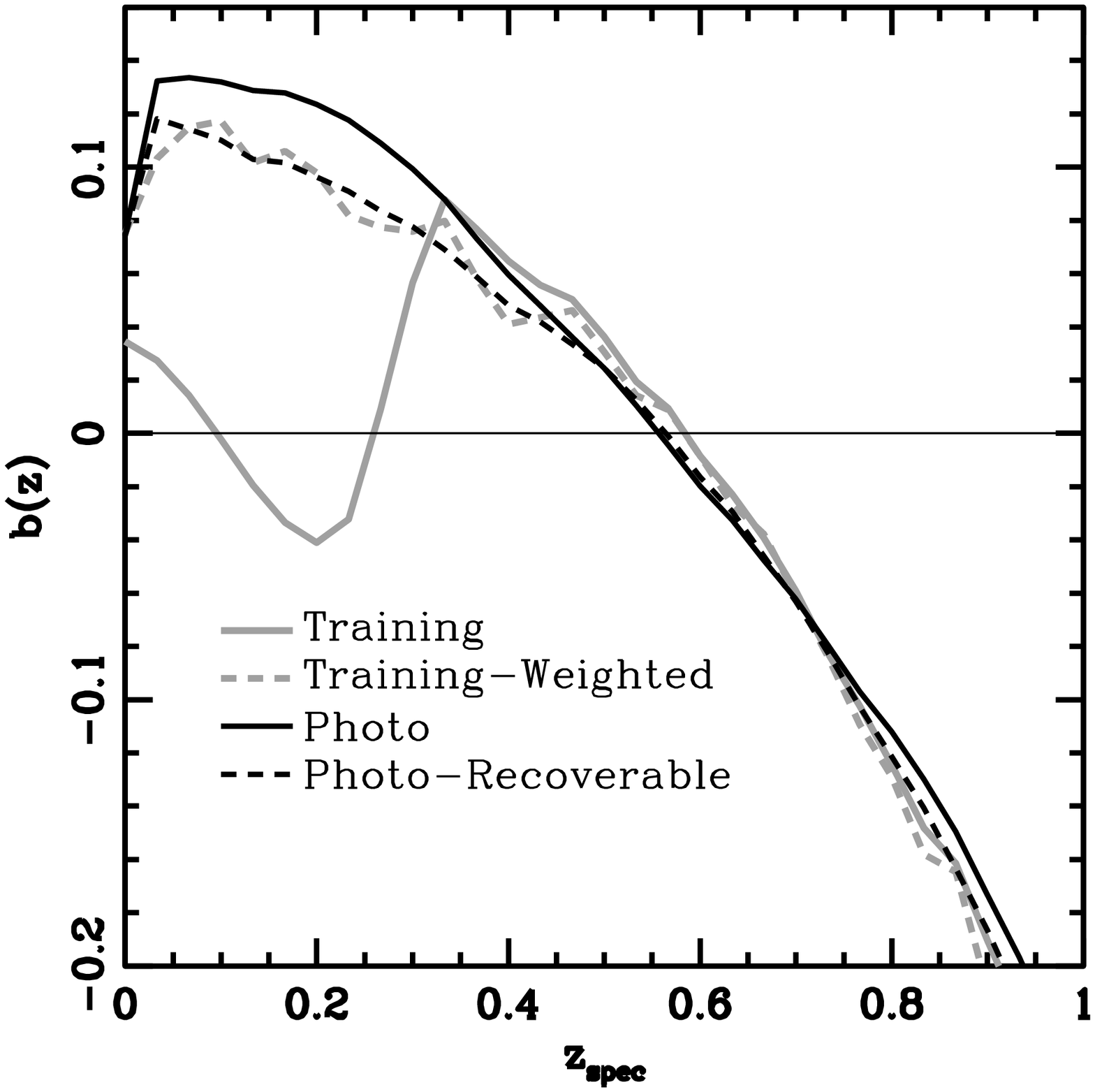}}
    \end{center}
  \end{minipage}
  \begin{minipage}[t]{85mm}
    \begin{center}
      \resizebox{85mm}{!}{\includegraphics[angle=0]{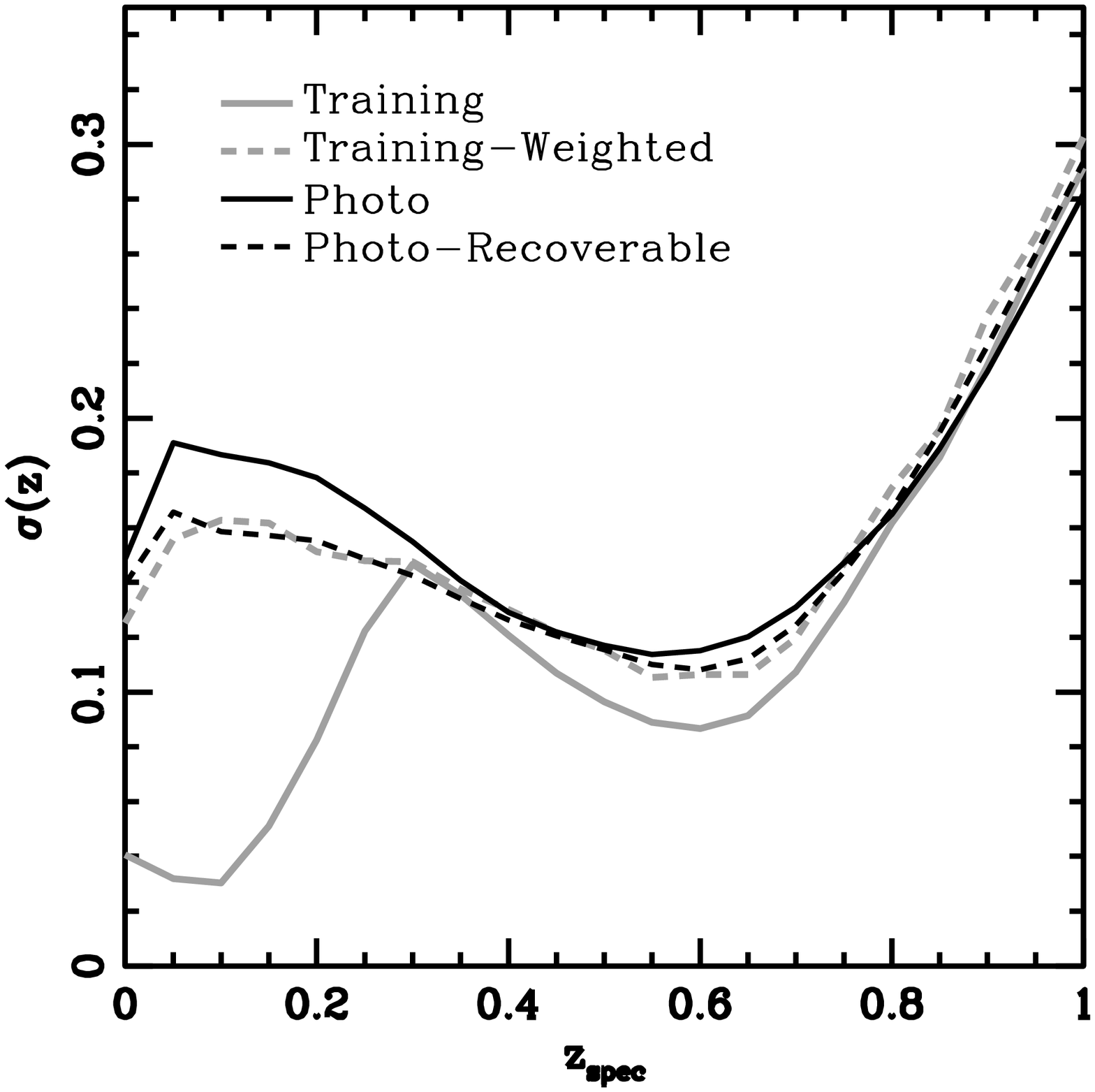}}
    \end{center}
  \end{minipage}
  \caption{({\it Left}) {\Photoz} bias vs. $z_{spec}$ and ({\it right}) scatter vs. $z_{spec}$
for the weighted and unweighted mock SDSS training set as well as for the mock photometric 
set and the recoverable photometric set.  The weighted training set results 
more accurately match those for the photometric set and very accurately 
match those for the recoverable photometric set.
}\label{fig.bias.sig.wei}
\end{figure*}

Even if one uses the average {\photoz} solution, comparison of the left and right 
panels of Fig. \ref{fig.bias.sig.variability} demonstrates the qualitative 
point made above, that the scatter and bias vs. 
redshift for the training set are not accurate estimators of the scatter and bias 
over the full redshift range for the photometric sample. As shown 
more explicitly in Fig. \ref{fig.bias.sig.wei}, the training 
set scatter and bias tend to underestimate those measures for the photometric 
sample, particularly at redshifts $z_{\rm spec} < 0.3$. 
This is simply because the training-set objects are generally brighter than 
those in the photometric set at similar redshift, 
which implies that the training-set galaxies have smaller photometric errors
and consequently smaller {\photoz} errors.
 
The weighting procedure provides a straightforward avenue for addressing this problem of 
estimating the {\photoz} scatter and bias for the photometric sample.
Since the {\it weighted} training set has, by construction, magnitude distributions
similar to those of the photometric set, we can instead
use weighted versions of $\sigma(z)$
 and $b(z)$ for the training set as estimates of the scatter and bias for 
the photometric set, i.e., 
\begin{equation}
\sigma^2_w(z_j)\equiv(1/N_j)\Sigma_{i=1}^{N_j}w_i|z_{{\rm phot},i} - z_{{\rm spec},i}|^2,
\end{equation}
\begin{equation}
b_w(z_j)\equiv(1/N_j)\Sigma_{i=1}^{N_j}w_i(z_{{\rm phot},i} - z_{{\rm spec},i}),
\end{equation}
where the weights $w_i$ are given by Eqn. (\ref{eqn:wei}) and the sums are over all 
objects in the training set.
Fig.~\ref{fig.bias.sig.wei} shows the scatter and bias for the training set, the weighted 
training set, and the full photometric set, where the average {\photoz} of the 
five neural network solutions has been used. We see that the weighted training 
set yields estimates of scatter and bias that are much closer to those of the photometric set over 
the entire redshift range. Moreover, as noted in \S \ref{sec:select}, we expect 
the weighting method to work best for the {\it recoverable} portion of the 
photometric sample. Fig. ~\ref{fig.bias.sig.wei} also shows the scatter 
and bias vs. redshift for the recoverable photometric sample, showing that 
the weighted training-set estimates are very accurate in this case.

Since the weights can be used to improve the estimates of 
{\photoz} scatter and bias for the photometric set, one 
might hope that the weights could also be used to improve the {\photoz} solution itself.
However, because of the large number of degrees of freedom of the ANN, most of 
the information for the {\photoz} solution comes from 
small regions in the space of photometric observables around each 
training-set object.
The weights do not vary strongly over those small regions,
and therefore the {\photoz} solution does not change
significantly between the unweighted and weighted cases.

\subsection{Estimating Photo-$z$ Errors}

As demonstrated above, the weighting procedure improves the 
estimates of {\photoz} scatter and bias for a photometric sample but 
does not improve the {\photoz} accuracy itself. Another issue, which 
we now discuss, is the accuracy of {\photoz} {\it error} estimates.  

We estimate {\photoz} errors for objects in the photometric catalog using 
the Nearest Neighbor Error (NNE) estimator \citep{oya08b}. 
The NNE method is training-set based and 
associates {\photoz} errors to photometric objects by considering the 
errors for objects with similar multi-band magnitudes in a spectroscopic
sample, hereafter termed the ``validation set''. The validation set is 
chosen to be independent of the training set in order to avoid the issue of 
over-fitting, i.e., so that the ANN is not trained to fit the statistical 
fluctuations of the training set, which would result in 
NNE underestimating the {\photoz} errors.

The NNE procedure to estimate the redshift error $\sigma_{\rm NNE}$ 
for a galaxy in the photometric
sample is as follows. Using 
the distance measure of Eqn. (\ref{eqn:dist_gen_def}),
we find the validation-set nearest neighbors in magnitude space to the galaxy of 
interest. 
Since the selected nearest neighbors are in the spectroscopic sample, 
we know their {\photoz} errors, $\delta z = z_{\rm phot}-z_{\rm spec}$, where 
$z_{\rm phot}$ has been estimated using the neural network method. 
We calculate the $68\%$ width of the $\delta z$ distribution 
for the neighbors and assign that number as the {\photoz} error 
estimate for the photometric galaxy. 
Here we select the nearest $100$ neighbors of each object to estimate its 
{\photoz} error. 
In studies of {\photoz} error estimators applied to mock and real galaxy catalogs, 
we found that NNE accurately predicts the {\photoz} error when the training set is 
representative of the photometric sample \citep{oya08b}. Here we 
investigate what happens when the training set is {\it not} representative, 
and we also consider the impact of weighting the neighbors using 
Eqn. (\ref{eqn:wei_def_num}) 
in computing the NNE estimate.

\begin{figure*}
  \begin{minipage}[t]{85mm}
    \begin{center}
      \resizebox{85mm}{!}{\includegraphics[angle=0]{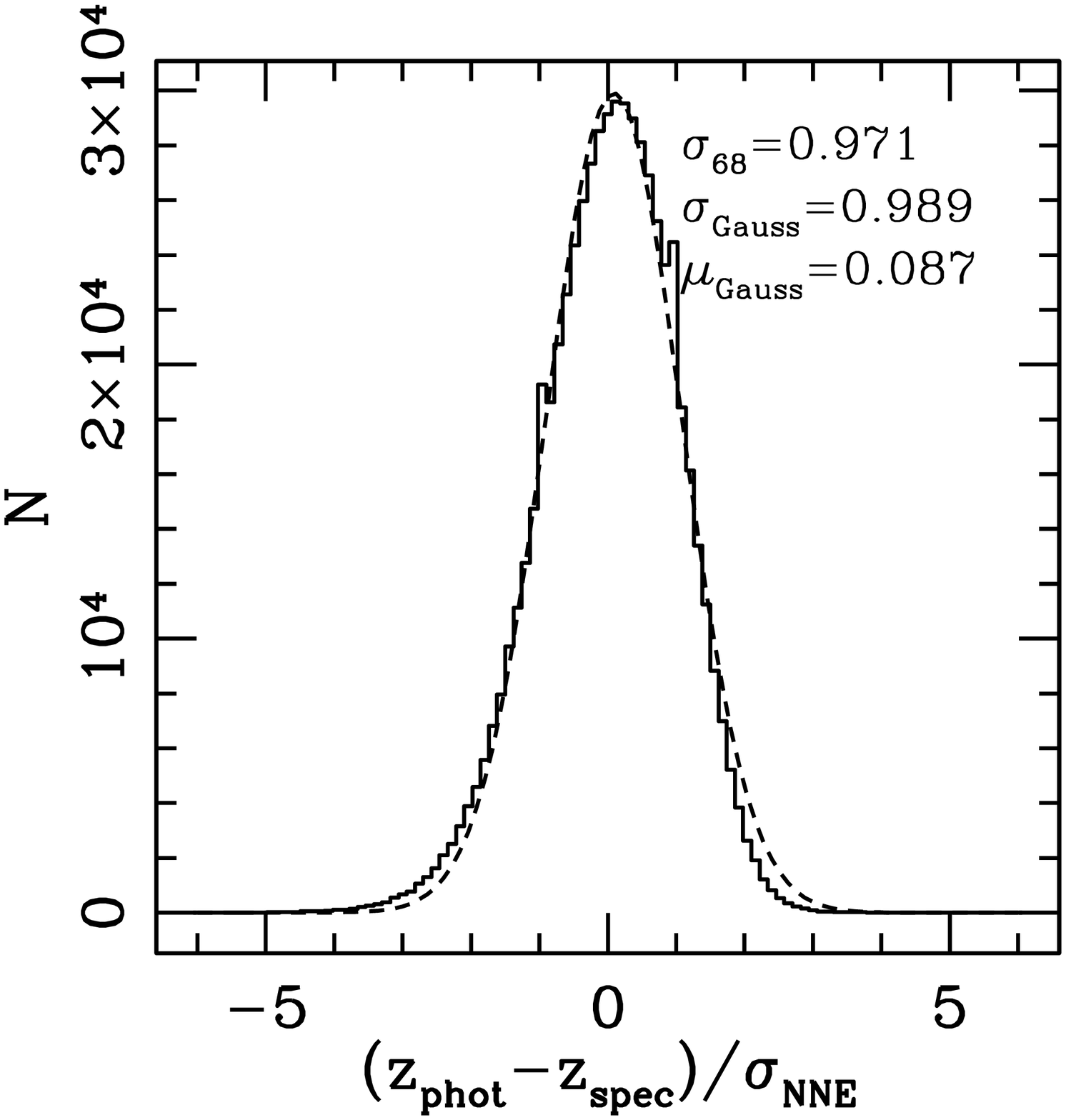}}
    \end{center}
  \end{minipage}
  \begin{minipage}[t]{85mm}
    \begin{center}
      \resizebox{85mm}{!}{\includegraphics[angle=0]{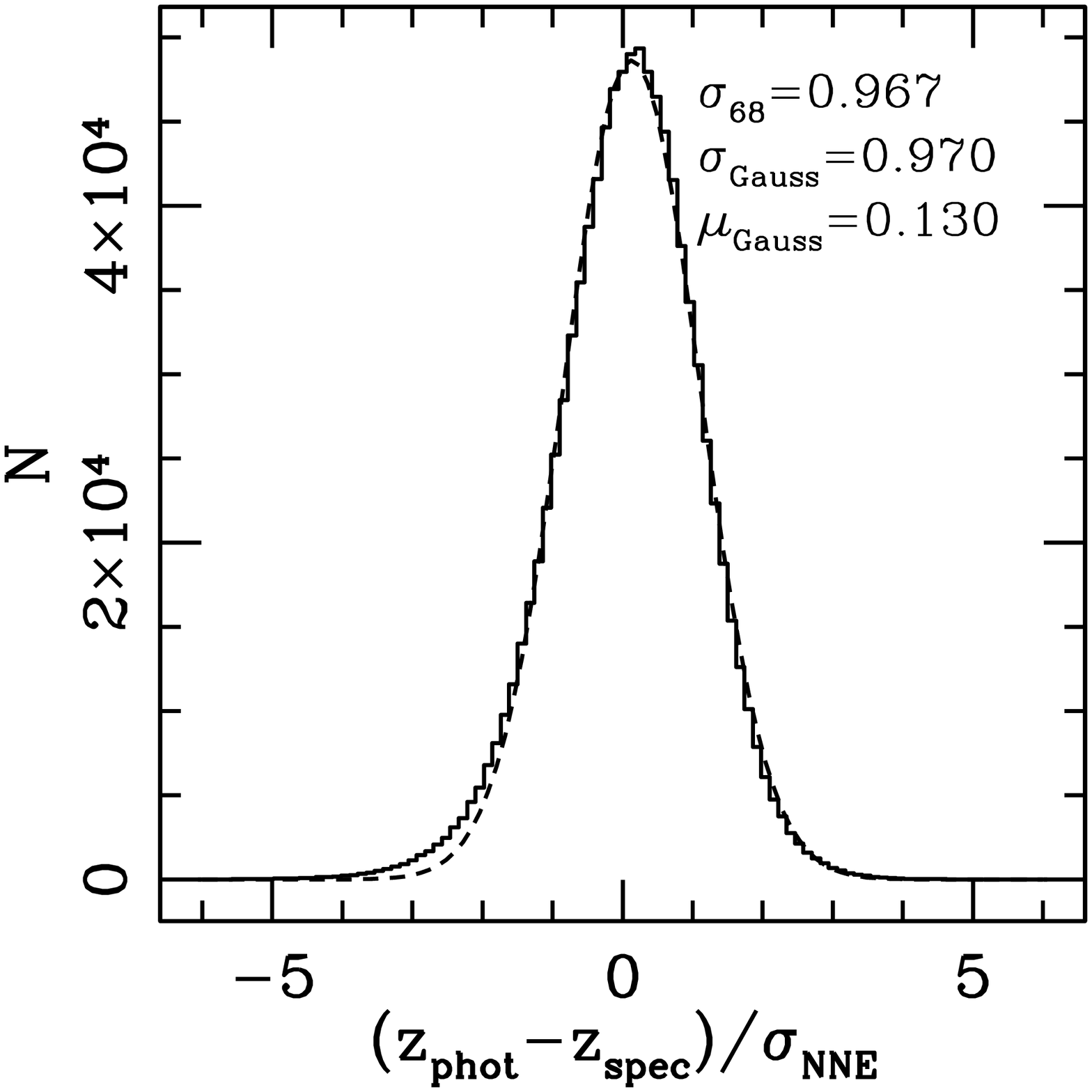}}
    \end{center}
  \end{minipage}
  \begin{minipage}[t]{85mm}
    \begin{center}
      \resizebox{85mm}{!}{\includegraphics[angle=0]{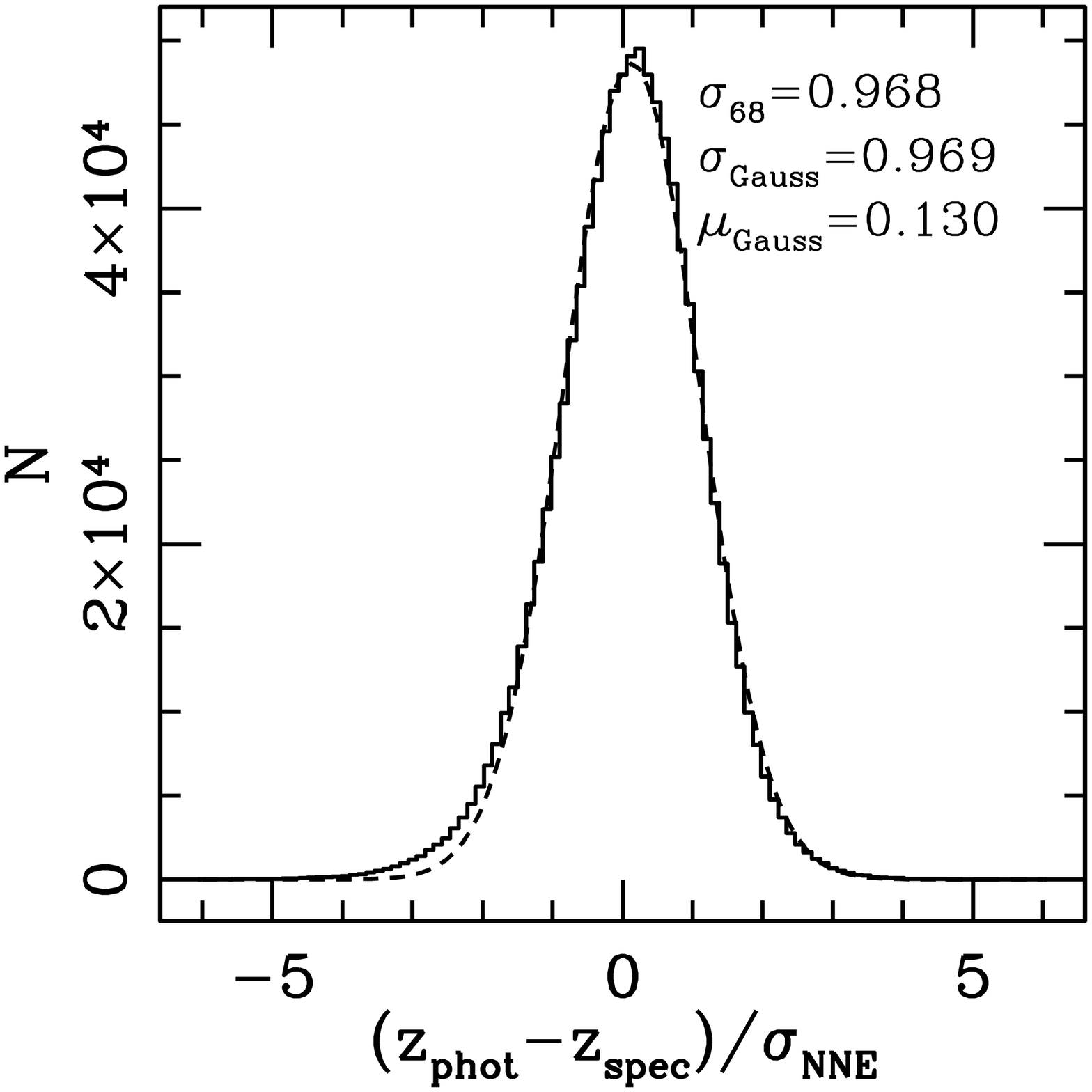}}
    \end{center}
  \end{minipage}
  \begin{minipage}[t]{85mm}
    \begin{center}
      \resizebox{85mm}{!}{\includegraphics[angle=0]{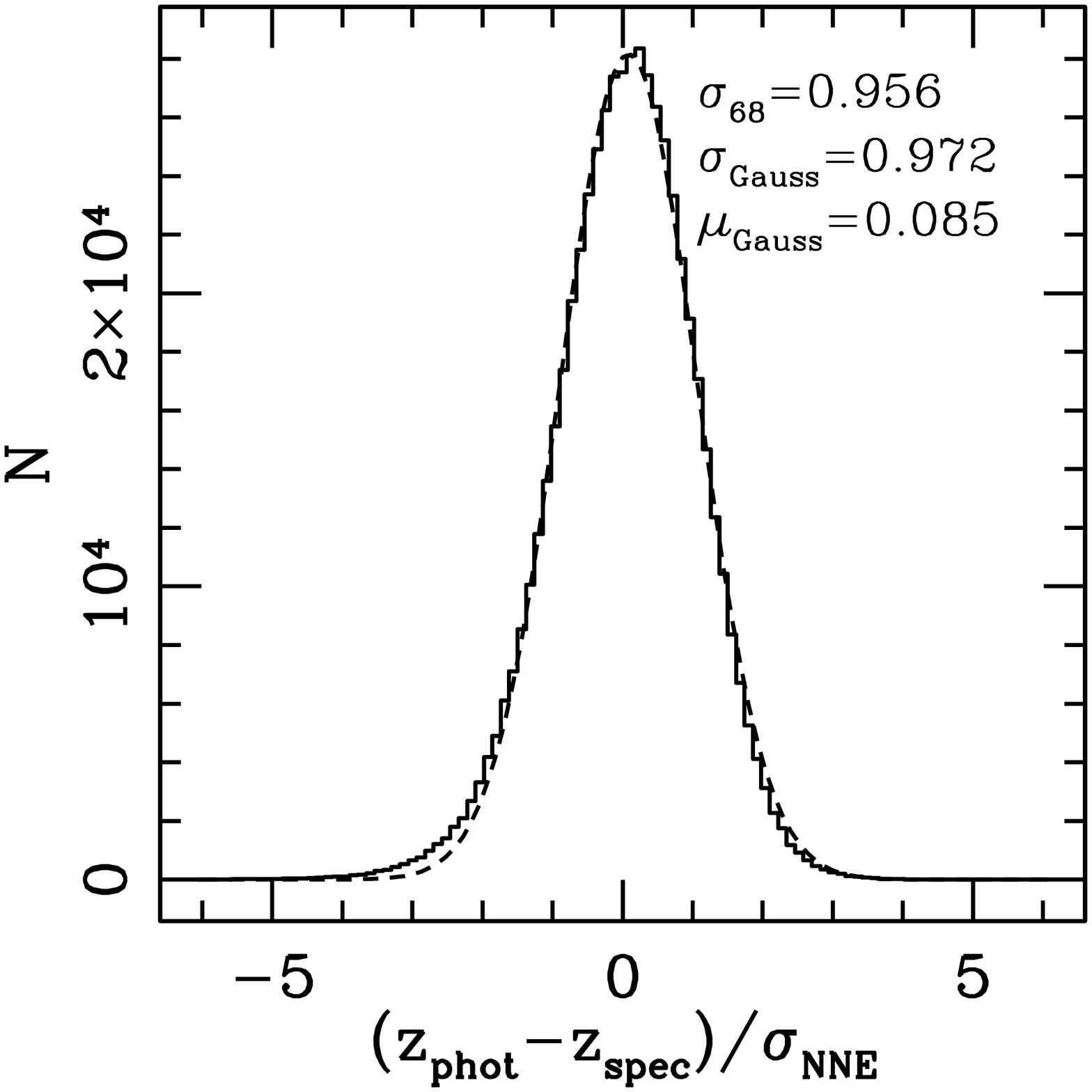}}
    \end{center}
  \end{minipage}
  \caption{Distributions of $(z_{\rm phot}-z_{\rm spec})/\sigma_{NNE}$ for the
({\it top left}) training set, 
({\it top right}) photometric set (using unweighted validation set), 
({\it bottom left}) photometric set (using weighted validation set), 
and ({\it bottom right}) recoverable photometric set 
(using weighted validation set). 
}\label{fig.gs}
\end{figure*}

Figure \ref{fig.gs} shows 
the distributions of $(z_{\rm phot}-z_{\rm spec})/\sigma_{\rm NNE}$,
i.e., the {\photoz} error distribution normalized by the NNE error 
estimate $\sigma_{\rm NNE}$, for the training set (upper left panel), for 
the photometric set using unweighted (upper right) and weighted (lower left) 
validation-set objects, and for the recoverable photometric set (lower 
right) using the weighted validation set. 
The dashed curves in these panels show 
Gaussian fits to the error distributions; we also 
indicate the best-fit Gaussian
means ($\mu_{\rm Gauss}$) and standard deviations 
($\sigma_{\rm Gauss}$), as well as the $\sigma_{68}$ widths (about zero) of the
distributions (not of the fits). 
The Gaussian fits give equal weight to each bin of the distributions and 
ignore objects for which $\sigma_{\rm NNE}=0$.
We see that the overall normalized error distributions are close to Gaussian 
for all the catalogs and that there is little difference among
the four cases. We conclude that the NNE error estimate is robust even 
when the training set is not representative and that  
the weights do not significantly affect
the NNE estimator.  In retrospect the latter is not too surprising since  
the NNE estimate is derived from a typically 
small nearest neighbor region, over which  
the weights do not vary strongly.

\subsection{Deconvolving the Photo-$z$ Distribution} \label{sec:dec}

The photometric redshift distribution is the convolution of the true 
redshift distribution $N(z_{\rm spec})$ 
with the distribution of photometric redshift errors.
For discrete distributions we can express this as
\begin{equation}
N(z_{\rm phot})_i=\sum_j P(z_{\rm phot}|z_{\rm spec})_{ij}N(z_{\rm spec})_j \label{eqn:dec}
\end{equation}
\noindent where the indices $i$ and $j$ refer to bins of $z_{\rm phot}$ and $z_{\rm spec}$, 
respectively, and  $P(z_{\rm phot}|z_{\rm spec})_{ij}$ is the probability that 
a galaxy has {\photoz} in bin $i$ given that its 
spectroscopic redshift is in bin $j$.

 As noted in \cite{pad05}, we can solve Eqn. (\ref{eqn:dec}) for 
$N(z_{\rm spec})$ by inverting  
$P(z_{\rm phot}|z_{\rm spec})_{ij}$. However, 
the inversion problem is ill-conditioned for two reasons.
First, the convolution is a smoothing operation, and some of the information in
$N(z_{\rm spec})_j$ is irretrievably lost in that process.
Second, small errors in $P(z_{\rm phot}|z_{\rm spec})_{ij}$ are magnified by the matrix inversion.

Both problems can be alleviated by using prior information to regularize the inversion and restore
some of the lost information. 
Following \cite{pad05}, we use a forward difference operator, defined as 
\begin{equation}
S=\sum_{j=0}^{N_{bin}-1}\Big (\Big [N(z)\Big ]_{j+1}-\Big [N(z)\Big ]_{j}\Big ),
\end{equation}
\noindent as a prior on the smoothness of the reconstruction.
To incorporate the prior information into the deconvolution procedure, 
we must represent the deconvolution as a minimization problem.
If we define 
\begin{equation}
E_0\equiv \sum_{i} \Big |P^{-1}(z_{\rm phot}|z_{\rm spec})_{ij} \Big [N(z_{\rm spec})\Big ]_j - \Big [N(z_{\rm phot})\Big ]_{i} \Big |^2 ~,
\end{equation}
\noindent then the deconvolution can be stated as the problem of minimizing $E_0$ with respect to
$N(z)$. To incorporate the prior, we define
\begin{equation}
E=E_0 + \lambda S ~, 
\end{equation}
\noindent and the regularized deconvolution is achieved by minimizing $E$. 
The parameter $\lambda$ sets how much importance is given to the smoothing 
and is often chosen {\it ad hoc}.  
Here, following \cite{pre92}, we set
\begin{equation}
\lambda=\frac{Tr\Big (P^T(z_{\rm phot}|z_{\rm spec}) \cdot P(z_{\rm phot}|z_{\rm spec})\Big )}{Tr\Big (B^T\cdot B\Big )} ~,
\end{equation}
\noindent where $B$ is the $(N_{bin}-1) \times (N_{bin})$ first difference matrix given by
$B=\delta_{(i+1)j}-\delta_{ij}$.
This choice of $\lambda$ gives comparable weight to both parts of the minimization.

The preceding discussion summarizes the ``standard'' {\photoz} deconvolution 
method for estimating the redshift distribution.  
The weighting method can provide a better estimate of 
$P(z_{\rm phot}|z_{\rm spec})_{ij}$ for the photometric sample,
reducing the need for regularization and thereby improving 
the deconvolution estimate of $N(z_{\rm spec})$. 
We can incorporate the weights into the estimation of 
$P(z_{\rm phot}|z_{\rm spec})_{ij}$ by 
calculating, for each $z_{\rm spec}$ bin, the 
$z_{\rm phot}$ distribution for the weighted training-set galaxies.

We postpone discussion of the performance of the deconvolution and 
weighted deconvolution methods to the next section, where we 
compare them with direct application of the weighting method to 
estimation of $N(z_{\rm spec})$. 

\section{Applications of the Weighting Method II: Estimates of $N(z)$ and $p(z)$ in mock photometric 
samples}

\label{sec:app2}

\subsection{The redshift distribution $N(z)$}\label{sec:appdist}

We now have at hand a number of methods for estimating the 
true redshift distribution $N(z)$ for a photometric galaxy sample. 
Using {\photoz}'s, one can simply use the {\photoz} distribution 
itself, $N(z_{\rm phot})$, as an estimator, or the deconvolved 
{\photoz} distribution described in \S \ref{sec:dec}, or the weighted,  
deconvolved {\photoz} distribution mentioned at the end of 
\S \ref{sec:dec}. Alternatively, one can use the weighted 
spectroscopic redshift distribution of the training-set galaxies 
to directly estimate $N(z)$, i.e., Eqn. (\ref{eqn:Nzest}), without 
recourse to {\photoz}'s. Finally, we can sum the redshift 
probability distributions $p(z)$ for each galaxy in the 
photometric sample (again estimated from the weighted 
training set) to estimate $N(z)$, using Eqn. (\ref{eqn:pzest}). 
In this section, we compare results of these different 
estimates of $N(z)$ using the mock SDSS DR6 sample. The 
results are summarized in Tables \ref{tbl:stats} and \ref{tbl:stats.20bins} 
and the best results
for each method are shown in Figs. \ref{fig.dists.full}, \ref{fig.dists.rec}, 
and \ref{fig.dists.wei}.

\subsubsection{Measures of Reconstruction Quality} \label{sec:qua}
To compare the different methods, we need a statistical 
measure of the quality of the reconstruction of the estimated 
redshift distribution. We use two.
The first is a $\chi^2$ statistic (per degree of freedom and 
per galaxy), defined here as

\begin{eqnarray}
(\chi^{2})^{\rm X}&\equiv&\frac{1}{N_{\rm bin}-1}\sum_{i=1}^{N_{\rm bin}}
            \frac{\left[N(z^{i})^{\rm X}-N(z_{\rm spec}^{i})^{\rm P}\right]^2}
                 {N(z_{\rm spec}^{i})^{\rm P} \Delta z}~. 
\end{eqnarray}

\noindent Here $N_{\rm bin}$ is the number of redshift bins used, $\Delta z$ is the width of 
the bins, and $N(z^{i})^{\rm X}$ is equal to 
$N(z_{\rm spec}^{i})^{\rm T}_{\rm wei}$ if the weighting procedure is used 
or to 
$N(z_{\rm phot}^{i})^{\rm P}$ if the redshift distribution is instead 
estimated using {\photoz}'s. The usual definition of $\chi^2$
uses the numbers  of objects in given bins instead
of the normalized probability $N(z^i)$; multiplying our
$\chi^2$ by $N_{P,{\rm tot}} \Delta z$ gives the usual definition. 
We chose the above statistic 
so that it is independent of the number of galaxies 
and the number of redshift bins, allowing 
us to more fairly compare reconstruction quality
across different data sets.
Since the probabilities are
normalized, the number of degrees of freedom is $N_{\rm bin}-1$. 

The second measure we employ is a binned version of 
the Kolmogorov-Smirnov (KS) statistic, defined as 
the maximum difference between the two cumulative redshift distributions 
being compared, for example, the cumulative distributions corresponding 
to  $N(z_{\rm spec}^{i})^{\rm T}_{\rm wei}$ and $N(z_{\rm spec}^{i})^{\rm P}$. 
The KS statistic is more sensitive to differences in the medians of the two 
distributions being compared, whereas the $\chi^2$ statistic tends to stress the 
regions of the distribution that are least well sampled, i.e., regions
where $N(z^i)$ is small.
In our implementation, we use binned cumulative distributions instead of
unbinned cumulative distributions, so this statistic is not strictly
the KS statistic.

Note that we do not use the absolute values of these statistics 
as formal goodness-of-fit measures. Rather, we use their relative 
values for the different estimators to compare the quality of the different 
reconstructions---see Table \ref{tbl:stats}.

\subsubsection{{\Photoz} Estimates of $N(z)$}

\begin{figure}
  \begin{minipage}[t]{85mm}
    \begin{center}
      \resizebox{85mm}{!}{\includegraphics[angle=0]{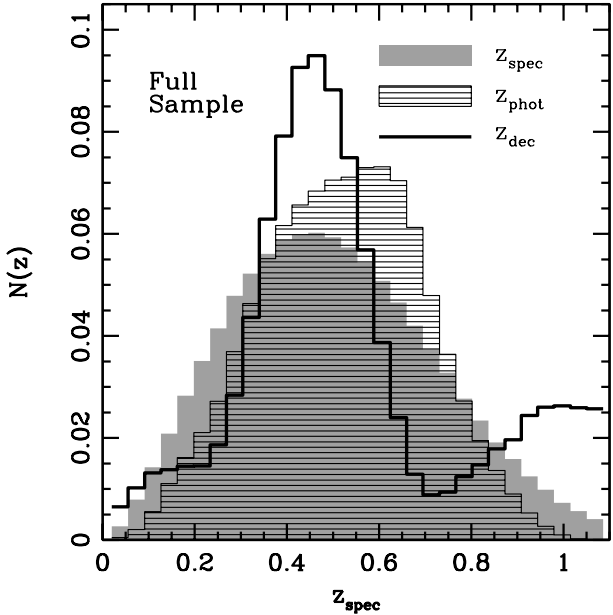}}
    \end{center}
  \end{minipage}
   \caption{True spectroscopic redshift distribution ({\it solid grey})
of the mock SDSS photometric sample, and estimates of the redshift distribution
using the {\photoz} distribution ({\it hatched}) and deconvolved {\photoz} 
distribution ({\it black line}).
}\label{fig.dists.full}
\end{figure}

\begin{figure}
  \begin{minipage}[t]{85mm}
    \begin{center}
      \resizebox{85mm}{!}{\includegraphics[angle=0]{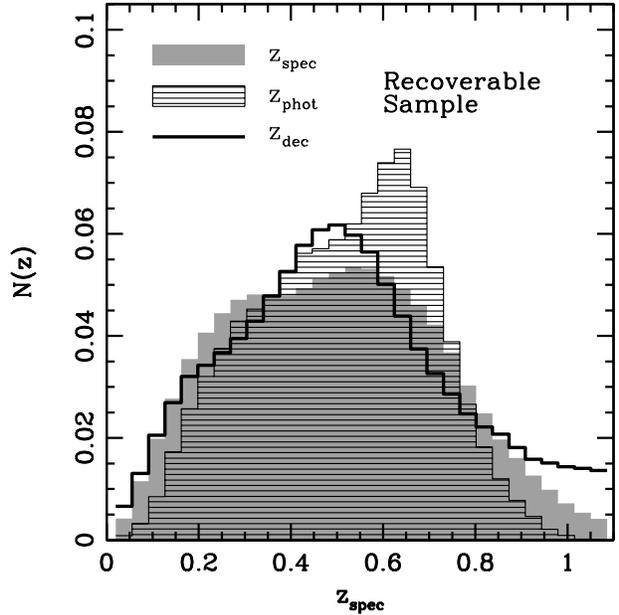}}
    \end{center}
  \end{minipage}
   \caption{True spectroscopic redshift distribution ({\it solid})
of the recoverable mock photometric sample, and estimates of the redshift distribution
using the {\photoz} distribution ({\it hatched}) and deconvolved {\photoz} 
distribution ({\it black line}).
}\label{fig.dists.rec}
\end{figure}

The {\photoz} estimate $N(z_{\rm phot})$ of the true redshift
distribution for the mock SDSS photometric 
sample is shown in Fig. \ref{fig.dists.full} (hatched histogram).
We can see that $N(z_{\rm phot})$ underestimates the true distribution, 
$N(z_{\rm spec})$ (grey histogram), 
at both low and high redshifts and overestimates it at intermediate redshifts,  
$0.4<z_{\rm spec}<0.8$.
In addition, the peak of $N(z_{\rm phot})$ is biased with respect to the 
peak of $N(z_{\rm spec})$. Comparing the two distributions, 
we find that $\chi^2=0.107$ and ${\rm KS}=0.0848$. 
The {\photoz} and true redshift distributions for the recoverable photometric 
sample are shown in Fig. \ref{fig.dists.rec} (hatched and grey histograms).
Again, $N(z_{\rm phot})$ 
underestimates $N(z_{\rm spec})$ at low and high redshifts and 
overestimates it in between. 
The reconstruction statistics are similar to those for the full 
photometric sample, $\chi^2=0.105$ and ${\rm KS}=0.0674$. This 
indicates that the faithfulness of $N(z_{\rm phot})$ as an 
estimate of the true redshift distribution 
is not very sensitive to whether the training set is representative 
of the photometric sample: the errors in the recovered redshift 
distribution are dominated by a systematic effect.
The fact that $N(z_{\rm phot})$ is more sharply peaked than 
$N(z_{\rm spec})$ is a common feature of 
training-set-based {\photoz} estimates and results from the breakdown  
of the fundamental {\photoz} assumption that a single {\zphot} can 
represent a full redshift distribution \footnote{Maximum-likelihood 
template-fitting {\photoz} methods suffer from a similar problem but with 
opposite consequences. 
Because of the different way in which $p(z|{\rm observables})$ is 
estimated in those cases, $N(z_{\rm phot})$ tends to be flatter than 
the true redshift distribution \citep{bro06}}.
For the full photometric sample the peak in $N(z_{\rm phot})$ is not as  
pronounced as it is for the recoverable photometric sample,  
because the larger {\photoz} scatter in regions not covered 
by the training set smoothes out the peak.

We have also tested the {\photoz} deconvolution method of \S \ref{sec:dec} 
as an estimate of the redshift distribution. 
The standard (unweighted) deconvolution was not successful at recovering 
$N(z)$, with $\chi^2=0.577$, ${\rm KS}=0.124$ for the 
full photometric sample and $\chi^2=0.499$, ${\rm KS}=0.140$ for the recoverable
photometric sample. The result for the {\it weighted} deconvolution method, 
where the weights have been estimated using the five nearest 
neighbors, is shown by 
the black line in Fig. \ref{fig.dists.full}; it is also not very effective for the
full photometric sample, with $\chi^2=0.521$ and ${\rm KS}=0.989$. 
Although the peak of the deconvolved redshift distribution is at the 
correct redshift, the distribution shows an oscillatory behavior with redshift.
However, as shown in Fig. \ref{fig.dists.rec} (black line vs. grey histogram), 
the weighted deconvolution performs much better for the recoverable 
photometric sample, with
$\chi^2=0.0648$ and ${\rm KS}=0.0266$.

The deconvolution estimate of the redshift distribution 
oscillates about the true distribution.
This kind of behavior is typical of the inversion techniques used to 
perform the deconvolution. It 
can be alleviated by either increasing the training-set size, decreasing the number
of redshift bins, or using prior knowledge to improve the 
estimate of $P(z_{\rm phot}|z_{\rm spec})$.
We briefly investigate the second of these possibilities. 
As Table \ref{tbl:stats.20bins} shows, using only 20 as opposed to 30 
redshift bins improves the deconvolution estimate,
$\chi^2=0.0509$ and ${\rm KS}=0.0235$ (here 
with weights calculated using the two nearest neighbors). However, 
fewer bins means coarser redshift information, 
so it would be preferable to find a 
method that can accommodate a large number of redshift bins.
Table \ref{tbl:stats.20bins} also shows that the other methods are 
not as sensitive to the number of bins.
The deconvolution can also be improved by Monte Carlo resampling the training 
set \citep{pad05}.
Ideally, the resampling should be done in the space of observables used to 
calculate the {\photoz}'s. 
However, this approach is prohibitively time consuming for large datasets,
and it requires accurate knowledge of the magnitude errors -- which may be hard
to obtain.

\subsubsection{Weighting Method Estimates of $N(z)$}\label{sec:weires}

\begin{figure}
  \begin{minipage}[t]{85mm}
    \begin{center}
      \resizebox{85mm}{!}{\includegraphics[angle=0]{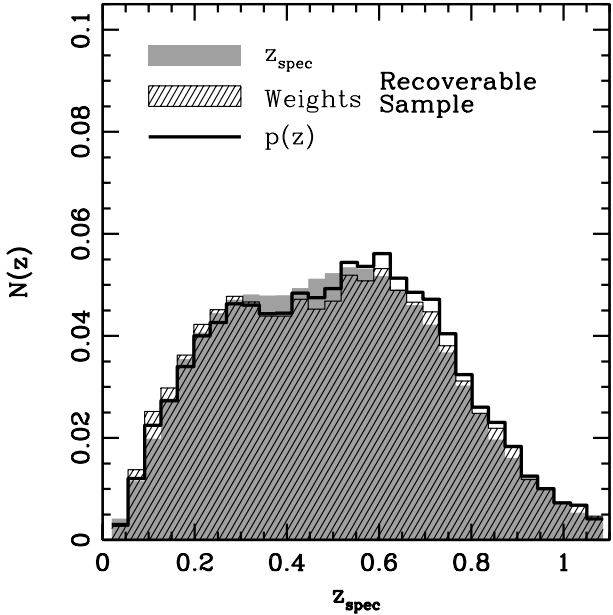}}
    \end{center}
  \end{minipage}
   \caption{True spectroscopic redshift distribution ({\it solid grey})
of the recoverable mock photometric sample, 
and estimates of the redshift distribution
using the weights ({\it hatched}) and $p(z)$ ({\it line}) methods. 
}\label{fig.dists.wei}
\end{figure}

The direct estimate of the redshift distribution for the photometric 
sample using the weighting method of Eqn. (\ref{eqn:Nzest}), $N(z)_{\rm wei}$, 
is shown by the 
hatched region in Fig. \ref{fig.dists.wei}. By construction, this 
estimate is the same for both the full and the recoverable photometric 
samples, that is, the weighting method in practice provides an estimate 
of the redshift distribution for the recoverable photometric sample. 
Comparison with the true redshift distribution of the recoverable 
sample (solid grey histogram in Fig. \ref{fig.dists.wei}) shows that the 
weighting method provides the best redshift distribution 
estimate of any of the methods under consideration here. For 
the full photometric sample, 
$\chi^2=0.0341$ and ${\rm KS}=0.0456$ (using 100 nearest neighbors), 
and for the recoverable sample, $\chi^2=0.00571$ and ${\rm KS}=0.0145$ 
(with 5 nearest neighbors). 
As shown in Table \ref{tbl:stats}, $N(z)_{\rm wei}$ is relatively insensitive to 
the number of neighbors used in the calculation.

Finally, using the sum of the $p(z)$ estimates for each galaxy 
in the photometric sample is almost 
identical to using the weights to estimate $N(z)$.
The estimate $N(\sum p(z))$ of the redshift distribution is shown 
by the solid black line in Fig. \ref{fig.dists.wei}, 
using 5 nearest neighbors to estimate $p(z)$. For this case, from  
Table \ref{tbl:stats} we have  $\chi^2=0.00493$, ${\rm KS}=0.0241$ 
for the 
recoverable photometric set, quite close to the values for the 
$N(z)_{\rm wei}$ estimate. Table \ref{tbl:stats} also shows that 
using fewer nearest neighbors slightly improves the KS statistic but not 
the $\chi^2$ statistic. Moreover, 
by using fewer neighbors, one is unable to accurately 
characterize $p(z)$, so we caution
against using fewer than 100 neighbors in the weighted estimate of $p(z)$.

\begin{table}
\caption{Redshift Distribution Reconstruction Statistics - 30 bins}
\begin{center}
\leavevmode
\begin{tabular}{ l l l } \hline \hline
\multicolumn{1}{c}{Full Photometric Set} & \multicolumn{1}{c}{$\chi^2$} & \multicolumn{1}{c}{KS parameter}\\
\hline 
{\Photoz} & 0.107 &0.0848 \\
{\Photoz} deconvolution (no weights) &0.577 &0.124 \\
{\Photoz} deconvolution ($100 ~{\rm nb}$) &0.521 &0.0989  \\
Weights ($100 ~{\rm nb}$)&0.0341 &0.0456 \\
\hline
\multicolumn{1}{c}{Recoverable Photometric Set} & \multicolumn{2}{c}{}  \\
\hline
{\Photoz} &0.105 & 0.0674 \\
{\Photoz} deconvolution (no weights) &0.499 &0.140 \\
{\Photoz} deconvolution ($2 {\rm nb}$) & 0.0682 & 0.0295 \\
{\Photoz} deconvolution ($5 {\rm nb}$) &0.0648 &0.0266 \\
{\Photoz} deconvolution ($100 {\rm nb}$) &0.102 & 0.0351 \\
Weights ($2 ~{\rm nb}$)& 0.00624 &0.0129\\
Weights ($5 ~{\rm nb}$)& 0.00571 &0.0145 \\
Weights ($100 ~{\rm nb}$)&0.00643 &0.0246 \\
$p(z)$ ($2 ~{\rm nb}$) &0.00540 &0.0219 \\
$p(z)$ ($5 ~{\rm nb}$) &0.00493 & 0.0201 \\
$p(z)$ ($100 ~{\rm nb}$) &0.00534 &0.0241 \\
\hline \hline
nb = neighbors
\end{tabular}
\end{center}
\label{tbl:stats}
\end{table}

\begin{table}
\caption{Redshift Distribution Reconstruction Statistics - 20 bins}
\begin{center}
\leavevmode
\begin{tabular}{ l l l } \hline \hline
\multicolumn{1}{c}{Recoverable Photometric Set} & \multicolumn{1}{c}{$\chi^2$} & \multicolumn{1}{c}{KS parameter}\\
\hline
{\Photoz} &0.105 & 0.0674 \\
{\Photoz} deconvolution (no weights) &0.404 &0.125 \\
{\Photoz} deconvolution ($2 ~{\rm nb}$) & 0.0509 & 0.0235 \\
{\Photoz} deconvolution ($5 ~{\rm nb}$) &0.0566  & 0.0232 \\
{\Photoz} deconvolution ($100 ~{\rm nb}$) & 0.0971 &0.0290 \\
Weights ($2 ~{\rm nb}$)& 0.00484 &0.0129 \\
Weights ($5 ~{\rm nb}$)& 0.00467 &0.01327 \\
Weights ($100 ~{\rm nb}$)&0.00547 &0.0232 \\
\hline \hline
nb = neighbors
\end{tabular}
\end{center}
\label{tbl:stats.20bins}
\end{table}

\subsubsection{Error estimate for $N(z_{\rm wei})$}
\label{subsubsec:errest}

From Eqns. \ref{eqn:wei} and \ref{eqn:den_m_gen_def}, the 
errors in the weights depend upon the uncertainties  
in determining the volumes of the training-set and photometric-set 
regions around an object and upon the uncertainties 
in the number of nearest neighbors for both the training and photometric sets.  
All of these quantities are correlated, making error estimation for the weighting method  
a challenge. Instead, we apply a bootstrap resampling procedure to directly estimate
the errors in the quantity of interest, in this case the 
weighted estimate of the redshift distribution, $N(z)_{\rm wei}$.  
We sample with replacement from 
the training and photometric sets to generate resampled
training and photometric sets of the same sizes as the originals. 
Then, for each pair of resampled training and photometric sets, 
we calculate the weights using Eqn. (\ref{eqn:wei_def_num}) 
and generate $N(z)_{\rm wei}$ using Eqn. (\ref{eqn:Nzest}).
We repeat this procedure $10,000$ times 
and estimate the covariance matrix by
\begin{eqnarray}
C(z_\alpha,z_\beta) &=& {1 \over n_s-1} \times \nonumber \\
\sum_{i=1}^{n_s} && \hspace{-0.8cm} (\hat{N}_i(z_\alpha)-\langle 
\hat{N}(z_\alpha)\rangle)(\hat{N}_i(z_\beta)-\langle \hat{N}(z_\beta)\rangle)~,
\label{eqn:boot}
\end{eqnarray}
\noindent where $n_s$ is the number of bootstrap samples, $\hat{N}_i(z)$ 
is the weighted estimate of the redshift distribution in the $i$th bootstrap 
sample, and $\langle \hat{N}(z)\rangle$ is the mean of the bootstrap estimates.
The correlation matrix is defined in the usual way by $\rho(z_\alpha,z_\beta) = 
C(z_\alpha,z_\beta)/\sigma(z_\alpha)\sigma(z_\beta)$. 

\begin{figure}
  \begin{minipage}[t]{85mm}
    \begin{center}
      \resizebox{85mm}{!}{\includegraphics[angle=0]{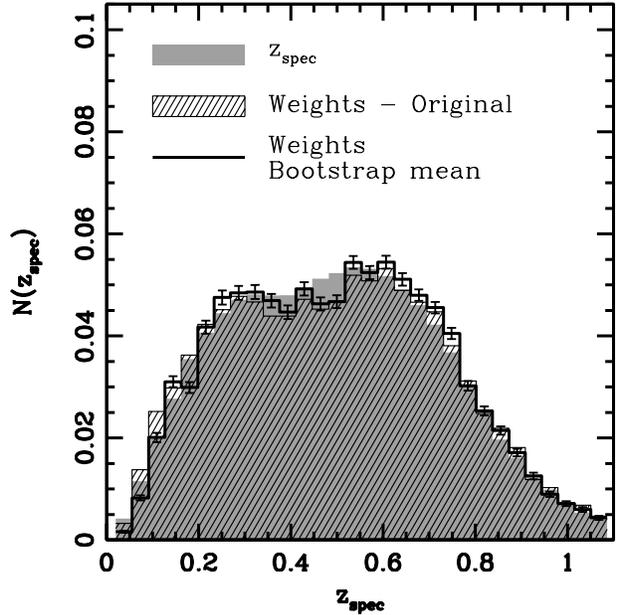}}
    \end{center}
  \end{minipage}
 \caption{
True spectroscopic redshift distribution ({\it solid grey})
of the recoverable photometric set, the estimated redshift distribution
using the weighting method ({\it hatched region}),  
and the mean of the bootstrap samples for the weighting method ({\it black line}).
The error bars are given by the square root of the diagonal terms of the
covariance matrix calculated from the bootstrap samples. 
}\label{fig:errs}
\end{figure}

Fig. \ref{fig:errs} shows $N(z)_{\rm wei}$ (hatched), 
the mean of the bootstrap estimates (solid black), and 
error bars given by the square root of 
the diagonal elements of the covariance matrix. 
There are small anti-correlations between nearby redshift bins, of at most -0.2.
Correlations between non-adjacent bins are smaller by at least an order of magnitude.

\subsubsection{Correcting Systematic Errors in the $N(z)$ Estimate}\label{sec:syst.ind}

From Fig. \ref{fig.dists.wei}, 
we note that the $N(z_{\rm wei})$ distribution is slightly 
flatter than $N(z_{spec})$, a feature that 
also shows up in other catalogs \citep[see, e.g.,][]{lim08}. This smoothing 
of the redshift distribution is 
a consequence of using non-negligibly 
small regions in magnitude space around the training-set 
galaxies to estimate the weights. This is especially 
problematic for regions where 
the training set is sparse, for then the ``neighbor volume'' used 
to calculate the weights 
may be large compared to the typical 
scale of change of the redshift/observable hypersurface. 
The problem is compounded when photometry errors 
are large, because large errors broaden 
the redshift distribution in a bin of observables.
Broader distributions require a larger number of training-set objects in order to 
be well characterized, but increasing the number of 
training-set nearest neighbors in the weights calculation 
increases the non-locality of the estimate.
The ideal solution would be to increase the total number of 
training-set objects in the sample, or at least the number in 
sparsely covered regions, but that is 
not always an option. 
The poor man's alternative is to develop ways to 
characterize and correct for the systematic errors. 

\begin{figure}
  \begin{minipage}[t]{85mm}
    \begin{center}
      \resizebox{85mm}{!}{\includegraphics[angle=0]{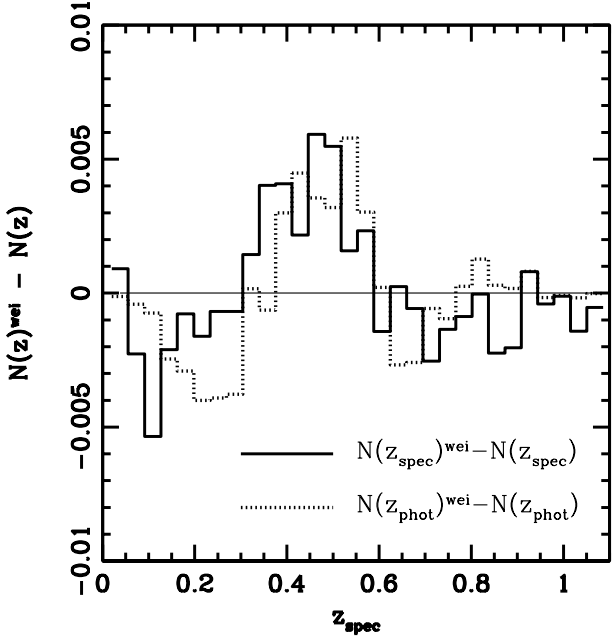}}
    \end{center}
  \end{minipage}
 \caption{Bias in the weighting-method estimate of the 
redshift distribution for the recoverable photometric set. Solid line 
shows the bias in the true redshift distribution. 
Dotted line shows the bias in the weighted {\photoz} 
distribution, also for the recoverable photometric set. Since they 
approximately match, we can use the 
bias in the weighted {\photoz} distribution, which is an observable, 
to estimate the bias in
the weighted true redshift distribution.
} \label{fig.systematic}
\end{figure}

An empirical approach we have developed makes use of the photometric 
redshifts in the following way. Starting with the training set, compute the {\photoz} 
distribution of the {\it weighted} training set, $N(z_{\rm phot})_{\rm wei}$,  
i.e., use Eqn. (\ref{eqn:Nzest}) but with $z$ replaced by {\zphot} 
everywhere. The difference between $N(z_{\rm phot})_{\rm wei}$ and 
the {\photoz} distribution for the photometric sample, $N(z_{\rm phot})$, 
is shown by the dotted line in Fig. \ref{fig.systematic}. The bias 
we are actually interested in is $N(z_{\rm spec})_{\rm wei}-N(z_{\rm spec})$, 
shown by the solid line in Fig. \ref{fig.systematic}. We see that these 
two differences have similar behavior with redshift, presumably due 
to similar non-locality of the weight solution in regions where the 
training set is sparse. We can therefore use $N(z_{\rm phot})_{\rm wei}-N(z_{\rm phot})$, 
the bias in the weighted {\photoz} distribution and 
which is an observable for the photometric sample, to estimate 
 $N(z_{\rm spec})_{\rm wei}-N(z_{\rm spec})$, the systematic 
error in the weighted estimate of the true redshift distribution. 
The redshift distribution estimate can then be approximately corrected 
for this bias.

To reduce the effect of random errors in the estimation of the bias, we smooth  
$N(z_{\rm phot})_{\rm wei} - N(z_{\rm phot})$ using 
a ``moving window'' method. Each redshift window 
has width greater than half of the separation 
between window centroids. The smoothing factor is the 
ratio of the window size to the redshift bin size when no smoothing is used. 
We have used smoothing factors of 1, 2, 3, and 5 to calculate  
$N(z_{\rm phot})_{\rm wei} - N(z_{\rm phot})$.
A smoothing factor of 1 corresponds to a window size of 0.0367 in redshift.
We picked the other smoothing factors based on the natural scales set by the $\sigma$ 
and $\sigma_{68}$ of the {\photoz}'s in the training and photometric sets.

\begin{figure}
  \begin{minipage}[t]{85mm}
    \begin{center}
      \resizebox{85mm}{!}{\includegraphics[angle=0]{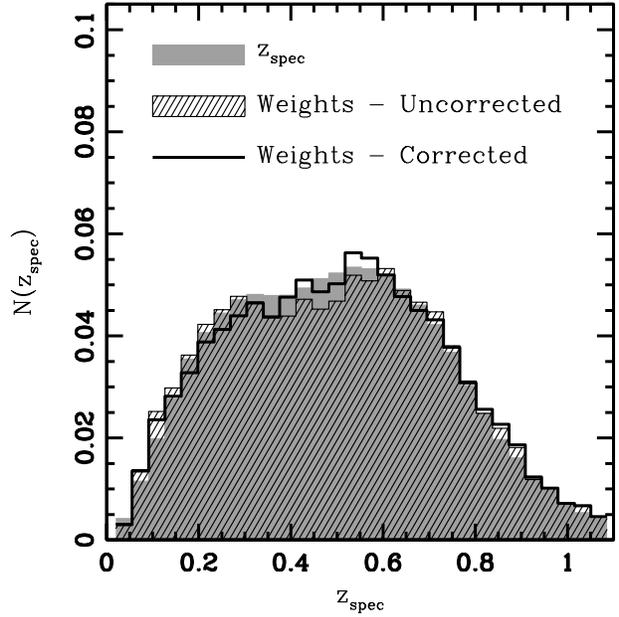}}
    \end{center}
  \end{minipage}
 \caption{True spectroscopic redshift distribution ({\it solid grey})
of the recoverable photometric set, and the estimated redshift distribution
using the weighting method, showing both the uncorrected results ({\it hatched}) 
and results corrected for systematic errors ({\it black line}) as described
in the text.
}\label{fig.sys.cor}
\end{figure}

Table \ref{tbl:stats.syst} shows the recovery statistics for the distributions corrected 
for systematics in this way, and Fig. \ref{fig.sys.cor} shows the improvement in 
the $N(z)$ estimate when the correction with smoothing factor of 2 is applied.
While these results are suggestive, more testing should be done before 
adopting this method as a correction for systematic errors in practice.

\begin{table}
\caption{Redshift Distribution Reconstruction Statistics - Correction of Systematics}
\begin{center}
\leavevmode
\begin{tabular}{ l l l } \hline \hline
\multicolumn{3}{c}{Recoverable Photometric Set - 5 neighbors - $0<z<1.1$}  \\
\hline
\multicolumn{1}{c}{Smoothing factor} & \multicolumn{1}{c}{$\chi^2$} & \multicolumn{1}{c}{KS parameter}  \\
\hline
No correction & 0.00571 & 0.0145\\
Unsmoothed & 0.00487 & 0.0151\\
2 &  0.00351 & 0.0127 \\
3 &  0.00349 & 0.0134 \\
5 &  0.00355 & 0.0131 \\
\hline
Bootstrap mean (no correction) &0.00600 &0.0189 \\
\hline \hline
\end{tabular}
\end{center}
\label{tbl:stats.syst}
\end{table}

\subsection{The Probability Distribution $p(z)$}

In this section we examine the effectiveness of the weighted training set in 
estimating the redshift probability distribution $p(z)$ for 
individual galaxies,
and the relation between $p(z)$, \zphot, and \zspec. For this study, 
we have increased the size of the mock 
photometric set to $9,000,000$ galaxies in order to improve the 
statistics.  
As before, we calculate the training-set estimate of $p(z)$, hereafter 
$p(z_{\rm train})$, for a training-set galaxy by selecting its 100 nearest 
neighbors in the training set. 
The spectroscopic 
redshift distribution of these objects is $p(z_{\rm train})$.
We then select all the galaxies in the photometric sample that are 
closer to the given galaxy  
in magnitude space than its $100^{th}$-nearest training-set neighbor.
The spectroscopic redshift distribution of the selected 
photometric galaxies is, barring 
statistical fluctuations  and non-locality, 
the true redshift distribution, hereafter 
$p(z_{\rm true})$, of the region of observable 
space centered about the selected 
galaxy.

\begin{figure*}
  \begin{minipage}[t]{58mm}
    \begin{center}
      \resizebox{58mm}{!}{\includegraphics[angle=0]{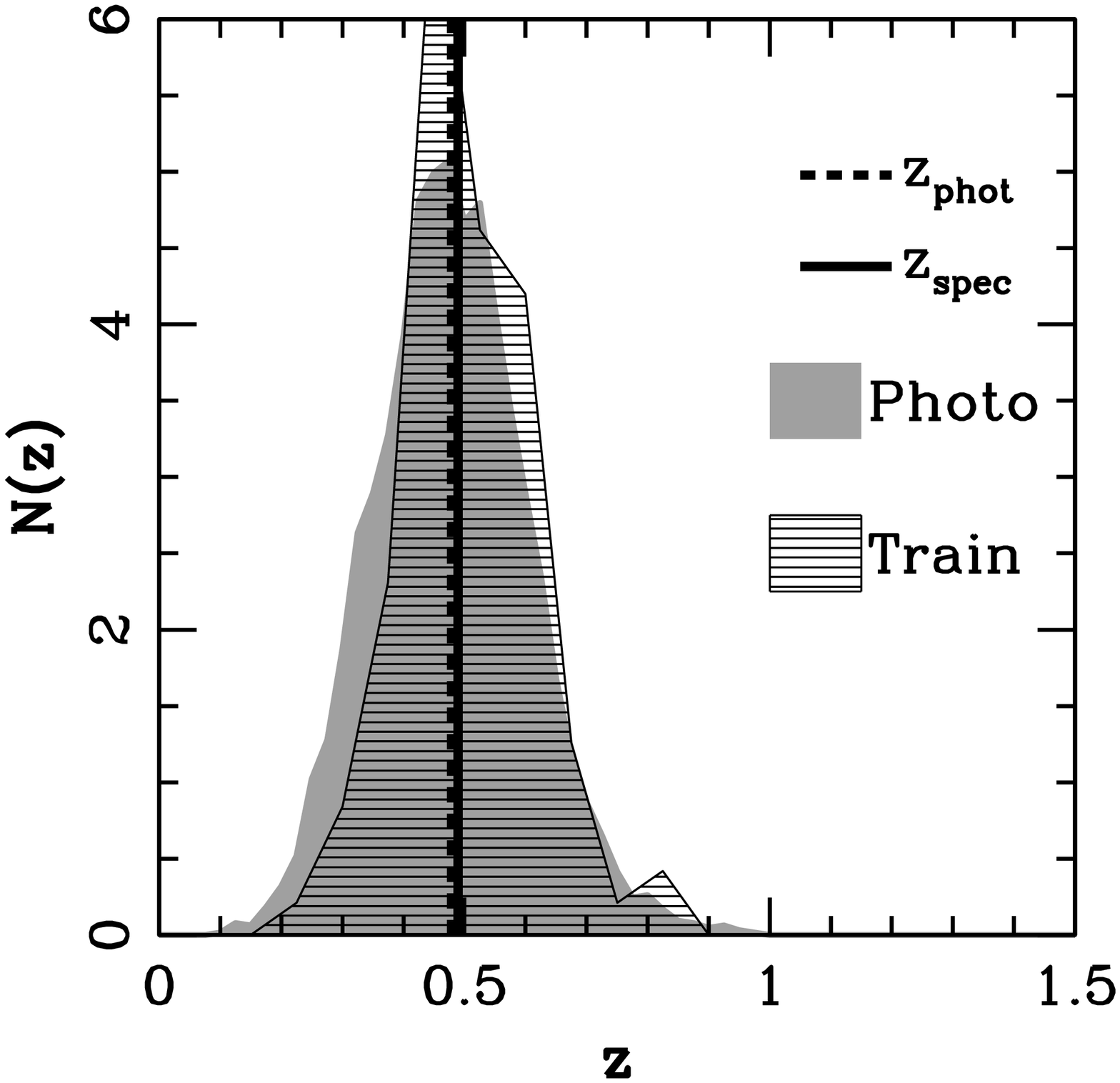}}
    \end{center}
  \end{minipage}
  \begin{minipage}[t]{58mm}
    \begin{center}
      \resizebox{58mm}{!}{\includegraphics[angle=0]{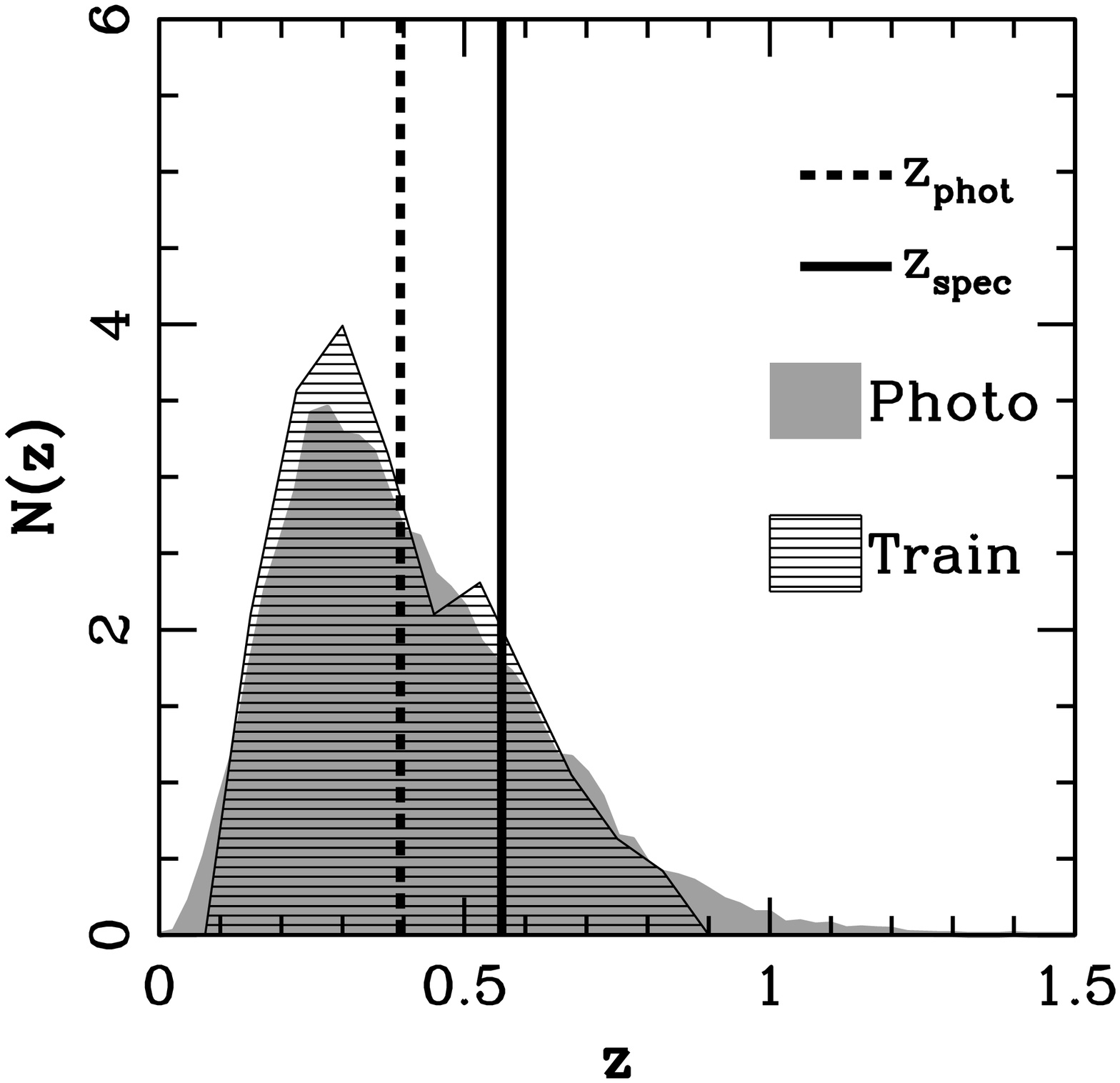}}
    \end{center}
  \end{minipage}
  \begin{minipage}[t]{58mm}
    \begin{center}
      \resizebox{58mm}{!}{\includegraphics[angle=0]{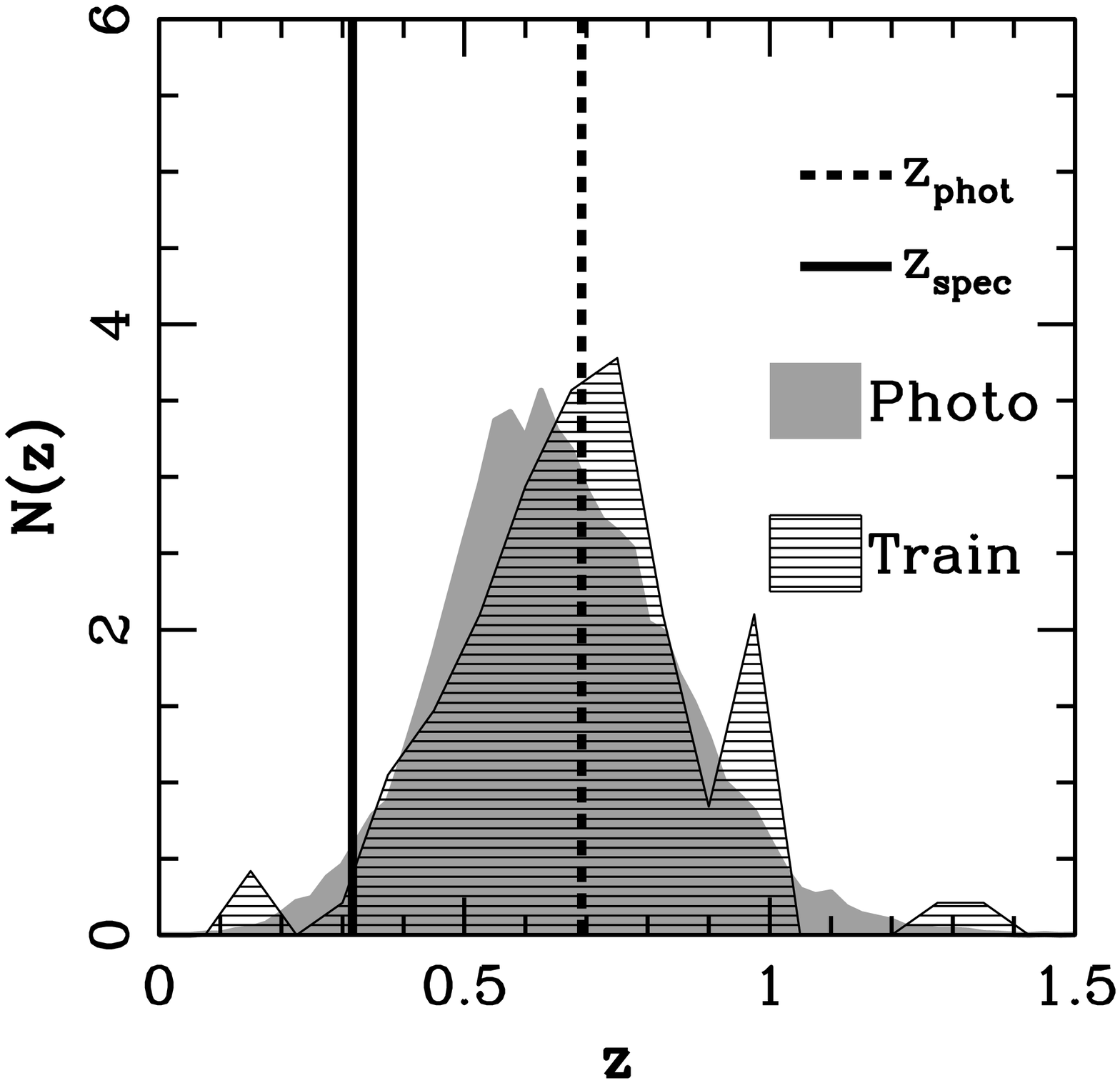}}
    \end{center}
  \end{minipage}
  \caption{Distributions of $p(z_{\rm true})$
({\it solid grey histograms}) and  $p(z_{\rm train})$ 
({\it hatched histograms}) 
for three training-set galaxies in the mock SDSS sample. 
The vertical solid (dashed) lines indicate \zspec (\zphot) 
for each galaxy. {\it Left:} an early-type galaxy at $z=0.48$; 
{\it middle:} a late-type galaxy at $z=0.56$; {\it right:} a 
faint, early-type galaxy at $z=0.31$.} \label{fig.pz}
\end{figure*}

In Figure \ref{fig.pz} we show the redshift distributions for three galaxies.
In each panel, $p(z_{\rm true})$ is shown as a grey histogram with 60 bins, and 
$p(z_{\rm train})$ is shown as the hatched histogram with 20 bins. 
We have rescaled the histograms by multiplying each by the 
width of the histogram bin for easier comparison of the distributions.
The {\it solid} vertical line indicates the true redshift of the galaxy and the {\it dashed}
vertical line indicates its ANN \zphot\ estimate. 
The {\it left} panel of the figure is for an early-type ($T=1.5$) galaxy with $r$-mag of 20.67 and $z_{\rm spec}=0.48$.
This galaxy has 4,006 neighbors in the photometric sample, i.e., 
that many photometric objects are as close to it in magnitude space 
as its 100 nearest training-set neighbors. 
In this example, 
the true redshift distribution of this region of observable space 
is narrow, $p(z_{\rm train})$ is a quite 
accurate estimate of $p(z_{\rm true})$, \zphot\ is very near {\zspec}, and both are at the peak of the $p(z)$ distributions. 

The {\it middle} panel shows the distributions 
for a late-type ($T=3.1$) galaxy with $r =21.1$ and $z_{\rm spec}=0.56$.
There were 14,606 neighbors to this galaxy in the photometric sample. 
With the exception of the extreme tails of the distribution,  
$p(z_{\rm train})$ provides an accurate  
estimate of $p(z_{\rm true})$. The redshift PDF 
$p(z_{\rm true})$ for this galaxy is much broader than that for 
the galaxy in the {\it left} panel, in part because 
the magnitudes of late-type galaxies do not correlate with
redshift as well as those of early types.
 The neural network \photoz\ is 0.39 for this object, higher than the 
peak of $p(z_{\rm train})$
at $z=0.3$ or its median at $z=0.34$.
The true redshift of this object, $z_{\rm spec}=0.56$, 
is far removed from the peak of its 
redshift distribution. However, 
the photo-z error, $z_{\rm phot}-z_{\rm spec} =0.16$, 
is comparable to the photo-z scatter 
at this redshift, $\sigma(z_{\rm spec}=0.56)\sim 0.13$ (see bottom right plot
of Fig. \ref{fig.bias.sig.variability}), 
which shows that this example is not atypical.
The broader $p(z_{\rm true})$ is, the more likely it 
is that \zspec\ will be far from the peak 
of the distribution. In that case, 
the \photoz estimator cannot zero in on the correct redshift, and a 
single-point \zphot\ estimate will be a poor redshift 
estimate for a large fraction of the objects in this region 
of observable space.

The {\it right} panel of Fig. \ref{fig.pz} shows the distributions for another 
early-type ($T=1.4$) galaxy with $r =21.8$ and $z_{\rm spec}=0.31$, with 18,366 neighbors
in the photometric set. 
This is the most pathological of the three examples. 
The large width of  $p(z_{\rm true})$ for this galaxy is due 
to its faintness, which results in large magnitude errors.  
The peaks of $p(z_{\rm train})$ and $p(z_{\rm true})$ are offset by $\sim 0.1-0.2$, and  $p(z_{\rm train})$ shows a spurious second 
peak at $z\sim 1$. 
Such fluctuations are not uncommon when one uses 100 galaxies to estimate $p(z)$. 
The true redshift of this galaxy is at the low-redshift tail of $p(z_{\rm true})$, and 
\zphot\ for this object is 
catastrophically wrong even though it is near the peak of $p(z_{\rm true})$.
The catastrophic error results from 
using a single number to represent a very broad distribution, 
and in this case the galaxy in question is quite different 
from most of its neighbors in magnitude space.
For a photometric survey, the redshift distribution is typically 
broad near the photometric 
limit of the survey.  
To avoid catastrophic errors and biases, 
one should work with the full redshift probability distribution per object.

\section{Application to SDSS DR6 Data} \label{sec:res.real}

Now that we have tested the weighting method on mock SDSS photometric 
samples, we apply it to the actual SDSS DR6 photometric sample.

\subsection{Bias and Scatter in SDSS {\Photoz}'s} 

\begin{figure*}
  \begin{minipage}[t]{85mm}
    \begin{center}
      \resizebox{85mm}{!}{\includegraphics[angle=0]{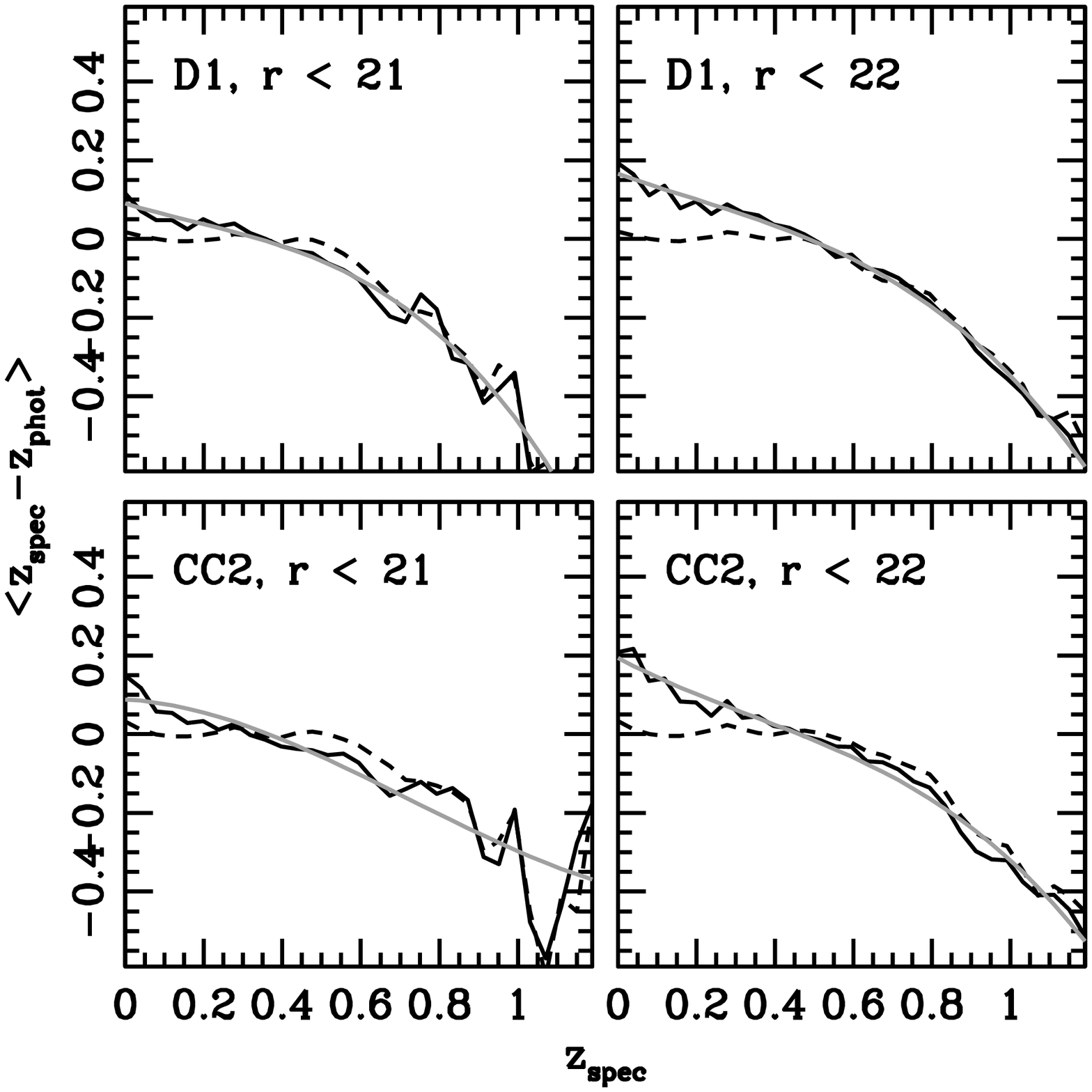}}
    \end{center}
  \end{minipage}
  \begin{minipage}[t]{85mm}
    \begin{center}
      \resizebox{85mm}{!}{\includegraphics[angle=0]{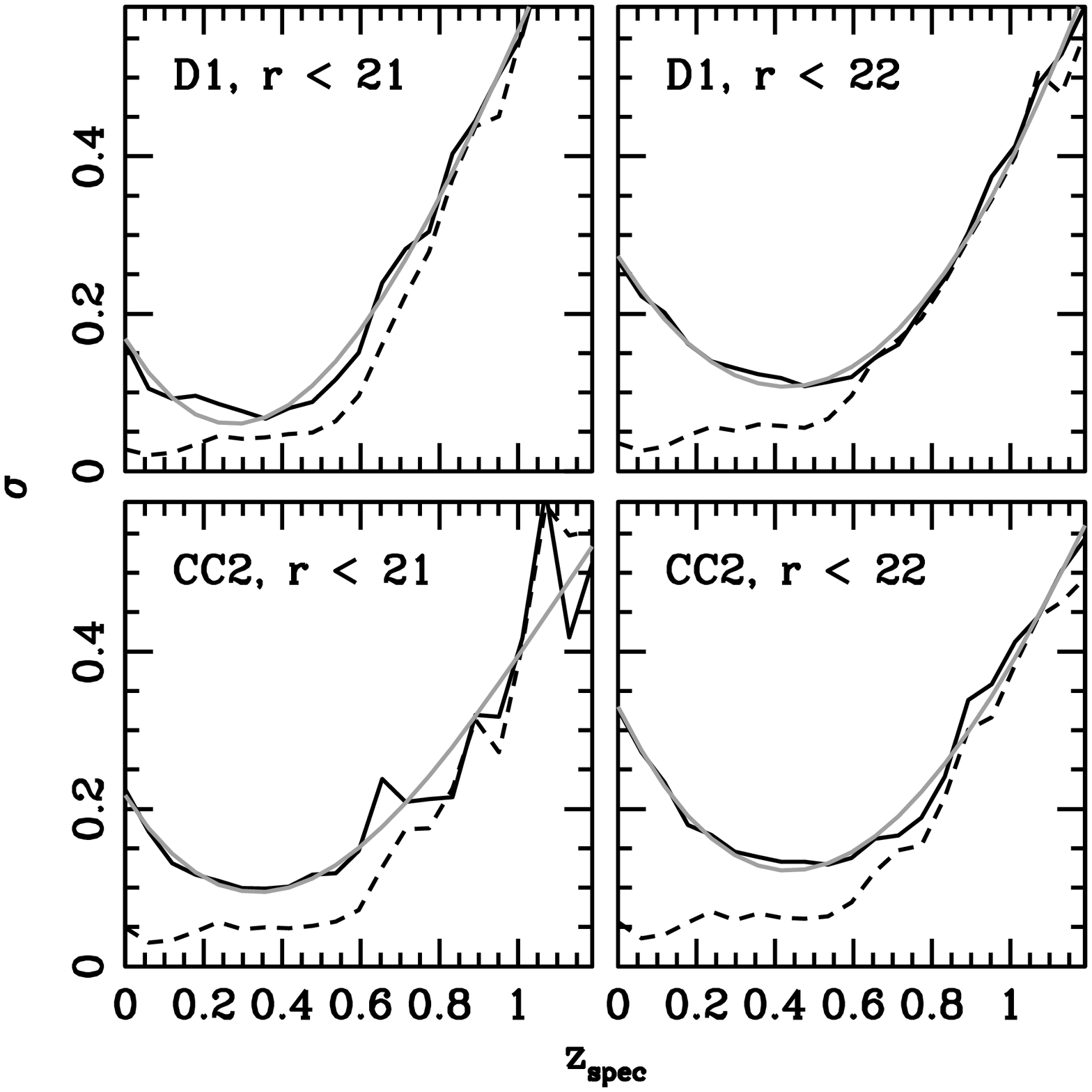}}
    \end{center}
  \end{minipage}
  \caption{{\it Left panels}:
Estimated {\photoz} bias vs. redshift for the weighted and unweighted training set of the 
SDSS DR6 catalog for four cases: 
({\it top left}) D1 photo-z's with $r<21$, 
({\it top right}) D1 photo-z's with $r<22$, 
({\it bottom left}) CC2 photo-z's with $r<21$, and 
({\it bottom right}) CC2 photo-z's with $r<22$.
{\it Right panels}: Estimated {\photoz} 
 scatter vs. redshift for the weighted and unweighted training 
set of the SDSS DR6 catalog for the same cases depicted in the left panels.
In each plot the dashed line corresponds to the unweighted result, 
the solid dark line to the weighted result, 
and the solid red line is a 3rd order polynomial fit to the weighted result.
The fit coefficients are given in Table~\ref{tbl:biassigfits}.
} \label{fig.bias.sig.wei.real}
\end{figure*}

\citet{oya08a} estimated {\photoz}'s for the SDSS DR6 photometric 
sample using an artificial neural network (see Appendix \ref{app:neu}) 
and several different combinations of photometric observables. One 
version, denoted there by D1, used as input observables the five magnitudes 
$ugriz$ and five concentration indices, also splitting the training set and 
the photometric sample into 5 bins of $r$ magnitude and performing 
separate ANN fits in each bin. Version CC2 used as inputs the 
four colors $u-g$, $g-r$, $r-i$, $i-z$, plus the concentration indices in 
$g$, $r$, and $i$. Here, as in \S \ref{sec:biasig.mock}, we use the 
weighting method to obtain improved estimates of the bias and scatter 
of these {\photoz} estimates. 
Figure \ref{fig.bias.sig.wei.real} shows the weighted and unweighted $b(z)$ and 
$\sigma(z)$ estimates derived from the training set, along with third-order polynomial
fits to the weighted estimates. 
The polynomial fit coefficients are given in Table~\ref{tbl:biassigfits}.
The differences between the weighted and unweighted  $b(z)$ and 
$\sigma(z)$ curves are qualitatively consistent with the results on the 
mock sample (Fig. \ref{fig.bias.sig.wei}), but the real data have larger
scatter and bias than the mocks.

\begin{table*}
\caption{Fit coefficients to the weighted estimates of {\photoz} bias and scatter 
vs. redshift for SDSS DR6 catalog}
\begin{center}
\leavevmode
\begin{tabular}{ l l l} \hline \hline
\multicolumn{3}{c}{D1 {\Photoz}'s}\\
\multicolumn{1}{c}{} & \multicolumn{1}{c}{$r<21$} & \multicolumn{1}{c}{$r<22$} \\
\hline
$b(z)$ &[0.0900269,-0.293255,0.262842,-0.523857]  & [0.16574,-0.35082,0.192806,-0.355683]\\
$\sigma(z)$ &[0.167949,-0.82395,1.69819,-0.484006] &[0.273305,-0.788055,0.951591,-0.0426683] \\
\hline
\multicolumn{3}{c}{CC2 {\Photoz}'s}\\
\hline
$b(z)$ &[0.0884344, -0.0574277, -0.607687, 0.279678] &[0.193711, -0.527042, 0.421479, -0.408717]\\
$\sigma(z)$ &[0.217213, -0.77692, 1.36055, -0.406967] &[0.329747, -1.0009, 1.31667, -0.262655] \\ 
\hline \hline
\multicolumn{3}{l}{}\\  
\multicolumn{3}{l}{All fits are 3rd order polynomials of the form $a_1+a_2z+a_3z^2+a_4z^3$}.\end{tabular}
\end{center}
\label{tbl:biassigfits}
\end{table*}

\subsection{The SDSS Redshift Distribution}\label{sec:sdssreal.dist}

\begin{figure*}
  \begin{minipage}[t]{85mm}
    \begin{center}
      \resizebox{85mm}{!}{\includegraphics[angle=0]{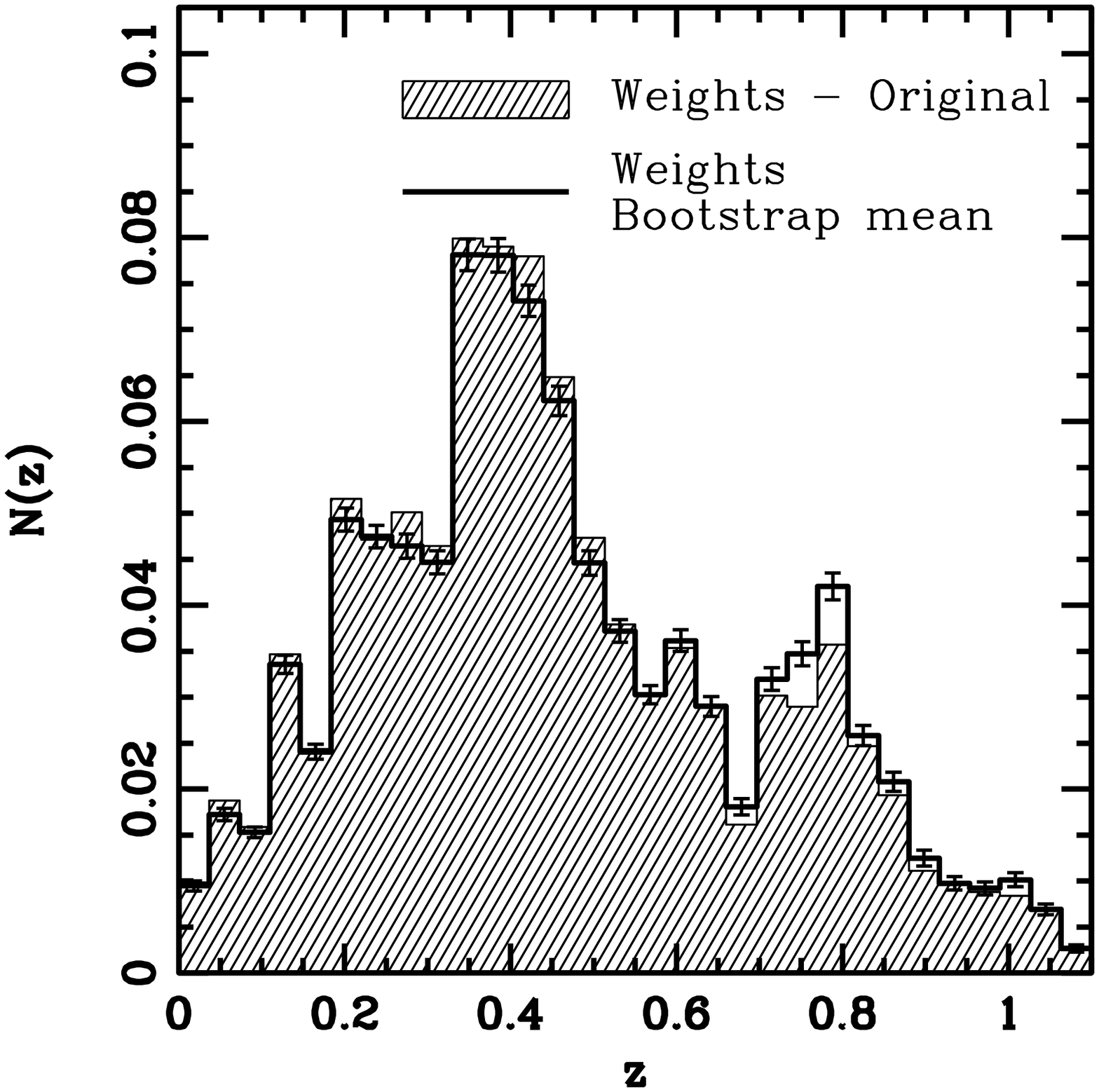}}
    \end{center}
  \end{minipage}
 \caption{
Estimated redshift distribution for the 
SDSS DR6 sample (with $r<22$),
computed using the weighting method ({\it hatched}) 
and the mean of the bootstrap samples ({\it solid line}).
The error bars are the diagonal bootstrap errors. 
}\label{fig.errs.real}
\end{figure*}

Fig. \ref{fig.errs.real}  shows the weighting method estimate, 
$N(z)_{\rm wei}$, 
for the redshift distribution of the SDSS DR6 photometric sample with $r<22$. 
The error bars on $N(z)_{\rm wei}$ are given by the 
square root of the diagonal elements of the covariance matrix 
obtained by the bootstrap resampling procedure 
described in \ref{subsubsec:errest}.

The coarse-grained structure of the redshift distribution is similar to 
that of the mock SDSS sample (Fig. \ref{dist.dndz}). However, the fine-grained  
structure shows peaks and dips that the study of 
\S \ref{sec:syst.ind} suggests are indications of systematic error. 
As noted there, large photometric errors, combined with sparseness of the 
training set, can lead to distortions of the inferred redshift distribution. 
This effect is likely present in the weighted estimate of the SDSS DR6 
redshift distribution for galaxies with $r<22$. 
The bump in $N(z)_{\rm wei}$ around $z =0.75$ is the result of the magnification
of the sampling errors in the training set caused by the lack of redshift
information in the photometry of faint galaxies, 
combined with the lack of training-set coverage in that redshift range.

\begin{figure*}
  \begin{minipage}[t]{85mm}
    \begin{center}
      \resizebox{85mm}{!}{\includegraphics[angle=0]{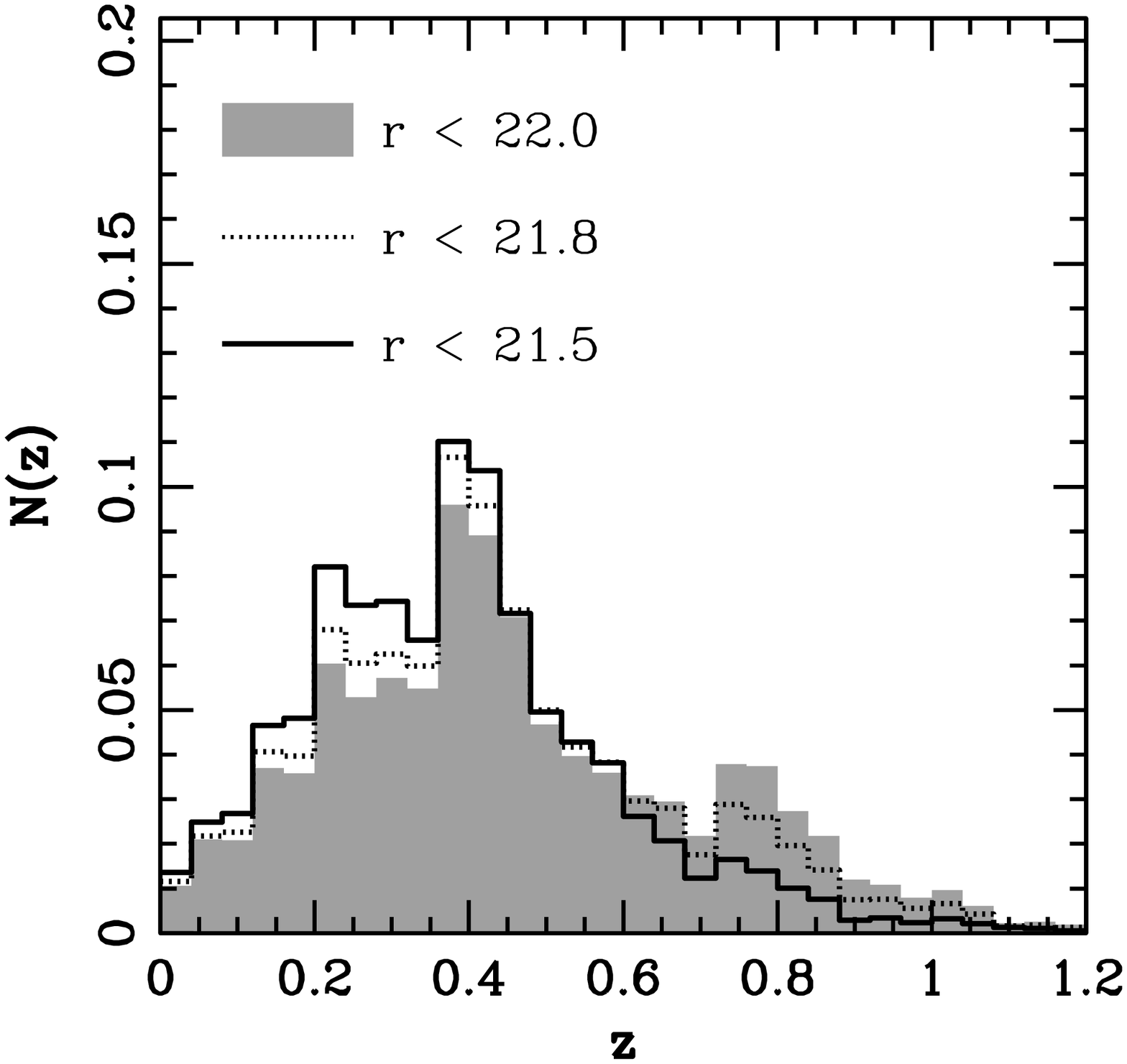}}
    \end{center}
  \end{minipage}
  \begin{minipage}[t]{85mm}
    \begin{center}
      \resizebox{85mm}{!}{\includegraphics[angle=0]{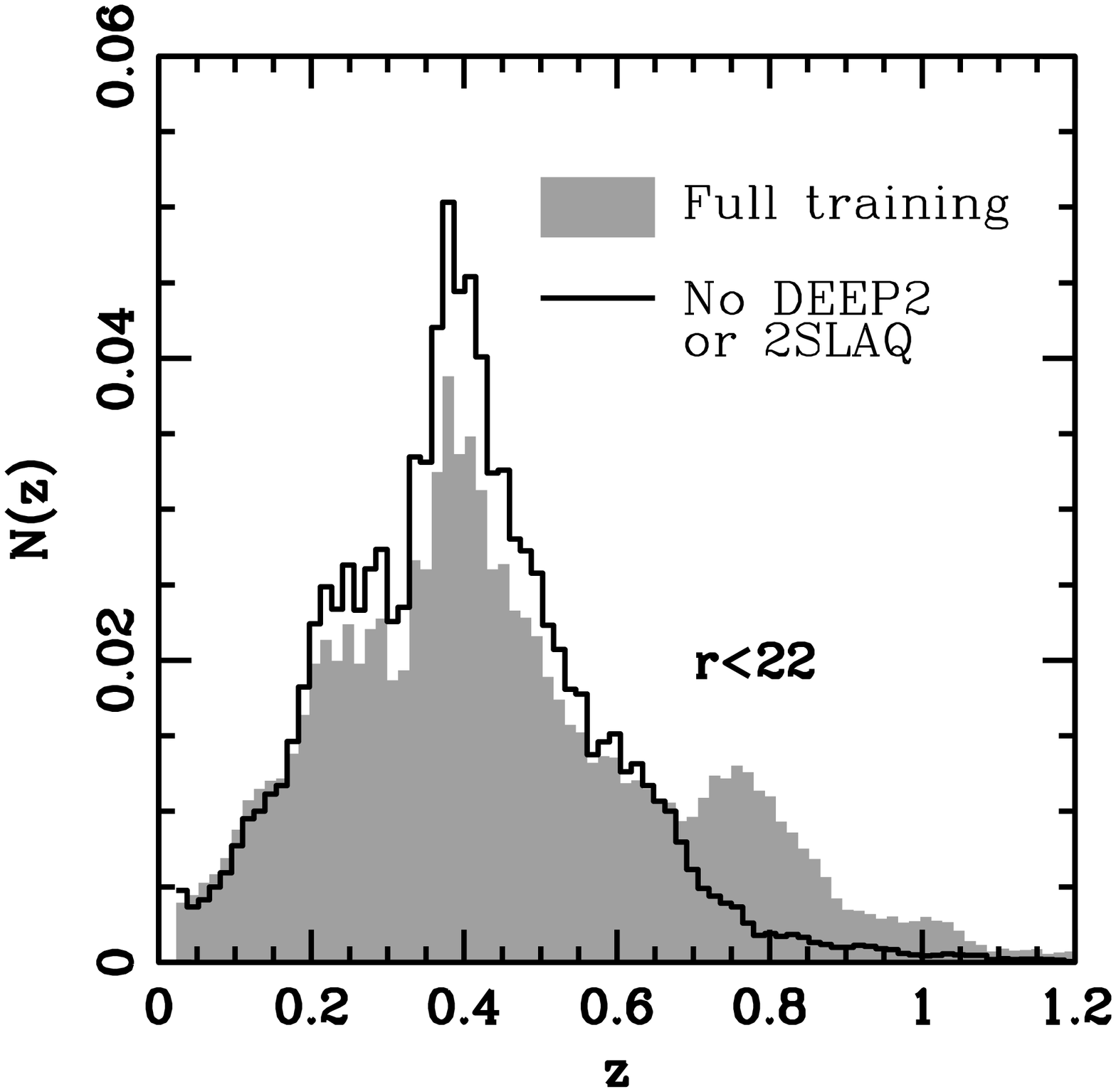}}
    \end{center}
  \end{minipage}
  \caption{({\it Left})Weighted estimates of the redshift distribution for the SDSS DR6 photometric sample, with $r<21.5$, $r<21.8$, and $r<22$. ({\it Right}) p(z) estimates of redshift distribution for SDSS DR7 photometric sample using the full training set as well as a training set without the DEEP2 (the non-EGS part) and without 2SLAQ.}\label{fig.dists.real}
  \end{figure*}

When we impose more stringent $r$-magnitude cuts, Fig. \ref{fig.dists.real} ({\it left}) shows 
that the feature disappears.
In Fig. \ref{fig.dists.real} ({\it right}) we show the $N(z)$ distribution estimated using
$p(z)$ using two different training sets.
In one case we use the full training set to estimate the $p(z)$'s while in the other we
remove all galaxies from DEEP/DEEP2 and 2SLAQ (totalling 84,568 galaxies) from the training
set and we add 6,069 from two approximately flux-limited samples, DEEP2-EGS \citep{dav07} and
zCOSMOS \citep{lil07}, which we describe in further detail in \S \ref{sec:pz.dr7}. 
The bump at $z =0.75$ disappears when DEEP/DEEP2 is not included, showing that the selection
in DEEP/DEEP2, which was done to target $z\sim0.7$ galaxies and in a different photometric
system from SDSS is responsible for the bump. 
The effects of 2SLAQ are much less pronounced, and consist in a small overall shift of the
distribution. 
2SLAQ has morphology cuts (in addition to the SDSS ugriz magnitude cuts) which could
have yielded some systematic biases.
As mentioned previously, the selection effects are amplified by the photometry errors, so
that the systematics are reduced if one imposes more stringent magnitude cuts.
If one is primarily interested in the overall redshift distribution, the $N(z)$ estimate 
using the training set without DEEP/DEEP2 or 2SLAQ is more reliable.
However, if one requires redshift information for individual galaxies, the estimate
using the full training set is still preferable. 
Without DEEP/DEEP2 and 2SLAQ, the training set is too sparse at faint magnitudes.
As a result, the individual $p(z)$ estimates are derived using training set objects 
spread out over a large region of observable space, which makes the $p(z)$'s poor 
representations of the {\it local} redshift distributions around the corresponding galaxies.
Given the training sets available, the best way to reduce the effects of selection 
issues while having reliable $p(z)$ estimates is to perform magnitude cuts.

We see another feature in  $N(z)_{\rm wei}$ in the range $0.2<z<0.4$ 
that does not go away with tighter 
$r$-mag cuts (see Fig. \ref{fig.dists.real}).
Similar features can be seen in the zCOSMOS+DEEP2/EGS redshift 
distribution used by \cite{man07} \citep[see the 
bottom right panel of Fig. 4 of][]{man07}, in the CNOC2 distribution used in our 
training set \citep[see Fig. 2 in][]{lim08}, and in the full CNOC2 sample shown in \cite{lin99}.
The feature in the DEEP2 data appears to be caused, at least partially, by 
spectroscopic failures affecting both early- and late-type galaxies in 
that redshift range (J. Newman, private communication), 
and it is possible that this is affecting the weighted estimate. 
In general, one should not expect that the redshift distribution of 
a sample flux-limited in one filter will be smooth, 
due to  k-correction-like effects. 
The complex shape of the spectral energy distributions of 
galaxies implies that a flux 
limit based on a single filter will preferentially select 
certain galaxy types at certain redshifts. 
We do not see such a feature in the mock SDSS catalog, because the mock was 
{\it created} with a smooth $r$-magnitude distribution and redshift distribution,
and we only applied a cut in the $r$-band.

\subsection{A p(z) Catalog for SDSS DR7}\label{sec:pz.dr7}

We calculated p(z)'s for the full SDSS DR7 sample satisfying the selection
cuts of the Photoz2 photometric redshift table described in Appendix \ref{sec:real.photo} 
and in \cite{oya08a} - a total of 78,135,961 galaxies.
We added a sample of 4,241 galaxies with spectra from zCOSMOS with quality flags 2.5, 3.4, 3.5, 4.4, 4.5, 9.3, 9.4, 9.5 \citep{lil07}
and 1,828 galaxies from DEEP2-EGS \citep{dav07} with $z_{\rm quality} \geq 3$  to the training set.
We do not use the ubercalibrated magnitudes \citep{pad08} available for DR7 because these
were not available for most of our training set galaxies.
The catalog is available from the SDSS DR7 value-added catalogs 
website\footnote{\tt http://www.sdss.org/dr7/products/value\_added/index.html}.
There are 240 files, ordered by $RA$, one for every $0.1$ hour of $RA$ from $0-23h$
Thus, the file named pofz.ra12h3.dat has photo-z's and $p(z)$'s for objects
with $12.3h \leq RA < 12.4h$, and so forth.
The p(z) values are tabulated for 100 redshift bins, centered
at $z = 0.03$ to $1.47$, with redshift spacing $dz = 1.44/99$.
To reduce effects of Poisson noise we adopt a 'moving window' smoothing technique.
Each entry for a given p(z) is calculated based on a bin of width $4*dz$.
As discussed in \S \ref{sec:sdssreal.dist}, the quality of the estimates degrades
rapidly for $r>21.5$.
We therefore recommend a cut in brightness of at least $r<21.8$.

\section{Discussion and Future Work} \label{sec:conclusion}
We have extended and applied the weighting technique of estimating redshift
distributions \citep{lim08}. 
The weighting procedure allows one to use a spectroscopic training set 
to accurately estimate the bias and scatter of {\photoz}'s as a function of 
redshift. In addition, the weighting method provides a natural, robust way
to select galaxies in the photometric sample that are well represented 
in the training set. Moreover, we have shown that 
the weighting technique provides a precise estimate of the redshift distribution 
of a photometric sample in the region of observable space where 
the training set and the photometric sample intersect. The estimate 
$N(z)_{\rm wei}$ more accurately estimates the redshift distribution for a 
photometric sample than methods based on {\photoz}'s. 
We have also extended the weighting method to estimate the redshift 
probability distribution function for individual galaxies, $p(z)$. Use of this 
PDF can substantially reduce biases associated with the use of single-point 
{\photoz}'s, and we recommend its use in the analysis of 
future photometric galaxy surveys.

We have outlined the potential different sources of error of the 
weights technique and we have demonstrated how to use information 
from the photo-z distribution to reduce systematic errors 
in the weights. 
We have shown that for the SDSS DR7, selection 
effects in the training set are the dominant source of error in 
the estimation of N(z), and that this systematic increases 
sharply with $r$-magnitude. 
In particular, we have found that the selection of the DEEP2 survey, 
which uses a different set of filters from the SDSS, is the dominant source of systematic 
errors. 

We have made public a catalog of $p(z)$ for $\sim 78$ million SDSS DR7 galaxies.
We have also provided fitting-functions for the weights-based estimates of the bias and scatter 
of {\photoz}'s as a function of redshift for the D1 and CC2 {\photoz's} of the SDSS DR6. 

For the future, investigations of the weighting method should include 
study of optimizing the weights estimation, e.g., with a variable number of 
nearest neighbors in different regions of observable space, and inclusion of 
systematic effects, e.g., associated with large-scale structure and 
spectroscopic failures, in the mock catalogs.

\smallskip

\noindent {\it Acknowledgements:}

We acknowledge useful conversations with Jeff Newman.
CC would like to thank Rachel Mandelbaum and Reiko Nakajima for 
extensive testing of the SDSS DR7 p(z) catalog.
This work was supported by the KICP 
 under NSF No. PHY-0114422 and NSF PHY-0551142,
by NSF grants AST-0239759, AST-0507666, 
and AST-0708154 at the University of Chicago,
by the DOE at the University of Chicago and Fermilab, and by DOE 
contract number DE-AC02-07CH11359. 

Funding for the SDSS and SDSS-II has been provided by the Alfred P. 
Sloan Foundation, the Participating Institutions, the National Science 
Foundation, the U.S. Department of Energy, the National Aeronautics and
 Space Administration, the Japanese Monbukagakusho, the Max Planck 
Society, and the Higher Education Funding Council for England. The 
SDSS Web Site is {\tt http://www.sdss.org/}.

The SDSS is managed by the Astrophysical Research Consortium for the
 Participating Institutions. The Participating Institutions are the
 American Museum of Natural History, Astrophysical Institute Potsdam,
 University of Basel, University of Cambridge, Case Western Reserve
 University, University of Chicago, Drexel University, Fermilab, the
 Institute for Advanced Study, the Japan Participation Group, Johns
 Hopkins University, the Joint Institute for Nuclear Astrophysics,
 the Kavli Institute for Particle Astrophysics and Cosmology, the 
Korean Scientist Group, the Chinese Academy of Sciences (LAMOST), 
Los Alamos National Laboratory, the Max-Planck-Institute for Astronomy 
(MPIA), the Max-Planck-Institute for Astrophysics (MPA), New Mexico 
State University, Ohio State University, University of Pittsburgh, 
University of Portsmouth, Princeton University, the United States 
Naval Observatory, and the University of Washington.

\appendix

\section{SDSS DR6 Data Sample}

\subsection{Photometric set}\label{sec:real.photo}
The Sloan Digital Sky Survey (SDSS) comprises a large-area 
imaging survey of the north Galactic cap, a multi-epoch imaging survey of  
an equatorial stripe in the south Galactic cap, and a spectroscopic survey of 
roughly $10^6$ galaxies and $10^5$ quasars 
\citep{yor00}. 
The survey used a dedicated, wide-field, 2.5m telescope \citep{gun06} at 
Apache Point Observatory, New Mexico. 
Imaging was carried out in drift-scan mode using a 142 mega-pixel camera 
\citep{gun06} that gathers data in five broad bands, $u g r i z$, spanning 
the range from 3,000 to 10,000 \AA \, \citep{fuk96}, with an effective exposure 
time of 54.1 seconds per band. 
The images were processed using specialized 
software \citep{lup01,sto02} and were 
astrometrically \citep{pie03} and photometrically \citep{hog01,tuc06} 
calibrated using observations of a set of primary standard stars 
\citep{smi02} observed on a neighboring 20-inch telescope.

The imaging in the sixth SDSS Data Release \citep[][hereafter DR6]{dr6} covers a nearly  
contiguous region of the north Galactic cap.
In any region where imaging runs overlap, one run was 
declared primary\footnote{For the precise definition of primary objects see 
{\tt http://cas.sdss.org/dr6/en/help/docs/glossary.asp\#P}} 
and was used for spectroscopic target selection; 
other runs were declared secondary. 
The area covered by the DR6 primary imaging survey, including the 
southern stripes, is $8520 \textrm{ deg}^2$, but  
DR6 includes both the primary and secondary observations of 
each area and source \citep{dr6}.

In this paper, we use a random $1\%$ subset of the SDSS DR6 Photoz2 catalog described 
in \cite{oya08a} as our photometric sample. 
The Photoz2 catalog contains  all primary objects from DR6 
(drawn from the SDSS CasJobs website\footnote{{\tt http://casjobs.sdss.org/casjobs/}}) 
that have the TYPE flag equal to $3$ (the type for galaxy) and that do not 
have any of the flags BRIGHT, SATURATED, SATUR\_CENTER, or 
NOPETRO\_BIG set. 
For the definitions of these flags we refer the reader to the 
PHOTO flags entry at the SDSS 
website\footnote{{\tt http://cas.sdss.org/dr6/en/help/browser/browser.asp}}.
The full Photoz2 photometric sample comprises $77,418,767$ galaxies. 
The $r$ magnitude, $g-r$, and $r-i$ color distributions 
are shown in the bottom panels of Figs.~\ref{dist.rmag}a and ~\ref{dist.col}a.

\subsection{Spectroscopic training samples}\label{sec:real.train}

As noted in the text, the spectroscopic training sample we use for SDSS DR6 is drawn 
from a number of spectroscopic galaxy catalogs that overlap 
with SDSS imaging. 
Each survey providing spectroscopic redshifts defines a redshift 
quality indicator; we refer the reader to the respective publications listed 
below for their precise definitions.
For each survey, we chose a redshift quality cut roughly corresponding
to 90\% redshift confidence or greater. 
The SDSS spectroscopic sample 
provides $531,672$ redshifts, principally from the MAIN and 
Luminous Red Galaxy (LRG) samples, with confidence level 
$z_{\rm conf} > 0.9$. The remaining redshifts are:
$21,123$ from the Canadian Network for Observational Cosmology (CNOC) 
Field Galaxy Survey \cite[CNOC2;][]{yee00},
$1,830$ from the Canada-France Redshift Survey \cite[CFRS;][]{lil95}
with Class $> 1$,  
$31,716$ from the Deep Extragalactic Evolutionary Probe \cite[DEEP;][]{deep2}
with $q_z$ =  A or B and from DEEP2  
\citep{wei05}\footnote{{\tt http://deep.berkeley.edu/DR2/ }}
with $z_{\rm quality} \geq 3$, 
$728$ from the Team Keck Redshift Survey \cite[TKRS;][]{wir04}
with $z_{\rm quality} > -1$, and 
$52,842$ LRGs from the 
2dF-SDSS LRG and QSO Survey 
\cite[2SLAQ;][]{can06}\footnote{{\tt http://lrg.physics.uq.edu.au/New\_dataset2/ }}
with $z_{\rm op} \geq 3$.

We positionally matched the galaxies with spectroscopic redshifts against photometric 
data in the SDSS {\tt BestRuns} CAS database, which allowed us
to match with photometric measurements in different SDSS imaging runs.
The above numbers for galaxies with redshifts count independent photometric 
measurements of the same objects due to multiple SDSS imaging of the same 
region; in particular SDSS Stripe 82 has been imaged a number of times.  
The numbers of {\em unique} galaxies used from these surveys are 
$1,435$ from CNOC2, 
$272$ from CFRS, 
$6,049$ from DEEP and DEEP2,  
$389$ from TKRS, and  
$11,426$ from 2SLAQ.
The SDSS spectroscopic samples were drawn from the SDSS primary galaxy sample 
and therefore are all unique.

\section{SDSS DR6 Mock catalog}

\begin{table*}
\caption{Schechter luminosity function parameters \citep{zuc06}  
used to derive polynomial fits to the relationships between the Schechter luminosity 
function parameters, redshift, and galaxy spectral type. 
The parameters in \citet{zuc06} were derived using the $B$-band of the VVDS survey; here 
we use them to generate the $r$-band magnitude distributions, using the
appropriate k-corrections by galaxy type \citep{bla05}}.
\begin{center}
  \leavevmode
  \begin{tabular}{c c c c c } \hline \hline
\multicolumn{1}{c}{Type} & \multicolumn{1}{c}{z-bin} & \multicolumn{1}{c}{$\alpha$} & \multicolumn{1}{c}{$M^{*}_{\rm AB}-5log(h)$}
& \multicolumn{1}{c}{$\phi^{*}(10^{-3}h^3Mpc^{-3})$}\\
\hline
$1$ & 0.0 - 0.2 & $-0.15^{+0.30}_{-0.30}$ & $-20.00^{+0.30}_{-0.30}$ & $6.15^{+0.70}_{-0.70}$ \\
$1$ & 0.2 - 0.4 & $-0.04^{+0.28}_{-0.27}$ & $-20.27^{+0.27}_{-0.31}$ & $5.15^{+0.64}_{-0.64}$ \\ 
$1$ & 0.4 - 0.6 & $-0.40^{+0.20}_{-0.20}$ & $-20.49^{+0.17}_{-0.18}$ & $3.12^{+0.30}_{-0.30}$ \\ 
$1$ & 0.6 - 0.8 & $-0.22^{+0.17}_{-0.17}$ & $-20.22^{+0.09}_{-0.10}$ & $3.53^{+0.25}_{-0.25}$ \\ 
$1$ & 0.8 - 1.0 & $-0.01^{+0.25}_{-0.24}$ & $-20.73^{+0.11}_{-0.12}$ & $2.36^{+0.18}_{-0.18}$ \\ 
$1$ & 1.0 - 1.2 & $-1.23^{+0.34}_{-0.34}$ & $-20.53^{+0.11}_{-0.12}$ & $2.39^{+0.22}_{-0.22}$ \\ 
$1$ & 1.2 - 1.5 & $-1.30^{+0.40}_{-0.40}$ & $-20.50^{+0.30}_{-0.30}$ & $2.3 ^{+0.30}_{-0.30}$ \\
\hline
$2$ & 0.0 - 0.2& $-0.60^{+0.20}_{-0.20}$ & $-20.00^{+0.20}_{-0.20}$ &   $7.60^{+0.90}_{-0.90}$ \\ 
$2$ & 0.2 - 0.4 & $-0.67^{+0.13}_{-0.13}$ & $-20.13^{+0.19}_{-0.21}$ &   $6.50^{+0.56}_{-0.56}$ \\   
$2$ & 0.4 - 0.6& $-0.50^{+0.15}_{-0.14}$ & $-19.97^{+0.12}_{-0.12}$ &   $4.35^{+0.31}_{-0.31}$ \\ 
$2$ & 0.6 - 0.8& $-0.57^{+0.13}_{-0.13}$ & $-20.39^{+0.09}_{-0.10}$ &   $4.58^{+0.26}_{-0.26}$ \\ 
$2$ & 0.8 - 1.0& $-0.60^{+0.20}_{-0.20}$ & $-20.55^{+0.10}_{-0.11}$ &   $3.54^{+0.22}_{-0.22}$ \\ 
$2$ & 1.0 - 1.2& $-0.76^{+0.34}_{-0.33}$ & $-20.77^{+0.12}_{-0.13}$ &   $3.01^{+0.23}_{-0.23}$ \\ 
$2$ & 1.2 - 1.5& $-1.57^{+0.61}_{-0.62}$ & $-20.82^{+0.13}_{-0.14}$ &   $2.19^{+0.22}_{-0.22}$ \\ 
\hline
$3$ & 0.0 - 0.2&  $-0.80^{+0.30}_{-0.30}$ & $-19.00^{+0.60}_{-0.60}$ &  $10.0^{+0.60}_{-0.60}$  \\
$3$ & 0.2 - 0.4&  $-0.84^{+0.10}_{-0.10}$ & $-19.14^{+0.12}_{-0.13}$ &  $9.82^{+0.54}_{-0.54}$  \\
$3$ & 0.4 - 0.6&  $-1.07^{+0.10}_{-0.10}$ & $-20.04^{+0.11}_{-0.11}$ &  $6.31^{+0.30}_{-0.30}$  \\
$3$ & 0.6 - 0.8&  $-0.79^{+0.13}_{-0.13}$ & $-20.10^{+0.09}_{-0.09}$ &  $7.11^{+0.29}_{-0.29}$  \\
$3$ & 0.8 - 1.0&  $-0.87^{+0.15}_{-0.15}$ & $-20.33^{+0.08}_{-0.08}$ &  $6.27^{+0.27}_{-0.27}$  \\
$3$ & 1.0 - 1.2&  $-1.39^{+0.26}_{-0.26}$ & $-20.38^{+0.10}_{-0.10}$ &  $5.57^{+0.33}_{-0.33}$  \\
$3$ & 1.2 - 1.5&  $-1.86^{+0.55}_{-0.59}$ & $-20.81^{+0.12}_{-0.13}$ &  $3.67^{+0.27}_{-0.27}$  \\
\hline
$4$ & 0.0 - 0.2&  $-1.55^{+0.20}_{-0.20}$ &   $-19.60^{+0.40}_{-0.40}$ &   $2.60^{+0.40}_{-0.40}$  \\
$4$ & 0.2 - 0.4&  $-1.59^{+0.11}_{-0.12}$ &   $-19.73^{+0.29}_{-0.33}$ &   $2.59^{+0.13}_{-0.13}$  \\
$4$ & 0.4 - 0.6&  $-1.53^{+0.18}_{-0.19}$ &   $-19.38^{+0.17}_{-0.18}$ &   $4.10^{+0.19}_{-0.19}$  \\
$4$ & 0.6 - 0.8&  $-1.35^{+0.15}_{-0.15}$ &   $-19.95^{+0.12}_{-0.12}$ &   $4.07^{+0.16}_{-0.16}$  \\
$4$ & 0.8 - 1.0&  $-1.68^{+0.20}_{-0.21}$ &   $-20.10^{+0.12}_{-0.12}$ &   $4.72^{+0.20}_{-0.20}$  \\
$4$ & 1.0 - 1.2&  $-1.99^{+0.33}_{-0.34}$ &   $-20.19^{+0.12}_{-0.12}$ &   $6.95^{+0.36}_{-0.36}$  \\
$4$ & 1.2 - 1.5&  $-2.50^{+0.52}_{-0.91}$ &   $-20.53^{+0.12}_{-0.12}$ &   $4.34^{+0.32}_{-0.32}$  \\
\hline
\end{tabular}
\end{center}
\label{tbl:lumparam}
\end{table*}

Using spectral template libraries and observational data on the redshift-dependent 
luminosity functions of galaxies of different types, 
we have constructed mock photometric and spectroscopic samples that 
reproduce the main features of the real SDSS DR6 samples.
In particular, 
we fit simple polynomial functions to the Schechter parameters of \cite{zuc06} to derive a 
continuous relationship between the Schechter parameters
$M^*, \alpha, \phi^*$, redshift $z$, and galaxy type $T$, using 
the centroid of each redshift bin for the fit.
To regularize the fits, we visually extrapolate the results of \cite{zuc06} to 
the $z=(0, 0.2)$ bin and, where needed, for the $(1.2, 1.5)$ bin. 

The Schecter luminosity function is defined as
\begin{eqnarray}
\phi(M)dM=\frac{2}{5}\phi^{*}(\ln 10)\left [ 10^{\frac{2}{5}(M^*-M)}\right ]^{\alpha+1} \nonumber \\  
\times \exp \left[-10^{\frac{2}{5}(M^*-M)}\right ]dM,
\end{eqnarray}
\noindent where $\phi(M)dM$ is the number of galaxies with absolute magnitudes between $M$ and $M+dM$.

The Schechter parameters we use are shown in Table~\ref{tbl:lumparam}.
The polynomials we derive are: 
\begin{eqnarray}
\alpha&=&b_1T^2 + b_2Tz +b_3z  + b_4z^2 + b_5 \\
M^{*}&=&c_1T^2 + c_2Tz+ c_3z  + c_4z^2 + c_5 \\
\phi^{*}&=&d_1T^2 + d_2Tz+ d_3z  + d_4z^2 + d_5 \nonumber \\
&& + d_6T^2z + d_7T^3
\end{eqnarray}
We find the best-fit coefficients to be:
\begin{eqnarray}
\mathbf{b}&=&[-0.087,0.050,0.998,-1.143,-0.383], \nonumber \\
\mathbf{c}&=&[0.068,-0.202,-0.806,0.227,-19.86], \nonumber \\
\mathbf{d}&=&[2.04,-5.20,-0.636,0.910,4.181,1.417,-0.536] .\nonumber
\end{eqnarray}

\section{Artificial Neural Network Photo-z's} \label{app:neu}

For comparison with the weighting method, we use an 
Artificial Neural Network (ANN) method to estimate photometric redshifts \citep{col04, 
oya08a}
We use a particular type of ANN called a Feed Forward Multilayer
Perceptron (FFMP), which 
consists of several nodes arranged in layers through which 
signals propagate sequentially. 
The first layer, called the input layer, receives the input photometric 
observables (magnitudes, colors, etc.). 
The next layers, denoted hidden layers, propagate signals until 
the output layer, whose outputs are the desired quantities, in this
case the photo-z estimate. 
Following the notation of \cite{col04}, we denote a network with 
$k$ layers and $N_i$ nodes in the $i^{th}$ layer as $N_1:N_2: ... :N_k$.

A given node can be specified by the layer it belongs to and the 
position it occupies in the layer. Consider a node in layer $i$ and 
position $\alpha$  with $\alpha=1,2,...,N_i$. 
This node, denoted $P_{i\alpha}$, receives
a total input $I_{i\alpha}$ and fires an output $O_{i\alpha}$ given by
\begin{eqnarray}
O_{i\alpha}=F(I_{i\alpha}) \,,
\end{eqnarray}  
where $F(x)$ is the activation function. 
The photometric observables are the inputs $I_{1\alpha}$ to the 
first layer nodes, which produce outputs $O_{1\alpha}$. 
The outputs $O_{i\alpha}$ in layer $i$ are
propagated to nodes in the next layer $(i+1)$, denoted $P_{(i+1)\beta}$,
with $\beta=1,2,..N_{i+1}$. 
The total input $I_{(i+1)\beta}$ is a weighted sum of the outputs 
$O_{i\alpha}$
\begin{eqnarray}
I_{(i+1)\beta} = \sum_{\alpha=1}^{N_i} w_{i\alpha\beta} O_{i\alpha},
\end{eqnarray}
where $w_{i\alpha\beta}$ is the weight that connects nodes 
$P_{i\alpha}$ and $P_{(i+1)\beta}$.
Iterating the process in layer $i+1$, signals propagate from hidden layer 
to hidden layer until the output layer.
There are various choices for the activation function $F(x)$ such as:
a sigmoid, a hyperbolic tangent, a step function, a linear function, etc.
The choice of the activation function typically has no important effect
on the final photo-z's, and different activation functions can be used
in different layers. 
In our implementation, we use a network configuration $N_m:15:15:15:1$, 
which receives $N_m$ magnitudes and outputs a photo-z. 
We use hyperbolic tangent activation  functions in the hidden layers and a 
linear activation function for the output layer.

\smallskip

\bibliographystyle{mn2e}
\bibliography{dist2}

\end{document}